\documentclass[reprint,superscriptaddress,amsmath,amssymb, aps, prx, nofootinbib]{revtex4-2}

\usepackage{graphicx}
\usepackage{booktabs}
\usepackage[utf8]{inputenc}
\usepackage{physics}
\usepackage{multirow}
\usepackage{placeins}

\usepackage{siunitx}[=v2]
\DeclareSIUnit{\gauss}{G}
\DeclareSIUnit{\mf}{m_F}
\usepackage{hyperref}

\usepackage{xcolor}

\begin{document}

\title{Spin-resolved microscopy of \textsuperscript{87}Sr \\ SU(\textit{N}) Fermi-Hubbard systems}

\author{Carlos Gas-Ferrer}%
\thanks{C.G. and A.R. contributed equally to this work}
\affiliation{ICFO - Institut de Ciencies Fotoniques, The Barcelona Institute of Science and Technology, 08860 Castelldefels (Barcelona), Spain}%
\author{Antonio Rubio-Abadal}%
\altaffiliation{Present address: Departament de Física Quàntica i Astrofísica, Facultat de Física, Universitat de Barcelona, 08028 Barcelona, Spain}
\affiliation{ICFO - Institut de Ciencies Fotoniques, The Barcelona Institute of Science and Technology, 08860 Castelldefels (Barcelona), Spain}%
\author{Sandra Buob}%
\altaffiliation{Present address: Ideaded, Carrer de la Tecnologia, 35, 08840 Viladecans (Barcelona), Spain.}
\affiliation{ICFO - Institut de Ciencies Fotoniques, The Barcelona Institute of Science and Technology, 08860 Castelldefels (Barcelona), Spain}%
\author{Leonardo Bezzo}%
\affiliation{ICFO - Institut de Ciencies Fotoniques, The Barcelona Institute of Science and Technology, 08860 Castelldefels (Barcelona), Spain}%
\author{Jonatan Höschele}%
\altaffiliation{Present address: Ideaded, Carrer de la Tecnologia, 35, 08840 Viladecans (Barcelona), Spain.}
\affiliation{ICFO - Institut de Ciencies Fotoniques, The Barcelona Institute of Science and Technology, 08860 Castelldefels (Barcelona), Spain}%
\author{Leticia Tarruell}%
 \email{Contact author: leticia.tarruell@icfo.eu}%
\affiliation{ICFO - Institut de Ciencies Fotoniques, The Barcelona Institute of Science and Technology, 08860 Castelldefels (Barcelona), Spain}%
\affiliation{ICREA, Pg. Llu\'{i}s Companys 23, 08010 Barcelona, Spain}

\date{\today}

\begin{abstract}
Quantum-gas microscopes provide direct access to the phases of the Hubbard model, bringing microscopic insight into the complex competition between interactions, SU(2) magnetism, and doping. Alkaline-earth(-like) fermions extend this spin-1/2 paradigm by realizing higher symmetries and giving access to SU($N$) Hubbard models, with rich phase diagrams to be unveiled. Despite its fundamental interest, a microscopic exploration of SU($N$) quantum systems has remained elusive. Here we report the realization of a quantum-gas microscope for fermionic \textsuperscript{87}Sr. Our imaging scheme, based on cooling and fluorescence on the narrow intercombination line at \SI{689}{nm}, enables spin-resolved single-atom detection. By implementing a spin-selective optical pumping protocol, we determine the occupation of each of the 10 spin states in a single experimental realization, a crucial capability for probing site-resolved magnetic correlations. We benchmark our method by observing single-particle Larmor precession across the full spin-9/2 ground-state manifold and use numerical simulations of SU($N$) Fermi-Hubbard chains, incorporating the measured detection fidelities, to demonstrate its capability to resolve many-body spin correlations. Our results establish \textsuperscript{87}Sr quantum-gas microscopy as a powerful approach to study exotic magnetism in the SU($N$) Fermi-Hubbard model, and provide a new detection tool for studies in quantum simulation, computation, and metrology.
\end{abstract}
\maketitle

\section{\label{sec:introduction}Introduction}
Spin degrees of freedom are fundamental to quantum science, serving as resources for information processing, building blocks of quantum magnetism, and sensitive tools in quantum metrology. In neutral-atom platforms, microscopic control of individual spins has significantly pushed the boundaries of quantum simulation~\cite{Browaeys2020, gross_quantum_2021}. Quantum-gas microscopes, for instance, have used two-component ground-state mixtures to investigate quantum magnetism in the Fermi-Hubbard~\cite{parsons_siteresolved_2016,  cheuk_observation_2016a, boll_spin_2016} and isotropic Heisenberg models~\cite{fukuhara_quantum_2013, wei_quantum_2022}. Another example is optical-tweezer arrays, which exploit Rydberg-mediated interactions between electronic states to realize quantum Ising~\cite{Labuhn2016Tunable, Bernien2017Probing} and XY models~\cite{deLeseleuc2019Observation}, among others. A key aspect of both platforms is the detection of individual atoms, which often cannot resolve spin states in a single experimental realization. Instead, experiments frequently extract spin information indirectly, for example by interpreting the loss of a particular state as an indicator. One popular approach to retrieve the full spin information relies on spin-to-position mapping techniques, either with magnetic-field gradients~\cite{boll_spin_2016, Koepsell2020, Yan2022} or state-dependent potentials~\cite{Wu_SternGerlach_Qubit_2019, bluvstein2025architect}.

So far, microscopic studies of Hubbard and quantum spin models have mostly focused on spin-1/2 systems. The realization and detection of spin models with more components face several challenges. The first one is the single-atom detection of the different spin states, since methods based on spin separation are hard to scale to atoms with larger spin. The second challenge is the stability of the spin populations in multicomponent mixtures, which are commonly subjected to spin-exchange collisions. Such collisions lead to rich spin dynamics, investigated for various species in optical lattices~\cite{Widera2005, Krauser2012, dePaz2013, Patcheider2020}, but usually lead to leakage from the original spin populations. An exception for alkali atoms is $^6\text{Li}$, for which spin-exchange collision rates can remain small in certain regimes~\cite{Ottenstein2008, Huckans2009}, as shown in the recent realization of a 3-component Hubbard model~\cite{mongkolkiattichai2025}.

Alkaline-earth(-like) atoms overcome these limitations thanks to their fermionic isotopes with large nuclear spin $I$, which provide unique opportunities arising from their distinctive internal structure. Firstly, they support narrow and ultranarrow optical transitions, which can be used to individually address the spin states within the ground-state manifold. These transitions have made them the leading choices for quantum metrology with optical atomic clocks~\cite{bothwell_resolving_2022}. Furthermore, they are strong candidates for neutral-atom quantum computing~\cite{daley_quantum_2008, barnes_assembly_2022, Huie2023, Jenkins2022, Ma2022, Norcia2023}, supporting novel paradigms such as fermionic~\cite{GonzalezCuadra2023, Zache2023} and qudit~\cite{Omanakuttan2021, Ahmed2025} quantum processing. Secondly, their 2-electron singlet ground state effectively decouples the nuclear spin from the electronic angular momentum. This leads to an SU($N$) interaction symmetry, with the $N$ spin states interacting with identical strength~\cite{Hofstetter2004, cazalilla_ultracold_2014, Ibarra-Garcia-Padilla_2025}. As a result, collisional spin relaxation processes are suppressed~\cite{StellmerDetection2011}. These properties open the door to the study of exotic models of quantum magnetism. In particular, several theoretical studies have explored the SU($N$) Heisenberg model~\cite{hermele_mott_2009, Toth2010,corboz_simultaneous_2011, Nataf2014, Nataf2016, Romen2020}, identifying a rich variety of magnetic phases of matter. These include phases with spontaneous dimerization~\cite{corboz_simultaneous_2011}, plaquette order~\cite{Nataf2016}, and chiral spin liquids for large $N$~\cite{hermele_mott_2009}. Recent works have also focused on models with itinerance, leading to an even richer phenomenology emerging from the competition between motion and magnetism. For example, zig-zag antiferromagnets and alternating long-range order have been found near the ground state of the SU(3) Fermi-Hubbard model~\cite{Feng2023, Bird2025, kleijweg_corboz_2025}, and the emergence of magnetic polarons was investigated in the doped SU(3) $t$-$J$ model~\cite{schlomer_tJ_SU(3)}.

To date, ultracold SU($N$) fermions with large $N$ have been experimentally realized mostly with \textsuperscript{173}Yb ($I=5/2$ with $N=6$)~\cite{FukuharaYb} and \textsuperscript{87}Sr ($I=9/2$ with $N=10$)~\cite{desalvo_degenerate_2010}. Experiments with bulk SU($N$) gases have explored systems with different dimensionalities and studied aspects such as their spectroscopic response~\cite{Zhang2014SrSUN}, the emergence of bosonization for large $N$~\cite{Pagano2014OneDimensional,Song2020},  or their thermodynamical properties~\cite{sonderhouse_thermodynamics_2020}. By loading the clouds into three-dimensional optical lattices, the SU($N$) Fermi-Hubbard model has been realized~\cite{taie_su_2012}. It has allowed the implementation of synthetic gauge fields using the nuclear spin states of the ground-state manifold as a synthetic dimension~\cite{mancini_observation_2015, han_band_2019}, the measurement of the equation of state of the system across the Mott crossover~\cite{hofrichter_direct_2016, pasqualetti_equation_2023}, and the observation of flavor-selective localization when breaking the SU($N$) symmetry~\cite{Tusi2022}. An exciting frontier for these systems is the detection of SU($N$) quantum magnetism. Recently, first signals of nearest-neighbour correlations have been observed in an SU(6) Hubbard system using \textsuperscript{173}Yb ~\cite{taie_observation_2022}. However, all experimental studies so far have relied exclusively on global measurements, since alkaline-earth(-like) quantum-gas microscopes have only been realized with bosonic isotopes~\cite{miranda_siteresolved_2015, yamamoto_ytterbium_2016, buob_strontium_2024}. Given that microscopic observables have been key to reaching low entropies~\cite{ChiuEngineering2018} and detecting long-range magnetism~\cite{mazurenko_coldatom_2017} in the SU(2) Hubbard model, an SU($N$) Fermi-gas microscope is highly desirable to unveil the exotic SU($N$) quantum magnetic phases.

In this work, we demonstrate quantum-gas microscopy of a fermionic alkaline-earth species, \textsuperscript{87}Sr. The backbone of our system is the narrow intercombination line, which we use for both cooling and imaging, similarly to recent works with fermionic \textsuperscript{171}Yb and bosonic \textsuperscript{88}Sr in optical tweezers~\cite{urech_narrowline_2022, Huie2023, Norcia2023}. Furthermore, we introduce spin-resolved optical pumping and demonstrate an imaging protocol which enables the sequential and independent site-resolved detection of all 10 spin states of strontium in a single experimental run. We experimentally benchmark our imaging protocol through single-particle coherent Larmor precession and numerically show that it will enable the measurement of spin correlations in SU($N$) Fermi-Hubbard systems. Our method could be applied to other atomic species featuring narrow-linewidth transitions and offers a versatile detection method for quantum computing schemes based on the nuclear spin of \textsuperscript{87}Sr~\cite{GonzalezCuadra2023, Omanakuttan2021}.

\FloatBarrier
\section{\label{sec:sec2}Experimental setup and narrow-line imaging}

\begin{figure}
    \centering
    \includegraphics[width=\linewidth]{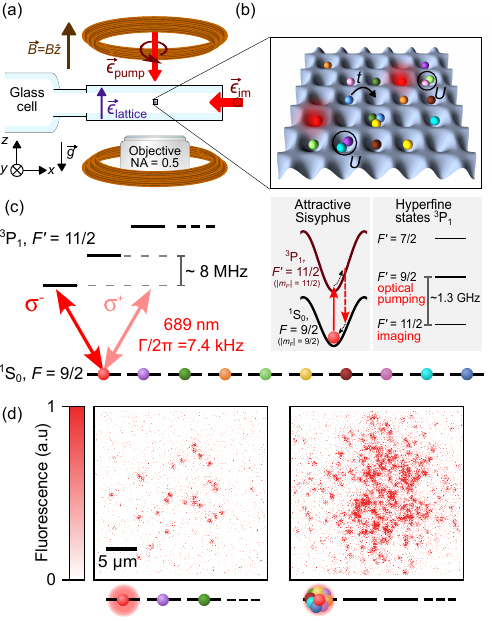}
    \caption{Microscopy of fermionic \textsuperscript{87}Sr atoms in an optical lattice. (a) Side-view schematic of the setup. Atoms are trapped in a square optical lattice at the center of a glass cell. A coil pair produces a vertical magnetic field $\vec{B}$ aligned with the lattice polarization, $\vec\epsilon_{\text{lattice}}$. A \SI{689}{nm} imaging beam enters horizontally with linear polarization, and scattered light is collected by a 0.5-NA objective. A circularly polarized optical-pumping beam is sent vertically. (b) Schematic of an SU(10) Fermi-Hubbard system with tunneling rate $t$ and on-site interaction energy $U$. Only $m_F=-9/2$ (red) atoms are addressed by the imaging light. (c) Energy level diagram of the ground and excited \textsuperscript{3}P\textsubscript{1} $F'=11/2$ state, showing the splitting of their Zeeman sublevels. The imaging light has $\sigma^+$ and $\sigma^-$ components but only resonantly addresses the $ ^1\text{S}_0 \ket{F=9/2, m_F=-9/2}\to \,^3\text{P}_1 \ket{F'=11/2, m_{F'}=-11/2}$ transition, which satisfies the condition for attractive Sisyphus cooling (inset, left panel). Two different transitions within the hyperfine structure of the \textsuperscript{3}P\textsubscript{1} manifold are used for optical pumping and imaging (inset, right panel). (d) Consecutive fluorescence images of a thermal cloud with 300 atoms. Left image: atoms in state $m_F=-9/2$. Right image: atoms in all states taken after a set of optical pumping pulses, showing a ten-fold increase in atom number. Both images share the same color scale.} 
    \label{fig:setup}
\end{figure}

In our experiment, we routinely prepare cold atomic clouds of \textsuperscript{87}Sr in the center of an ultra-high-vacuum glass cell, sketched in Fig.~\ref{fig:setup}(a). Such a cloud is, by default, spin-unpolarized, with all 10 Zeeman sublevels of the ground state \textsuperscript{1}S\textsubscript{0} $F=9/2$ equally populated. This results from the standard laser-cooling stages used for \textsuperscript{87}Sr, which require an additional ``stirring" laser for spin mixing during narrow-line cooling~\cite{Mukaiyama_redMOT_2003}. After laser cooling, the atomic cloud is loaded into a \SI{1064}{\nano\meter} optical dipole trap and then transferred to the optical lattice potential. It consists of a four-fold interfering square lattice with \SI{575}{nm} spacing and a vertically confining light sheet, both operating at $\lambda =$ \SI{813.4}{\nano\meter}. This clock-magic wavelength for strontium ensures that both the ground \textsuperscript{1}S\textsubscript{0} and excited clock \textsuperscript{3}P\textsubscript{0} states are equally trapped. Further details of the cooling process and trapping potentials can be found in App.~\ref{app:setup} and in our previous work with bosonic strontium~\cite{buob_strontium_2024}. 

The resulting system realizes the two-dimensional SU($N$) Fermi-Hubbard model, 
\begin{equation}\label{SUNFH}
    \hat{H}=-t\sum_{\substack{\langle i,j\rangle,\sigma}}\left({\hat{c}_{j}^{\sigma\dagger}}\hat{c}_{i}^\sigma+{\hat{c}_{i}^{\sigma\dagger}}\hat{c}_{j}^{\sigma}\right)+\frac{U}{2}\sum_{i, \sigma\not=\tau} \hat{n}_{i}^{\sigma}\hat{n}_{i}^{\tau},
\end{equation}
where fermions with $N$ internal spin states (up to 10) tunnel between neighboring lattice sites with amplitude $t$ and interact via an on-site interaction energy $U$ that is independent of the spin state. Here  $\hat{c}_{i}^{\sigma\dagger}$ ($\hat{c}_{i}^{\sigma}$) creates (annihilates) a particle of spin $\sigma$ at site $i$, and $\hat{n}_{i}^{\sigma}=\hat{c}_{i}^{\sigma\dagger}\hat{c}_{i}^{\sigma}$ is the corresponding number operator.

A schematic picture of this model is shown in Fig.~\ref{fig:setup}(b). In order to access the microscopic occupation of the system, the depth of the optical lattice is strongly increased to pin the positions of the atoms to their respective lattice sites, including during the imaging process.

Fluorescence imaging of strontium atoms is typically performed on the broad \SI{461}{\nano\meter} transition~\cite{cooper_alkalineearth_2018,norcia_microscopic_2018,buob_strontium_2024, tao_highfidelity_2024}. Most experiments combine imaging with simultaneous cooling to increase the number of scattered photons, although fast-exposure methods without cooling have also been recently demonstrated~\cite{tao_universalgates_2025}. Cooling is based on the narrow \SI{689}{\nano\meter} intercombination line (linewidth $\Gamma/2\pi = 7.4 $ kHz), which has enabled both narrow-line Sisyphus~\cite{cooper_alkalineearth_2018,covey_2000times_2019,urech_narrowline_2022,buob_strontium_2024, tao_highfidelity_2024} and resolved sideband cooling~\cite{cooper_alkalineearth_2018,norcia_microscopic_2018} in optical tweezers and lattices. Exploiting instead fermionic \textsuperscript{87}Sr considerably complicates the detection process. On the one hand, the hyperfine structure of the blue transition is not resolved, leading to spin mixing. On the other hand, cooling all 10 ground states on the narrow line is challenging due to their differing light shifts. Further details are provided in  App.~\ref{app:inlatticecooling}.

In this work, we overcome these challenges by performing fluorescence imaging of \textsuperscript{87}Sr directly on the narrow-line transition. This allows scattering of fluorescence photons that simultaneously cool the atoms, as recently demonstrated with bosonic \textsuperscript{88}Sr in optical tweezer arrays~\cite{urech_narrowline_2022}.  In \textsuperscript{87}Sr, the excited \textsuperscript{3}P\textsubscript{1} $F'=11/2$ state provides two cycling transitions suitable for attractive Sisyphus cooling: $ ^1\text{S}_0 \ket{F=9/2,m_F=\pm9/2}\to $ $ ^3\text{P}_1 \ket{F'=11/2,m_{F'}=\pm11/2}$. Here $F$ $(F')$ and $m_F$ ($m_{F'}$) denote the total angular momentum and magnetic quantum numbers of the ground (excited) state, respectively.

We exploit the narrow linewidth of the imaging transition to make the imaging process inherently spin-resolved. Our imaging beam resonantly addresses the $ ^1\text{S}_0 \ket{9/2,-9/2}\to$ $  ^3\text{P}_1 \ket{11/2,-11/2}$ cycling transition, while all other transitions remain off-resonant, see Figs.~\ref{fig:setup}(b) and (c). We reach this regime by setting a bias magnetic field of $B\approx$~\SI{20}{\gauss}, which results in a Zeeman splitting of the $F'=11/2$ excited state sublevels of $\Delta E = g_{F'}m_{F'} \mu_B B$. Here $g_{F'}$ is the g-factor of the excited state and $\mu_B$ is the Bohr magneton. Combined with the optical light shifts, it yields an overall energy splitting of approximately \SI{8}{\mega\hertz}, which is three orders of magnitude larger than the linewidth of the transition and ensures the independent addressing of a single spin state. Furthermore, the high efficiency of the cooling process makes it possible to work with relatively shallow trapping potentials. Specifically, we set the light sheet and the lattice depths to \SI{55}{\micro\kelvin} and \SI{80}{\micro\kelvin}, respectively.

Figure~\ref{fig:setup}(d) shows raw fluorescence images of a thermal cloud, taken with an exposure time of \SI{300}{\milli\second} and a near-resonant beam of intensity $400\,I_{\mathrm{sat}}$, where $I_{\mathrm{sat}}$ is the saturation intensity of the transition. To avoid populating the metastable states \textsuperscript{3}P\textsubscript{0,2} due to off-resonant Raman scattering induced by the trapping light (see Sec. \ref{sec:sec4} and App.~\ref{app:offResonant}), we use repumpers at $679$ and \SI{707}{nm}, respectively. The first image (left panel) is spin-resolved and contains approximately 30 detected atoms. Next, we optically pump all atoms into the stretched $ ^1\text{S}_0 \ket{9/2,-9/2}$ state (see  App.~\ref{app:pumpSpectrum} for details) and acquire a second image (right panel). We observe roughly a tenfold increase in the number of detected atoms. Averaging over 50 of such image pairs, we find an enhanced factor of 9.5(1.2), consistent with an even distribution of the atoms over the 10 spin states.

To characterize the performance of the imaging process, we compare two consecutive pictures of spin-polarized clouds. During the \SI{300}{\milli\second} exposure, we collect around 400 photons per atom, allowing us to reconstruct their occupation matrices, $n^\text{ref}$ and $n$ respectively. These matrices contain entries $n_{i,j}=1$ if the lattice site $(i,j)$ is occupied and $0$ otherwise. We extract the pinning fidelity of the imaging process, $\mathcal{F}_\text{pin}$, defined as the probability that an atom detected in the first picture is also detected in the second one, by computing the normalized overlap
\begin{equation}\label{eq:1}
\mathcal{O}=\frac{\langle n,n^\text{ref}\rangle}{N_{\text{ref}}}=\frac{\sum_{i,j}n_{i,j}n^\text{ref}_{i,j}}{\sum_{i,j} n^\text{ref}_{i,j}},
\end{equation}
where $N_{\text{ref}}$ is the total atom number detected in the first picture.  This analysis yields $\mathcal{F}_\text{pin}=92.5(7)\%$, where we identify a hopping rate of $\mathcal{H}=1.0(3)\%$ and a loss rate of $\mathcal{L}=6.5(7)\%$ (see App.~\ref{app:imaging} for further details). As we discuss in Sec.~\ref{sec:sec4}, a large fraction of the detected loss rate $\mathcal{L}$ does not arise from atoms physically leaving the lattice, but rather from spin depolarization processes. 
\FloatBarrier
\section{\label{sec:sec3}Spin-selective manipulation }

\begin{figure}
    \centering
    \includegraphics[width=\linewidth]{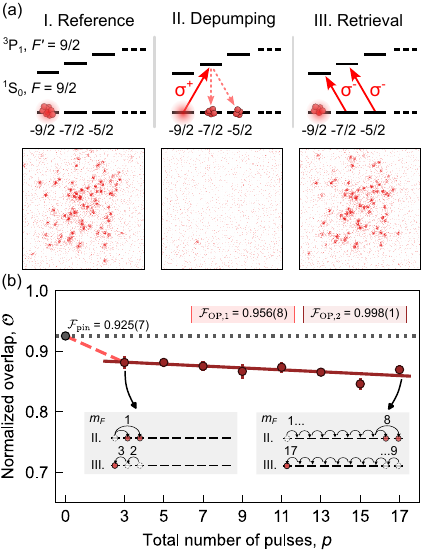}
    \caption{Narrow-line spin-resolved optical pumping. (a) Procedure to characterize the optical pumping fidelity. Left: a reference picture of a spin-polarized cloud is taken via narrow-line fluorescence, represented with a glowing spot. Center: the population of the initial state is depumped through an optical pumping pulse of $\sigma^+$ polarization addressing the $m_{F'} =-7/2$ state. A subsequent image shows practically no atoms, highlighting the spin-resolved nature of the detection. Right: two optical pumping pulses of $\sigma^-$ polarization retrieve the atomic population back to the $m_{F} =-9/2$ state and the cloud is imaged again.
    (b) Optical pumping with an increasing number of pulses $p$, which bring the atoms to different $m_F$ states and subsequently retrieve them in the stretched $m_F=-9/2$ state (insets). The normalized overlap $\mathcal{O}$ decreases with $p$ (circles) and yields two distinct optical-pumping fidelities $\mathcal{F}_\text{OP,1}$ and $\mathcal{F}_\text{OP,2}$ (boxes). $\mathcal{F}_\text{OP,1}$ is determined by comparing the $p=3$ value to the pinning fidelity $\mathcal{F}_\text{pin}$ (red dashed and gray dotted lines). $\mathcal{F}_\text{OP,2}$ is extracted from a linear fit to the data for $p\geq3$ (dark red solid line). Error bars denote the standard error of the mean.}
     \label{fig:opticalpumping}
\end{figure}

The presented narrow-line imaging enables the detection of the stretched state, but additional spin manipulation tools are required to measure all other spin states. A common option is optical pumping, which can be used to efficiently transfer atoms through the ground-state manifold. In recent works with optical tweezers~\cite{Huie2023, Norcia2023}, a combination of narrow-line imaging and optical pumping was used to retrieve the spin-1/2 populations of fermionic $^{171}$Yb atoms. In our case, the 10-state manifold of \textsuperscript{87}Sr adds significant complexity to the optical pumping scheme compared to a 2-state scenario. Furthermore, detection in quantum-gas microscopes faces additional challenges, such as hopping of the atoms during imaging and manipulation, as well as the requirement of higher fluorescence signal-to-noise ratio in order to resolve the closely spaced sites. In this section, we demonstrate and quantify the performance of spin-resolved optical pumping of \textsuperscript{87}Sr using its narrow line. In contrast to our imaging scheme, which relies on exciting the $F'=11/2$ state, we perform optical pumping via the $F'=9/2$ excited state, which has more favorable coupling properties, see App.~\ref{app:pumpSpectrum}.

To characterize the fidelity of the spin transfer, we analyze the performance of different sequences of optical pumping pulses (see parameters in App.~\ref{app:pumpSpectrum}). We begin by preparing all atoms in the $m_F = -9/2$ state and taking a reference image of their positions. From this reference, we obtain the initial occupation $n^{\text{ref}}$, see left panel of Fig.~\ref{fig:opticalpumping}(a). We then apply a $ ^1\text{S}_0 \ket{9/2,-9/2}\to {^3\text{P}_1} \ket{9/2,-7/2}$ optical pumping pulse, selectively depumping the atoms from the stretched state, and acquire a second image showing virtually no fluorescence signal, see central panel. Finally, we transfer the atomic population back to $m_{F} = -9/2$ state through two successive optical pumping retrieval pulses and image it again, see right panel. We compare the occupation before and after the optical pumping procedure using Eq.~\eqref{eq:1}, which yields the fraction of atoms detected in the first image that returned to the same lattice site in the last one.  We obtain a normalized overlap $\mathcal{O} = 88(1)\%$. The value is slightly below the pinning fidelity $\mathcal{F}_\text{pin}=92.5(7)\%$ due to the finite optical pumping fidelity to the stretched state $\mathcal{F}_{\text{OP},1}$. We model it as $\mathcal{O}=\mathcal{F}_\text{pin} \,\mathcal{F}_{\text{OP},1}$, which yields $\mathcal{F}_{\text{OP},1}=95.6(8)\%$.

The same optical pumping procedure can be performed for an increasing number of depumping and retrieval pulses $p$. First, a reference image is acquired; the atoms are then depumped into a pair of $m_F$ states in the ground-state manifold; finally they are retrieved to the initial state. An intermediate measurement after depumping confirms that the population remaining in the first $(p-1)/2$ states is negligible, demonstrating the reliability of the depumping. The final retrieval measurement, after pumping the atoms back to the initial state, yields the normalized overlap $\mathcal{O}$. We plot it as a function of the total number of pulses $p$ in Fig.~\ref{fig:opticalpumping}(b). For $p=5$, $\mathcal{O}$ decreases only slightly compared to $p=3$, and the overlap shows minimal reduction as $p$ increases further. These results indicate that the optical pumping fidelity for the stretched state, $\mathcal{F}_{\text{OP},1}$, differs appreciably from that for the central states (see App.~\ref{app:pumpSpectrum}). To quantify the overlap decrease for $p\geq3$, we perform a linear fit to the data. The linear dependence is consistent with a high optical-pumping fidelity $\mathcal{F}_{\text{OP},2}$ between the central states, since the total fidelity can be approximated as $(\mathcal{F}_{\text{OP},2})^{p}\approx 1-(1-\mathcal{F}_{\text{OP},2})\cdot p$. From the fit, we estimate $\mathcal{F}_{\text{OP},2} = 99.8(1)\%$. Overall, these results indicate that single-atom spin manipulation via optical pumping can be performed with high efficiency, without leading to appreciable losses or hopping (see also Sec. \ref{sec:sec5}).

\section[Spin-resolved detection of SU(10) fermions]{\texorpdfstring{\label{sec:sec4}Spin-resolved detection\newline of SU(10) fermions}{Spin-resolved detection of SU(10) fermions}}

\begin{figure*}
    \centering
    \includegraphics{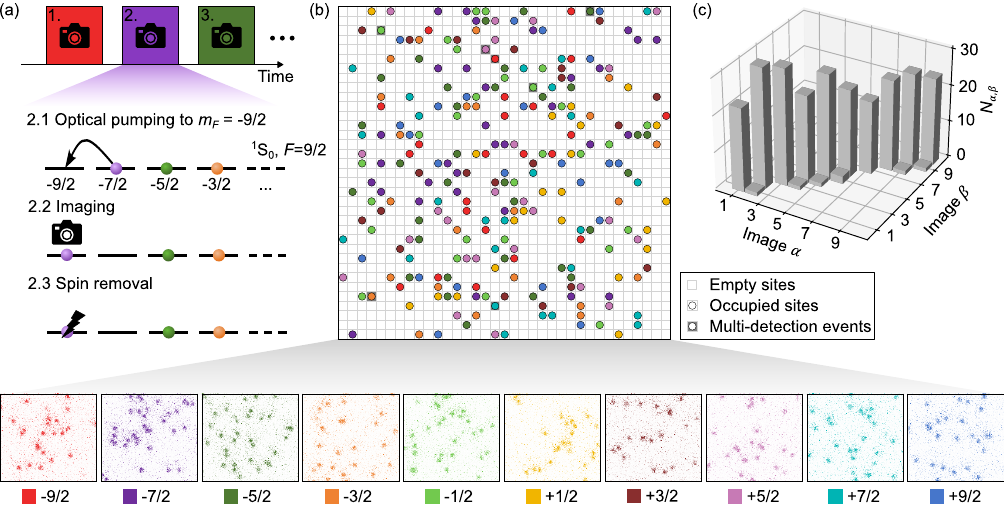}
    \caption{Spin-resolved microscopy of the 10 states of \textsuperscript{87}Sr. (a) The imaging protocol consists of 10 detection blocks, one per spin state, carried out sequentially from the $m_F=-9/2$ to the $m_F=+9/2$ state. Each detection block starts by optically pumping the target spin state to the $m_F=-9/2$ state, followed by spin-resolved fluorescence imaging, and concludes with spin-removal of the population in $m_F=-9/2$ by means of a blue-detuned pulse on the imaging transition. (b) Merging the reconstructed images yields the spin-resolved occupation for all 10 spin states in a single experimental sequence. Below are the 10 raw snapshots, with the colormap indicating the corresponding spin. The images correspond to $35\times35$ lattice sites containing $253$ atoms evenly distributed among the 10 spin states. (c) Image-to-image detection coincidences between the different images, $N_{\alpha,\beta}$. The diagonal indicates the different spin populations. The presence of off-diagonal coincidences corresponds to the multi-detection events, which are indicated with gray shaded squares in (b). They are primarily caused by off-resonant scattering of the trapping light leading to spin depolarization (see main text).}
    \label{fig:SU(10)}
\end{figure*}

In the Fermi-Hubbard model, quantum magnetism arises through the superexchange coupling between nearest-neighbour spins. Due to the Pauli exclusion principle, the emerging spin-spin interaction is antiferromagnetic, and leads to anticorrelated spins. SU(2) antiferromagnetism can be characterized by spin correlators such as ${C^{z}_{\boldsymbol{ij}} = \langle \hat{S}_{\boldsymbol{i}}^z \hat{S}_{\boldsymbol{j}}^z \rangle_c=\langle \hat{S}_{\boldsymbol{i}}^z \hat{S}_{\boldsymbol{j}}^z \rangle-\langle \hat{S}_{\boldsymbol{i}}^z\rangle \langle\hat{S}_{\boldsymbol{j}}^z \rangle}$, where $c$ indicates a connected correlator~\cite{ChalopinCRAS2026}. This quantity can be probed by accessing the site-resolved spin populations since the spin operator can be expressed as $\hat{S}_{\boldsymbol{i}}^z = \frac{1}{2}(\hat{n}_{\boldsymbol{i}}^{\uparrow}-\hat{n}_{\boldsymbol{i}}^{\downarrow})$, with $\hat{n}_{\boldsymbol{i}}^{\uparrow \,(\downarrow)}$ denoting the particle number operator in state $\uparrow (\downarrow)$ and site $\boldsymbol{i}$. Quantum-gas microscopes have enabled the measurement of such magnetic correlations by directly measuring $\hat{n}_{\boldsymbol{i}}^{\sigma}$. However, light-assisted collisions during fluorescence imaging lead to the detection of parity-projected occupation operators~\cite{gross_quantum_2021}. Additionally, spin-resolved detection often relies on spin removal techniques~\cite{parsons_siteresolved_2016,  cheuk_observation_2016a}, which make it impossible to distinguish holes, doublons, and one of the spins. An independent identification of both spins, needed to measure more advanced correlators~\cite{Chalopin2024}, requires more complex techniques, such as Stern-Gerlach splitting~\cite{boll_spin_2016, Koepsell2020, Yan2022, mongkolkiattichai2025, Hartke2025} or polarization-sensitive detection~\cite{jain2025, hammel2025}.

When dealing with SU($N$) systems with large $N$, measuring magnetic correlations becomes even more challenging. The spin correlations can be characterized analogously to the SU(2) case by
\begin{equation}\label{correlator}
{C^{z}_{\boldsymbol{ij}} = \sum_{\sigma\not=\tau}\left[ \langle \hat{n}_{\boldsymbol{i}}^{\sigma} \hat{n}_{\boldsymbol{j}}^{\sigma}  \rangle_c -\langle \hat{n}_{\boldsymbol{i}}^{\sigma} \hat{n}_{\boldsymbol{j}}^{\tau}  \rangle_c \right]},
\end{equation}
which involves the particle number operators $\hat{n}_{\boldsymbol{i}}^{\sigma}$ for each of the $N$ states, with $\sigma,\tau\in\{1,\dots,N\}$ \cite{ManmanaPRA2011}. For $N>2$, extracting $C^{z}_{\boldsymbol{ij}}$ therefore requires single-atom and full-spin resolution within a single experimental realization, a capability that has not yet been realized. Despite this, recent experiments have made progress towards the microscopic study of multi-component systems. Firstly, the detection of two spin states of \textsuperscript{87}Sr was achieved in optical tweezers, using the long-lived \textsuperscript{3}P\textsubscript{0} state to shelve one spin in a dark state while imaging the other~\cite{barnes_assembly_2022}. A second experiment achieved fluorescence detection of the occupation of \textsuperscript{173}Yb atoms in optical tweezers through broad-line imaging~\cite{Abdelkarim2025}. However, this led to mixing of all 6 spin states, i.e., no spin information could be retrieved. Finally, in optical-lattice systems, quantum-gas microscopy of 3 spin states of $^6\text{Li}$ has been realized by removing one of the spins and detecting only the remaining two independently~\cite{mongkolkiattichai2025}.

In this section, we go far beyond these previous experiments by probing the spin-resolved occupation of an SU(10) system. This is achieved by combining the techniques introduced in the preceding sections: spin-resolved imaging of the stretched state and spin-selective optical pumping. As schematically shown in Fig.~\ref{fig:SU(10)}(a), our protocol comprises 10 sequential detection blocks corresponding to the imaging of the 10 ground states. Each block starts with selective optical pumping of the target spin state into $m_F=-9/2$. In a second step, the atoms are imaged by addressing the $ ^1\text{S}_0 \ket{9/2,-9/2}\to {^3\text{P}_1} \ket{11/2,-11/2}$ transition. Finally, the detected atoms are discarded via a spin-removal pulse.

The bottom panel of Fig.~\ref{fig:SU(10)}(b) shows the 10 raw images captured in a single experimental realization, each corresponding to atoms detected in one of the 10 spin states. Reconstructing the occupation from these images yields the parity-projected spin-resolved occupation matrix. The result is displayed in the central panel of Fig.~\ref{fig:SU(10)}(b), with each spin state represented by a distinct color. From this matrix, we directly access the $\hat{n}_{\boldsymbol{i}}^{\sigma}$ operators for all $\sigma$ and consequently, the spin correlators. This capability is essential for revealing the magnetic properties of the SU($N$) Fermi-Hubbard model~\cite{Ibarra_metalinsulator_2023}. In particular, these measurements provide access to higher-order correlators~\cite{Chalopin2024}, which are crucial for understanding the many-body character of Hubbard systems.

From our measurements we observe that a few lattice sites, highlighted with a gray background in Fig.~\ref{fig:SU(10)}(b), are occupied in more than one image. We investigate these multi-detection events by computing the overlap between each possible pair of images, $\alpha$ and $\beta$, as $N_{\alpha, \beta} = \langle n^{\alpha}, n^{\beta} \rangle$. Here, $N_{\alpha, \alpha}=N_\alpha$ is the number of occupied sites in image $\alpha$, while $N_{\alpha, \beta}$ with $\alpha\not=\beta$ counts the sites that are occupied in both images. The results, shown in Fig.~\ref{fig:SU(10)}(c), reveal non-zero values between subsequent images, with the first off-diagonal terms $N_{\alpha, \alpha+1}$ being the largest contribution. In the specific example of Fig.~\ref{fig:SU(10)}, we measure 7 multi-detection events out of 253 atoms. Averaging over 40 repetitions of the experiment, the total off-diagonal counts correspond to $4.2(4)\%$ of the total atom number, with the first off-diagonal counts contributing $2.8(4)\%$. This effect cannot be explained by the inefficiency of the spin removal pulse, estimated at $<0.5(2)\%$, and must be caused by spin depolarization during imaging. We observe no multi-detection events beyond those explained by this effect, consistent with the expected parity projection caused by light-assisted collisions during imaging (see App.~\ref{app:inlatticecooling}).

The main mechanism contributing to spin depolarization is trap-induced off-resonant Raman scattering within the excited state manifold (see App.~\ref{app:offResonant}). As in optical atomic clocks~\cite{Dorscher_2018}, the \SI{813.4}{\nano\meter} light generating the optical lattice and light-sheet potential can drive two-photon off-resonant transitions within the \textsuperscript{3}P\textsubscript{$J$} manifold. Consequently, atoms in the imaging cycle may abruptly transition to a different state within the triplet manifold. Upon decay or repumping, these atoms can populate ground states with $m_F\not=-9/2$, opening the imaging cycle. This process is the main limitation to the observed pinning fidelity $\mathcal{F}_\text{pin}$ and causes the same atom to appear in consecutive images, appearing as a multi-detection event at the spin level (which is not a true double occupancy). Since depolarization occurs during the imaging sequence, the first detection reflects the original spin state of the atom. We therefore retain only this first assignment. Overall, these results show that our detection scheme not only is ready to detect SU($N$) quantum magnetism but also provides a sensitive diagnostic for identifying spurious processes that could be missed by other imaging methods.

\FloatBarrier
\section{\label{sec:sec5}Coherent Larmor precession}

\begin{figure*}
   \centering
   \includegraphics[width=17cm]{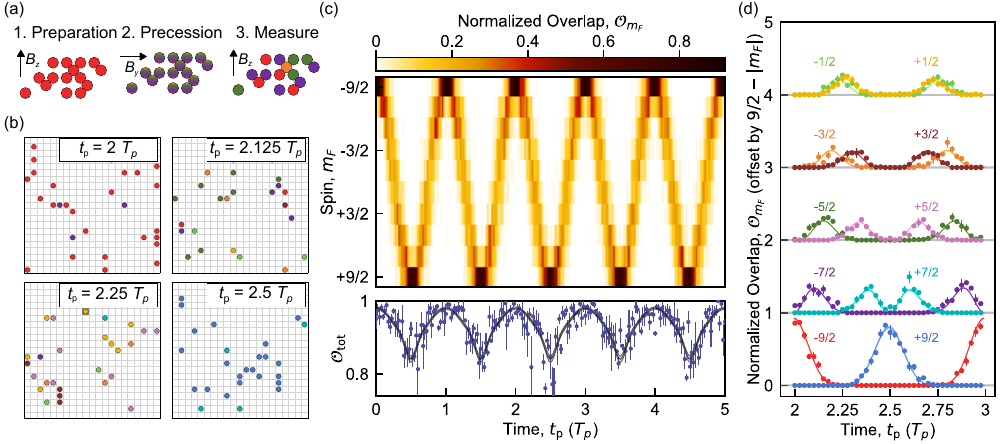}
   \caption{Microscopic observation of Larmor precession dynamics in a spin-9/2 system. (a) Scheme of the precession sequence. 1. Preparation: we optically pump all atoms in $m_F = -9/2$ and take a reference image. 2. Precession: a sudden rotation of the magnetic field initiates Larmor precession dynamics. 3. Measurement: after a certain precession time,  $t_{\text{p}}$, we quench back the field to its original orientation and perform the spin-resolved imaging protocol. (b) Sample snapshots of the reconstructed spin occupation at different precession times, $t_\text{p}$. The spin evolves from $-9/2$ to $+9/2$ over $T_p/2$ with $T_p=$~\SI{40.16(6)}{\milli\second}, from which we determine $B_y=$~\SI{135.0(2)}{\milli\gauss}.  (c) Top: Normalized overlap of the image corresponding to each spin, $\mathcal{O}_{m_F}$. Bottom: Total normalized overlap, $\mathcal{O}_\text{tot}$. The solid line shows theoretical predictions accounting for all relevant infidelities, with the period $T_p$ as the only fitting parameter. The shaded region indicates the effect of the uncertainties in the infidelities.  (d) Comparison of the measured $\mathcal{O}_{m_F}$ with theory (solid lines). Data for each $m_F$ are offset by $9/2-|m_F|$ for clarity. In all panels, error bars denote the standard error of the mean.}
   \label{fig:precession}
\end{figure*}

To benchmark our spin-resolved protocol, we study the Larmor precession dynamics of the ground-state nuclear spin, $F=9/2$. The experiment is performed in a deep optical lattice, which suppresses hopping and isolates single-particle spin dynamics. A bias magnetic field along the $z$ axis defines the quantization direction, splitting the nuclear Zeeman sublevels of the ground state by $\Delta E = g_Im_F \mu_B B\approx h \times$\SI{184}{\hertz/\gauss}, where ${g_I \approx -1.32 \cdot 10^{-4}}$~\cite{olschewski_messung_1972,thekkeppatt2025} is the nuclear spin g-factor of \textsuperscript{87}Sr and $h$ is the Planck constant. Following a sudden change in the direction of the quantization axis, the spin undergoes coherent precession with a period  $T_p=h/(g_I \mu_B B_p)$, where $B_p$ is the magnitude of the magnetic field during the precession. The evolution of the spin is then tracked using our spin-resolved imaging protocol.

Figure~\ref{fig:precession}(a) summarizes our spin-precession experiment. We begin with a spin-polarized cloud of atoms prepared in the $^1\text{S}_0 \ket{9/2,-9/2}$ state under a vertical magnetic field of $B_z\approx$~\SI{20}{\gauss}, from which we acquire a reference image of the atomic occupation, $n_\text{ref}$. We then rapidly rotate the magnetic field into the horizontal plane, setting it to $B_y$ and initiating Larmor precession. After a variable evolution time $t_\text{p}$, the magnetic field is switched back to $B_z$, freezing the spin dynamics and allowing us to perform the spin-resolved imaging protocol presented in Sec.~\ref{sec:sec4}. Figure~\ref{fig:precession}(b) shows four snapshots of the reconstructed spin occupations at different precession times $t_\text{p}$.

We present the spin-resolved precession dynamics in the top panel of Fig.~\ref{fig:precession}(c). The plot shows the normalized overlap $\mathcal{O}_{m_F}$ for each spin state as a function of the precession time $t_\text{p}$. The results provide clear evidence of nuclear spin coherence over the measured time range, which extends to hundreds of milliseconds. We observe coherent spin oscillations with revivals of the $m_F = -9/2$ state at integer multiples $t_\text{p}=nT_p$, and a maximum population transfer to the $m_F = +9/2$ state at $t_\text{p}=(n+1/2)T_p$, where $T_p=$~\SI{40.16(6)}{\milli\second}. A fit to the oscillations yields a magnetic field of $B_y=135.0(2)$~\SI{}{\milli\gauss}. These oscillations provide a stringent benchmark for our imaging protocol, as each evolution time $t_\text{p}$ produces a deterministic spin distribution. 
In fact, we observe no increase in hopping at any time, even at precession times when most of the atoms are detected in $m_F=+9/2$, the last spin state to be imaged in our sequential protocol. This indicates that the imaging and spin removal of $m_F=-9/2$ do not significantly heat up the atoms in other spin components.

In the previous section, we identified spin depolarization as the dominant contribution to the imaging infidelity. To isolate additional sources of infidelity, we extract a spin-insensitive observable from the data shown in Fig.~\ref{fig:precession}(c). Specifically, we combine the occupation matrices of all spin states, projecting out multi-detection events, which yields the total occupation $n^\text{tot}$. We then compute the corresponding normalized overlap, $\mathcal{O}_\text{tot}$. This analysis effectively eliminates the effect of spin depolarization (see App.~\ref{app:Precession}), and isolates the last remaining source of infidelity: vacuum-induced losses. We independently measure the vacuum lifetime in the lattice to be $\tau_\text{vac}=94(3)$~s, which translates into a loss of $\mathcal{L}_\text{vac}=0.95(3)\%$ for each spin-readout cycle of \SI{900}{\milli\second}. As a result, atoms in states with larger $m_F$ experience greater losses, since they remain in the lattice for longer before being imaged. This behavior is visible in the bottom panel of Fig.~\ref{fig:precession}(c) as a reduction of $\mathcal{O}_\text{tot}$ at precession times when higher-$m_F$ states are populated. The theory prediction (solid line) contains no fitting parameters other than the oscillation period, and its agreement with the data confirms that vacuum-limited loss alone accounts for the observed decay, with no additional loss induced by the imaging or spin-removal pulses on the spin states not yet addressed.
 
Figure~\ref{fig:precession}(d) compares the measured precession dynamics of each spin state, $\mathcal{O}_{m_F} (t_\text{p})$, to the theoretical predictions that include all identified limiting mechanisms. We again observe excellent agreement without additional fitting parameters, showcasing our solid understanding of all processes underlying the detection protocol.

\FloatBarrier
\section{Detection response and numerical study of spin correlations}
\label{sec:sec6}

\begin{figure*}
\hspace{-1cm}
\begin{minipage}{0.55\textwidth}
    \begin{tabular}{|l|l|l|l|}
    \hline
    Error type\rule{0pt}{3.5mm} & Quantity & Value & Section\\\hline\hline
    \multirow{3}{5em}{Occupation}
    & Hopping*, $\mathcal{H}$\rule{0pt}{3mm} & 1.0(3)\% & Sec.~\ref{sec:sec2}\\
    & Losses*, $\mathcal{L}$ & 6.5(7)\% & Sec.~\ref{sec:sec2}\\
    & Vacuum losses, $\mathcal{L}_\text{vac}$ & 0.95(3)\% & Sec.~\ref{sec:sec5}\\
     
    \hline
    \multirow{4}{4em}{Spin} 
    & Optical pumping stretched pulse, ($1-\mathcal{F}_\text{OP,1}$)\rule{0pt}{3.5mm} & 4.4(8)\% & Sec.~\ref{sec:sec3}\\
    & Other optical pumping pulses, ($1-\mathcal{F}_\text{OP,2}$) & 0.2(1)\% & Sec.~\ref{sec:sec3}\\
    & Spin depolarization & 4.2(4)\% & Sec.~\ref{sec:sec4} \\
    & Spin selectivity & $<$3.6(7)\% & App.~\ref{app:pumpSpectrum}\\
    
    \hline
    \multirow{2}{5em}{Reconstruc.} 
    & False negative & 1.1(2)\% & App.~\ref{app:imaging}\\
    & False positive & 0.51(5)\% & App.~\ref{app:imaging}\\
    \hline
    \end{tabular}
\end{minipage}
\hspace{1.5cm}
\begin{minipage}{0.3\linewidth}
    \includegraphics{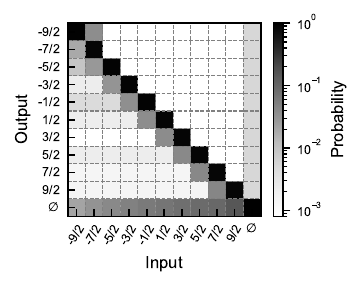}
\end{minipage}
    \caption{Readout protocol infidelity budget and detection-response matrix. Left: Summary of the imaging infidelity sources and their error probabilities, with the sections where they are characterized indicated. Right: Detection-response matrix constructed from these infidelities (processes marked with ``$*$" in the table are excluded, see App.~\ref{app:confusion}). Each column corresponds to an input spin-occupation state and each row to the detected output, including the empty state $\varnothing$. Double occupancies are excluded, as parity projection maps them to $\varnothing$. The matrix thus quantifies the stochastic mapping induced by the detection process.}
    \label{fig:confusion}
\end{figure*}

The Larmor precession measurements of the previous section experimentally benchmark our imaging protocol in the single-particle limit. In this section, we assess its performance in the many-body regime numerically by computing spin correlations in the SU($N$) Fermi–Hubbard model, applying to them the response of our imaging protocol, and evaluating how faithfully they could be recovered experimentally. We perform this study with SU(3) one-dimensional chains. As the simplest extension of SU(2), SU(3) provides a natural starting point for current experiments and has been extensively studied numerically \cite{Romen2020, Feng2023, Bird2025, kleijweg_corboz_2025}. Moreover, the one-dimensional geometry is amenable to exact diagonalization, enabling access to spin correlators at experimentally relevant finite temperatures.

We begin by constructing a realistic model of our imaging protocol. We compile the infidelity sources identified in the previous sections and summarize them in the table shown in the left panel of Fig.~\ref{fig:confusion}. We distinguish three types of errors: (i) occupation errors, which alter the lattice-site occupation; (ii) spin errors, which mix spin populations; and (iii) reconstruction errors, arising from imperfect identification of empty and occupied sites. These processes are encoded in a detection-response matrix, shown in the right panel of Fig.~\ref{fig:confusion}, which gives the probability for a given input spin-occupation state to be detected as a different output state. Double occupancy is not included, as parity projection maps it to an empty site $\varnothing$ (see App.~\ref{app:inlatticecooling}). This matrix provides a quantitative description of detection-induced stochastic processes.

We incorporate this detection model into numerical simulations of the many-body system using full exact diagonalization. We consider one-dimensional chains of length $L=8$ with periodic boundary conditions and interaction strength $U/t=4$. Using SU(3) symmetry, conservation of the spin populations $N_1,\,N_2$ and $N_3$ block-diagonalizes the Hamiltonian into sectors labeled by $(N_1,N_2,N_3)$, which are further resolved into momentum eigenstates by translation symmetry. Invariance under spin permutations relates different sectors, so that only one representative needs to be diagonalized in practice~\cite{taie_observation_2022}. We focus on the most balanced sectors accessible with $L=8$ and a filling of one atom per site, namely $(3,3,2)$ and its permutations. They are the most likely to occur and host the strongest magnetic correlations. 
 
\begin{figure}
    \centering
    \includegraphics{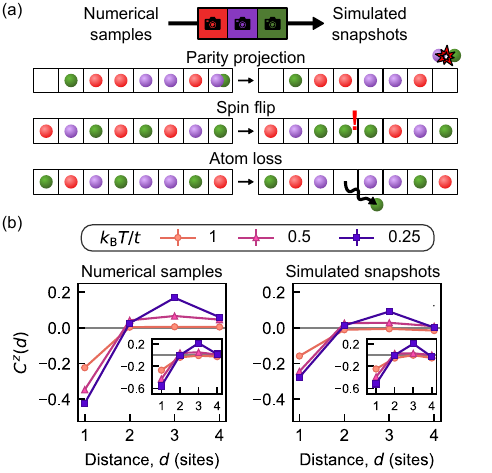}
    \caption{Numerical study of the robustness of SU(3) spin correlations under detection imperfections. (a) Schematic of the sampling procedure. Numerical samples drawn from exact diagonalization (left) are transformed into simulated snapshots (right) using the detection-response matrix (see Fig. \ref{fig:confusion} and App.~\ref{app:confusion}). The examples illustrate parity projection of doublons, spin-flip processes, and atom losses. (b) Spin correlator $C^z(d)$ at different temperatures, computed from numerical samples (left) and simulated snapshots (right). Detection imperfections reduce the correlation contrast while preserving sizable signals at experimentally relevant temperatures. Post-selection of snapshots excluding holes (insets) reduces the discrepancy, indicating that parity projection is the dominant source of signal reduction. Error bars denote the bootstrapped standard error and are smaller than the marker size.}
   
    \label{fig:SU(3)ED}
\end{figure}

We simulate individual experimental realizations as measured by our microscopy protocol. For each realization, we sample an eigenstate from the Boltzmann distribution at temperature $T$. We then project onto the occupation basis, yielding a numerical sample. Applying the detection model  produces a corresponding simulated snapshot by first removing doubly occupied sites and then stochastically altering the occupation at each site according to the detection-response matrix (adapted here to a three-spin detection scheme, see App.~\ref{app:confusion}). This procedure, illustrated in Fig.~\ref{fig:SU(3)ED}(a), accounts for parity projection of doublons, spin-flips induced by the imaging process, losses, and reconstruction errors. For each temperature, we generate 2000 samples and compute the spin correlator $C_{ij}^z$ (see Eq.~\eqref{correlator}) for both numerical samples and  simulated snapshots. Averaging over the $\mathcal{N}_d$ pairs of sites at distance $d$, and subtracting the corresponding infinite-temperature value to remove the finite-size offset, we obtain $C^z(d) = \sum_{i}C_{i,i+d}^z/\mathcal{N}_d - C^z_{T\to\infty}$.

Figure \ref{fig:SU(3)ED}(b) shows the resulting magnetic correlations for different temperatures $T$, expressed in units of the tunneling $t$ as $k_{\mathrm{B}}T/t$, where $k_{\mathrm{B}}$  is the Boltzmann constant. Numerical samples (left) exhibit clear antiferromagnetic correlations at short distances and, at low temperatures, a pronounced peak at $d=3$ characteristic of one-dimensional SU(3) systems \cite{BonnesPRL2012}. Simulated snapshots (right), which include detection imperfections, display reduced contrast but retain sizable correlations at experimentally relevant temperatures. Insets show post-selected data where simulated snapshots do not contain holes. In this case, the nearest-neighbor correlators at the lowest temperature differ by less than $10\%$. From this numerical study, we therefore conclude that parity projection is the dominant source of signal reduction, and other detection errors have a comparatively minor impact. These results demonstrate that the current performance of our detection protocol is sufficient to characterize magnetic correlations in SU($N$) Fermi–Hubbard systems at the single-site level.

\section{\label{sec:conclusion}Conclusion}

In this work, we demonstrate quantum-gas microscopy of \textsuperscript{87}Sr in a Hubbard-regime optical lattice using a narrow-line imaging scheme. By combining spin-selective optical pumping with spin-resolved single-atom fluorescence imaging, we implement a protocol that sequentially images all 10 nuclear spin states of \textsuperscript{87}Sr within a single experimental cycle. We benchmark the performance of our scheme at the single-particle level by probing the coherent precession of an $F = 9/2$ nuclear spin, finding excellent agreement with theoretical expectations. These results confirm that all processes underlying the detection protocol are well controlled and provide a clear route to further increase the pinning fidelity, which already exceeds $92\%$. Finally, numerical simulations show that parity projection sets the main constraint for many-body observables, while other detection imperfections play a minor role. Accordingly, our spin-resolved microscope is expected to already enable the measurement of magnetic correlations in SU($N$) systems.

Building on this capability, a natural next step is the microscopic characterization of SU($N$) degenerate Fermi gases of \textsuperscript{87}Sr with a tunable number of components~\cite{desalvo_degenerate_2010, sonderhouse_thermodynamics_2020}, as well as the preparation and site-resolved study of SU($N$) fermionic Mott insulators. An additional possibility at the Mott temperature scale is to engineer  artificial gauge fields by exploiting the nuclear spin states of the ground-state manifold as a synthetic dimension~\cite{Celi2014, han_band_2019, mancini_observation_2015,  zhou_observation_2023, Zhou2025_measuring_hall_quantum_simulator}, which open the door to microscopic investigations of quantum-Hall physics in strongly correlated regimes.

At lower temperatures, SU($N$) systems are expected to host a wealth of unconventional quantum-magnetic phases. In particular, nearest-neighbor spin correlations, which so far have only been accessed via global probes~\cite{taie_observation_2022}, should be readily accessible. Below the super-exchange energy scale, a realistic target for future studies is the SU(3) case, where a transition from diagonal stripe order to zigzag antiferromagnetism is expected upon decrease of the interaction strength~\cite{Toth2010,Feng2023}. There, site-resolved detection should allow discriminating between competing theoretical scenarios in the intermediate-coupling regime~\cite{Feng2023, Bird2025, kleijweg_corboz_2025}. Another promising direction is to study SU($N$) systems with tunable lattice geometries~\cite{tarruell_creating_2012, wei_observation_2023}, particularly in regimes where unconstrained magnetism and frustration govern the behavior of the system. For larger $N$, even more exotic magnetically ordered states have been predicted~\cite{Toth2010, corboz_simultaneous_2011, Nataf2016}, and quantum-gas microscopy may provide decisive insight into whether the SU($N$) Hubbard model can stabilize a chiral spin liquid phase for $N>5$~\cite{hermele_mott_2009}.

Beyond quantum simulation, the state-resolved detection technique developed here may have applications in other areas involving \textsuperscript{87}Sr. In optical lattice clocks, ground-state population detection could enable novel erasure-conversion strategies~\cite{ma2025} and improve coherence times. Combined with coherent control of the nuclear spin~\cite{Ahmed2025}, our approach also provides a scalable platform for qudit-based quantum computing~\cite{daley_quantum_2008, Omanakuttan2021}.

\textit{Note added.} During the writing of the manuscript we became aware of related experiments on free-space imaging of a single \textsuperscript{87}Sr atom in an optical tweezer \cite{plassmann_2026}. 

\FloatBarrier
\begin{acknowledgments}

We acknowledge insightful discussions with K. Hazzard, M. Robert-de-Saint-Vincent, F. Schreck, D. Wilkowski, as well as with the members of the ICFO Quantum Gases Experimental group. We are grateful to F. Faisant for his help adapting lattice occupation reconstruction algorithms and for numerous discussions. We also thank S. Hirthe, and Q. Redon for a careful reading of the manuscript. We acknowledge funding from the European Union (HORIZON-CL4-2022-QUANTUM-02-SGA through project PASQuanS2.1 No. 101113690 and ERC CoG project No. 101003295 SuperComp), the Spanish Ministry of Science and Innovation MCIU/AEI/10.13039/501100011033 (projects MAPS PID2023-149988NB-C22, and QuantERA DYNAMITE PCI2022-132919 with funding from European Union NextGenerationEU, PRTR-C17.I1 with funding from European Union NextGenerationEU and Generalitat de Catalunya, and Severo Ochoa CEX2024-001490-S), Fundació Cellex, Fundació Mir-Puig, and Generalitat de Catalunya (“Quàntica – Vall de la Mediterrània de les Ciències i les Tecnologies Quàntiques” Government Agreement GOV/51/2022 promoted by Secretariat of Digital Policies of the Government of Catalonia, and CERCA program). C.G. acknowledges support from MCIU/AEI/10.13039/501100011033/FEDER, EU and ESF+ (PREP2023-002109), A.R. from the “la Caixa” Foundation (ID 100010434) with fellowship code LCF/BQ/PI24/12050012, S.B. from MCIU/AEI/10.13039/501100011033 and ESF (PRE2020-094414), L.B. from Generalitat de Catalunya and ESF+ (FI-STEP 2025 STEP00008), and J.H. from the European Union (Marie Sk\l{}odowska-Curie–713729).

\end{acknowledgments}
\section*{Data Availability}
The data that support the findings of this article are openly available \cite{OnlineData}.

\appendix
\section{Experimental setup\label{app:setup}}

The atomic cloud is prepared using standard laser cooling techniques for \textsuperscript{87}Sr \cite{Mukaiyama_redMOT_2003}. We first load atoms into a continuously repumped magneto-optical trap (MOT) operating on the broad ${}^1\text{S}_{0} \rightarrow {}^1\text{P}_{1}$ transition at \SI{461}{\nano\meter}, following the procedure described in our previous works~\cite{hoschele_atomnumber_2023, buob_strontium_2024}. The atoms are then transferred to a narrow-line MOT operating on the ${{}^1\text{S}_{0} \rightarrow {}^3\text{P}_1}$ transition at \SI{689}{\nano\meter}, where temperatures on the order of \SI{1}{\micro\kelvin} are reached. During this stage, two hyperfine transitions are addressed simultaneously: $F' = 11/2$ for cooling and trapping, and $F' = 9/2$ to mix the $m_F$ states and enhance the cooling efficiency \cite{Mukaiyama_redMOT_2003}. This procedure yields an approximately equal population in the ten Zeeman sublevels.

The laser-cooled atoms are then loaded from the narrow-line MOT into a crossed optical dipole trap at \SI{1064}{\nano\meter}, formed by two perpendicular beams with a beam waist of approximately \SI{100}{\micro m} and a trap depth of $\approx$\SI{16}{\micro\kelvin}. A light-sheet beam at \SI{813.4}{\nano\meter}, with  waists of \SI{3.5}{\micro\meter} $\times$ \SI{60}{\micro\meter}, is superimposed on the optical dipole trap to provide tight vertical confinement. Next, the optical dipole trap is adiabatically turned off while the light-sheet trap depth is ramped from \SI{90}{\micro K} to \SI{55}{\micro K}. Finally, we ramp up the optical lattice potential, which is generated by the four-fold interference of a single laser beam operating at \SI{813.4}{nm}.

The final potential used for imaging is clock-magic and consists of the light sheet beam at a depth of \SI{55}{\micro\kelvin} and the optical lattice at a depth of \SI{80}{\micro\kelvin}, corresponding to 460\,$E_r$. Here $E_r = h^2/2 m \lambda^2 \approx \, h \times \SI{3.5}{kHz}$ is the recoil energy of the lattice beam photons, and $m$ is the mass of \textsuperscript{87}Sr. Both the lattice and light sheet are linearly polarized along the vertical direction, parallel to the bias magnetic field. This configuration results in a vertical confinement of $\omega_z \approx 2 \pi \times $\SI{6.6}{\kilo\hertz}, and on-site trap frequencies $\omega_{x\text{,site}} \approx 2 \pi \times$\SI{107}{\kilo\hertz} and $\omega_{y\text{,site}} \approx 2 \pi \times$\SI{98}{\kilo\hertz}. From the single-atom-resolved fluorescence images we measure lattice spacings of $a_x \approx$~\SI{606}{\nano\meter} and $a_y \approx$~\SI{549}{\nano\meter}. These values imply that the lattice beams intersect at an angle of $95.6^{\circ}$.

\section{In-lattice cooling\label{app:inlatticecooling}}

\begin{figure}
    \centering
    \includegraphics[width=83.2mm]{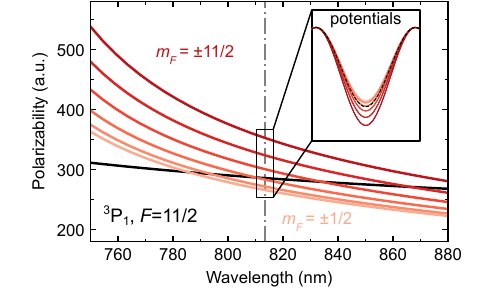}
    \caption{Atomic polarizabilities of the \textsuperscript{1}S\textsubscript{0} ground state (black line) and the \textsuperscript{3}P\textsubscript{1} $F=11/2$ manifold (red lines) as a function of wavelength for $\pi$-polarized trapping light. The gray dash-dotted line indicates the clock-magic trapping wavelength used in our experiment, \SI{813.4}{\nano\meter}. The inset schematically illustrates the trapping potential of a lattice site for the different internal states. Polarizabilities are given in atomic units (a.u.).}
    \label{fig:LevelsPolarizability}
\end{figure}

As discussed in Sec.~\ref{sec:sec2}, cooling is typically required to obtain a sufficient signal from each individual atom during fluorescence imaging. The specific cooling mechanism depends on the difference between the polarizability of the ground state, $\alpha_g$, and that of the excited state, $\alpha_e$. When $\alpha_e\not=\alpha_g$, the narrow linewidth of the transition can be exploited to excite the atom in specific regions of the trapping potential and allow it to evolve on the differently confined excited-state potential. On average, the atom then decays back to the ground-state potential at a potential energy different from the initial one, resulting in a net loss of kinetic energy per optical cycle. This process is known as attractive (repulsive) Sisyphus cooling for $\alpha_e>\alpha_g$ ($\alpha_e<\alpha_g$) and has been demonstrated in optical tweezer and lattice setups \cite{cooper_alkalineearth_2018,covey_2000times_2019,urech_narrowline_2022,buob_strontium_2024, tao_highfidelity_2024}. Instead, when $\alpha_g=\alpha_e$ the motional sidebands associated to different vibrational eigenstates can be spectroscopically resolved, enabling resolved sideband cooling \cite{cooper_alkalineearth_2018,norcia_microscopic_2018}. 

Figure~\ref{fig:LevelsPolarizability} shows the polarizabilities of the ground state and the excited \textsuperscript{3}P\textsubscript{1} $F'=11/2$ manifold of \textsuperscript{87}Sr. The optical potential at \SI{813.4}{\nano\meter} traps both manifolds, as indicated by the positive sign of their polarizabilities. However, in our configuration the \textsuperscript{3}P\textsubscript{1} manifold exhibits a tensor contribution to the polarizability. As a result, states with higher $|m_F|$ have a larger polarizability than the ground state, while those with smaller $|m_F|$ have a lower polarizability. We perform imaging and cooling on the cycling $ ^1\text{S}_0 \ket{9/2, -9/2}\to$ $ ^3\text{P}_1 \ket{11/2, -11/2}$ transition, for which $\alpha_g/\alpha_e \approx 0.8$, which enables efficient attractive Sisyphus cooling. We experimentally observe parity projection during this process. We attribute it to the vicinity of the narrow-line laser frequency to the red motional sideband of our lattice potential and to the length of the pulse ($300$ ms), which have been observed to result in efficient light-assisted collisions in tweezer experiments~\cite{cooper_alkalineearth_2018}.

By tuning the laser frequency, the same mechanism can also be used to induce heating. We exploit this effect to perform the selective spin removal employed in the imaging protocol described in Sec.~\ref{sec:sec4}. This spin-removal pulse consists of a \SI{6}{\milli\second} exposure to  blue-detuned light, during which the laser detuning is swept from 7 to $90~\Gamma$ with respect to the imaging transition. This induces heating of the atoms in the stretched state, which then escape the trapping potential.

\section{Image reconstruction \label{app:imaging}}

\begin{figure}
    \centering
    \includegraphics[width=\linewidth]{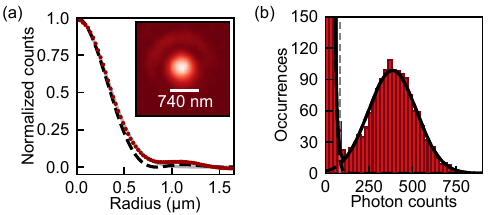}
    \caption{Imaging resolution and reconstruction of the atomic occupation. (a) Azimuthally averaged point-spread function (PSF) of the imaging system, compared with a Gaussian fit (gray line) and the diffraction-limited Airy disk (black dashed line). The inset shows the PSF obtained by averaging the fluorescence signal of 199 isolated atoms. (b) Histogram of photon counts per lattice site, accumulated over 64 images with an average filling of approximately 0.02 atoms per site in a $31\times31$ lattice-site region. Two well-separated peaks are visible: a background-noise peak near zero counts and an atom-signal peak centered at approximately 400 detected photons.}
    \label{fig:imagingPerformance}
\end{figure}

To extract the atomic occupations $n_{i,j}$ from each snapshot, we perform an image deconvolution. Specifically, we apply $50$ iterations of the Richardson-Lucy deconvolution algorithm~\cite{richardson_bayesianbased_1972, buob_strontium_2024}, using the measured point-spread function of the imaging system shown in Fig.~\ref{fig:imagingPerformance}(a). For images with a sparse filling of 0.02 atoms per site over a $31\times31$ lattice-site region, this procedure yields the histogram of photon counts per site shown in Fig.~\ref{fig:imagingPerformance}(b). Two well-separated peaks are visible, corresponding to the empty and occupied lattice sites. They can be distinguished using a threshold of 20 photons (gray dashed line). 

The quality of the reconstruction is quantified using the photon-count distributions for empty (noise) and occupied (signal) sites. We define the false negative probability as the fraction of occupied sites for which the photon count falls below the threshold, yielding $1.1(2)\%$. Conversely, the false positive probability is defined as the fraction of empty sites with counts above the threshold, yielding $0.51(5)\%$. These metrics provide a clear measure of the distinguishability of empty and occupied sites in the sparse-filling regime. At higher atomic densities, such as in the snapshot shown on the right of Fig.~\ref{fig:setup}(d), the reconstruction performance degrades due to the increased overlap of the fluorescence signals. In this regime, atom counting is instead performed using the reconstruction procedure described in Ref.~\cite{cheneau_reconstruction_2025}. For a filling of 0.4 atoms per site, we obtain false negative (positive) probabilities of approximately 2.4\% (1.6\%). Even under these conditions, the overall detection fidelity is limited by other error sources (see table in Fig.~\ref{fig:confusion}).

\section{Narrow-line optical pumping\label{app:pumpSpectrum}}

\subsection{Optical pumping pulses}

Narrow-line optical pumping is performed on the \textsuperscript{1}S\textsubscript{0},  $F = 9/2 \rightarrow$ \textsuperscript{3}P\textsubscript{1}, $F' = 9/2$ hyperfine manifold. This line is chosen because the Clebsch-Gordan coefficients are similar for all transitions, in contrast to the strongly asymmetric coefficients of the imaging line addressing the $F'=11/2$ manifold. The $F'=9/2$ manifold also exhibits a convenient Zeeman splitting, allowing all transitions to be spectrally resolved within the bandwidth of a double-pass acousto-optic modulator at the imaging magnetic field of \SI{20}{\gauss}. Optical pumping pulses are performed with an intensity of approximately $15$ $I$\textsubscript{sat} and a pulse duration of $2$ ms, and the optical pumping sequence is always repeated twice to maximize fidelity.

The results of Sec.~\ref{sec:sec3} reveal a significant difference between the optical pumping fidelities $\mathcal{F}_\text{OP,1}$ (pulse from $m_F=-7/2$ to the $m_F=-9/2$ state) and $\mathcal{F}_\text{OP,2}$ (pulses between the central states). We attribute this difference to experimental imperfections that affect the two processes differently. In the first case, a small fraction $\eta$ of $\pi-$polarized light in the optical pumping beam leads to a residual steady-state population of $\eta$ in the $m_F = -7/2$ state due to the two competing optical pumping processes between $m_F=-7/2$ and $m_F=-9/2$. In contrast, the same imperfection has negligible effect for the intermediate states because the optical pumping populates additional spin states, eventually emptying the initial one. 

\subsection{Spin selectivity}
To place a bound on the spin selectivity of our imaging scheme, i.e., on the fraction of atoms in  spin states other than the imaged one that are affected by the imaging pulse, we use the measurements of Fig.~\ref{fig:opticalpumping}(a) of Sec. \ref{sec:sec3}. Specifically, we analyze the residual signal in the image II taken after the depumping step. This signal corresponds to 0.5(1) atoms on average, compared with 25(2) atoms detected in the reference image I. 
After the depumping pulse, the atoms are expected to populate the $m_F=-7/2$ and $m_F=-5/2$ states in an approximate 60/40 ratio set by the corresponding Clebsch-Gordan coefficients. Therefore, any atom detected in image II
must either have remained in the stretched state after the depumping step or have been transferred off-resonantly to $m_F=-9/2$ from the other states by the imaging light. Because the off-resonant scattering rate decreases quadratically with detuning, most atoms undergoing this process are expected to originate from the $m_F=-7/2$ state. From these considerations, we infer that $<3.6(7)\%$ of the atoms in the $m_F=-7/2$ state are detected during imaging due to off-resonant scattering.

\subsection{Optical pumping spectrum}

\begin{figure*}
    \centering
    \includegraphics{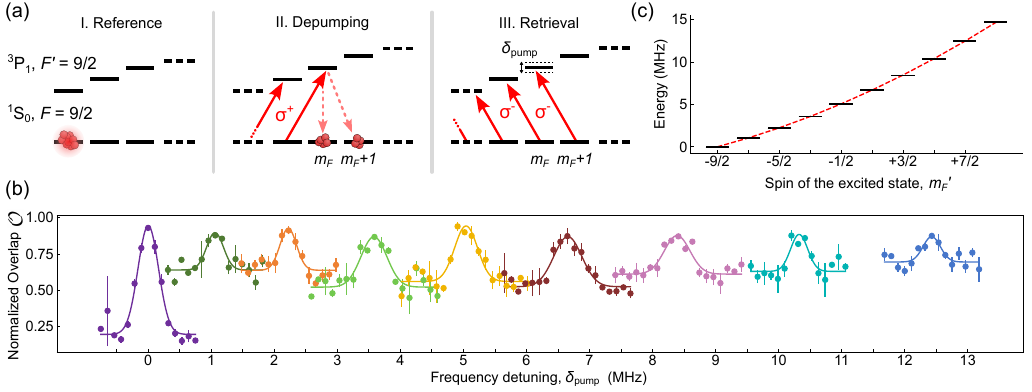}
    \caption{Hyperfine spectrum of the \textsuperscript{3}P\textsubscript{1}, $F'=9/2$ optical pumping transition. (a) We initialize the system in the $-9/2$ spin state and image its occupation in the lattice (left panel). We then populate the pair of spin states $m_F$ and $m_F + 1$ through optical pumping (central panel). Finally, we apply the retrieval pulses to bring the atomic population back to the imaging state and detect the atomic cloud again (right panel). In this last step, the frequency of the pulse that addresses the $m_F+1$ state is scanned with $\delta_\text{pump}$ to measure the corresponding resonance. (b) Measurement of the hyperfine spectrum using the method described in (a). The color coding matches the one used in the main text for the different $m_F$ states, considering here the ground $m_F$ state addressed by each transition. The resonances are well resolved, with a splitting that increases for higher $m_F$ values. Error bars denote the standard error of the mean. (c) Energy splitting of the excited-state manifold. The energies correspond to the peaks of the resonances measured in (b), extracted from Gaussian fits. The dashed red line is a fit to the theoretical Zeeman and AC Stark shifts, using the magnetic field as the only free parameter. }
    \label{fig:opticalspectrum}
\end{figure*}

To characterize the iterative optical pumping sequences used in our imaging scheme, we directly measure the spectrum of the optical pumping transitions. The procedure is similar to that described in Sec.~\ref{sec:sec3}, but here we scan the detuning of one of the optical pumping pulses (see Fig.~\ref{fig:opticalspectrum}(a)). We first prepare all atoms in the $m_F= -9/2$ spin state and take a reference image. We then pump the atoms sequentially to the $m_F$ and $m_F + 1$ states using pulses with $\sigma^+$ polarization. We finally return the atoms to the original state with $\sigma^-$ pulses and take a second image. During this final retrieval stage, we vary the frequency of the pulse addressing the $m_F+1$ state to locate the corresponding resonance.

From the two images, we extract the normalized overlap $\mathcal{O}$ and plot it as a function of the detuning $\delta_{\text{pump}}$ in Fig.~\ref{fig:opticalspectrum}(b). Repeating this procedure for different retrieval pulses yields the nine resonances shown, with colors indicating the specific $m_F+1$ state retrieved. We extract the resonant frequencies via Gaussian fits and plot them in the energy diagram of Fig.~\ref{fig:opticalspectrum}(c). The spacing between successive resonances increases for higher $m_F$ states, ranging approximately from $1-2$ MHz, reflecting the combined contributions of the Zeeman and AC Stark shifts. Fitting the data with the magnetic field as the only free parameter produces the red dashed curve, yielding a slightly corrected field of $B =$~\SI{19.2}{\gauss}.

Each resonance exhibits a finite baseline in $\mathcal{O}$, arising from atoms that remain in the $m_F$ state after the depumping stage and are successfully retrieved regardless of the pulse addressing the $m_F+1$ state. Using a similar approach, we also measure the resonance $ ^1\text{S}_0 \ket{9/2,+7/2}\to {^3\text{P}_1} \ket{9/2,+9/2}$, which is not shown in Fig.~\ref{fig:opticalspectrum}(b) because it can only be addressed with a $\sigma^+$ pulse. To probe this transition, we optically pump all atoms to the $m_F = +9/2$ state while scanning the frequency of the last depumping pulse, and then retrieve and image only the $m_F = +7/2$ atoms. For this resonance, the signal appears as a minimum in the normalized overlap, rather than a maximum. Although this measurement is omitted from Fig.~\ref{fig:opticalspectrum}(b), its corresponding resonance frequency is shown in the diagram of Fig.~\ref{fig:opticalspectrum}(c).   

\section{Spin populations and off-resonant Raman scattering \label{app:offResonant}}

Section~\ref{sec:sec4} demonstrates our ability to obtain quantum-gas microscopy images of all ten spin states of fermionic \textsuperscript{87}Sr in a single experimental realization. Here, we provide further details of the coincidence analysis between images of different spin states and discuss the origin of the spin-depolarization mechanism identified in the main text.

\begin{figure}
    \centering
    \includegraphics[width=\linewidth]{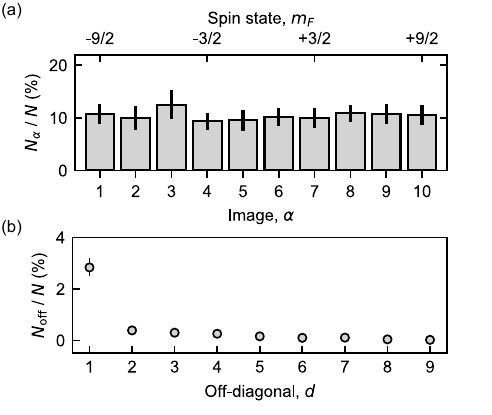}
    \caption{Spin population distribution in SU(10) systems. (a) Atom number $N_{\alpha}$ detected in each spin state, normalized by the total atom number $N$ and averaged over 40 experimental realizations. The error bars denote the standard deviation. (b) Plot of the number of multi-detection events in off-diagonal $d$, $N_\text{off} = \sum_{\alpha} N_{\alpha, \alpha + d}$ normalized by the total atom number, $N$. The main contribution arises from the first off-diagonal $d = 1$, where we find $2.8(4)\%$. The error bars represent a confidence interval of 95$\%$ obtained from 1000 bootstrap samples.}
    \label{fig:spinPopulation}
\end{figure}

In the main text, Fig.~\ref{fig:SU(10)}(c)  presents a coincidence analysis between images $\alpha$ and $\beta$ of different spin states acquired in a single experimental run, which allows us to link the multi-detection events observed in Fig.~\ref{fig:SU(10)}(b) to spin-depolarization during the imaging process. Here we present the same analysis, but averaged over 40 experimental realizations. Figure~\ref{fig:spinPopulation}(a) shows the ratio of atoms detected in image $\alpha$, $N_{\alpha}$, to the total atom number $N$. The distribution is approximately equal among the ten spin states, as expected. The diagonal axis of Fig.~\ref{fig:SU(10)}(c) shows a single realization of this dataset. Figure~\ref{fig:spinPopulation}(b) quantifies multi-detection events by plotting the summed overlaps along the $d$th off-diagonal, $N_\text{off} = \sum_{\alpha} N_{\alpha,\alpha+d}$. The dominant contribution occurs for $d=1$, corresponding to $2.8(4)\%$ of atoms detected in subsequent images. This nearest-image off-diagonal signal is consistent with spin depolarization during the \SI{300}{ms} imaging exposure. Summing all off-diagonal contributions yields a total depolarization probability of $4.2(4)\%$ per imaging cycle.

We identify off-resonant Raman scattering induced by the \SI{813.4}{\nano\meter} light generating the optical potential as the dominant mechanism behind this depolarization. Such scattering occurs from the excited ${{}^3\text{P}_1 \ket{11/2,-11/2}}$ state and populates the \textsuperscript{3}P\textsubscript{$J$} manifold via higher-lying excited states. Following Ref.~\cite{Dorscher_2018}, we estimate the scattering rate into other \textsuperscript{3}P\textsubscript{1} states to be on the order of \SI{1}{\per\second}. Atoms subsequently decaying to the ground-state manifold may populate states with $m_F \neq -9/2$, producing spin depolarization. A comparable scattering rate into the metastable \textsuperscript{3}P$_{0,2}$ manifolds ($\sim$\,\SI{1}{\per\second}) leads to population accumulation due to their long lifetimes. During imaging, repumpers (\SI{679}{\nano\meter} for \textsuperscript{3}P$_0$ and \SI{707}{\nano\meter} for \textsuperscript{3}P$_2$) return these atoms to the ground state, but the repumping process can still modify the final $m_F$ state. Since the final state depends on the number of repumping cycles, atoms that do not return to $m_F=-9/2$ preferentially populate nearby spin states. This naturally explains why the observed depolarization signal is dominated by the nearest-image contribution ($d=1$). The measured depolarization rate of $4.2(4)\%$ per imaging cycle is consistent with these estimated scattering rates.

\section{Spin-9/2 precession\label{app:Precession}}

\begin{figure*}
    \includegraphics{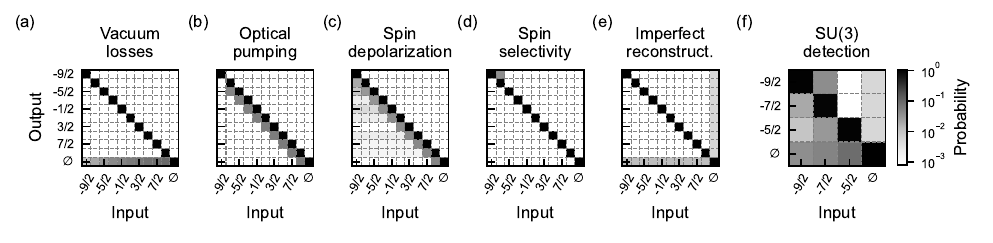}
    \caption{Partial detection-response matrices constructed from selected spurious processes listed in Fig.~\ref{fig:confusion}: (a) vacuum losses, (b) imperfect optical pumping, (c) spin depolarization, (d) spin selectivity, and (e) imperfect reconstruction. (f) Detection-response matrix for an SU(3) system with spin states $m_F=-9/2$, $-7/2$, and $-5/2$, relevant for the numerical study of Sec.~\ref{sec:sec6}. All matrices share the same color scale.}
    \label{fig:confusion_components}
\end{figure*}

Section~\ref{sec:sec5} of the main text benchmarks our spin-resolved imaging protocol through the observation of spin-9/2 Larmor precession. Here, we describe the dynamics within the ground-state manifold and how the measured populations are affected by finite imaging fidelities.

Consider an atom prepared in spin state $m_F=-9/2$ along the quantization axis $\hat{e}_z$. We suddenly rotate the magnetic field to point along $\hat{e}_y$ with magnitude $B_p$. After this quench, the initial state can be expressed in the new basis as

\begin{equation}
    |9/2,-9/2\rangle_z = \sum_k c_k |9/2,m_k\rangle_y,
\end{equation}
where $c_k$ are complex coefficients and $|9/2,m_k\rangle_y$ are the ten Zeeman states defined along the $\hat{e}_y$ axis. The unitary time evolution is then 
\begin{equation}
    |\Psi(t)\rangle = \sum_k c_k e^{-i E_k t/\hbar}|9/2,m_k\rangle_y,
\end{equation}
where $E_k$ is the energy of each spin state ${E_k = g_I  \mu_B B_p m_k}$. This evolution leads to Larmor precession, with periodic revivals of the magnetization at a period $T_p=h/(g_I \mu_B B_p)$. Projecting the state back along the original quantization axis $\hat{e}_z$, the population measured in a given $m_F$ state is $|_z\langle9/2,m_F |\Psi(t)\rangle|^2$.

To compare these theoretical predictions with our experimental measurements, we account for the finite fidelities of imaging process. For $m_F=-9/2$, the normalized overlap including the pinning fidelity is ${\mathcal{O}_{-9/2}=\mathcal{F}_\text{pin} |{}_z\langle9/2,-9/2|\Psi(t)\rangle|^2}$. For all other spin states, we also include optical pumping fidelity and vacuum losses:
\begin{equation*}
    \mathcal{O}_{m_F}=f_{m_F}|{}_z\langle9/2,m_F |\Psi(t)\rangle|^{2},
\end{equation*}
with 
\begin{equation*}
    f_{m_F}=\mathcal{F}_\text{pin}\mathcal{F}_\text{OP,1}(\mathcal{F}_\text{OP,2})^{7/2+m_F} (1-\mathcal{L}_{\text{vac}})^{9/2+m_F}.
\end{equation*}
This correction factor accounts for the combined effect of pinning, optical pumping, and vacuum losses. For a given $m_F$ state, the vacuum losses accumulate over the number of imaging cycles preceding its detection. Each cycle includes \SI{300}{\milli\second} of imaging exposure and \SI{600}{\milli\second} for camera readout, spin removal, and optical pumping. For the last spin state, $m_F=+9/2$, these losses accumulate to $(1-\mathcal{L}_{\text{vac}})^{9}=9.2(3)\%$.

The total occupation matrix $n^\text{tot}$, introduced in Sec.~\ref{sec:sec5}, is largely insensitive to spin-specific effects. This is because atoms not detected in the image corresponding to their spin state due to failed optical pumping or spin depolarization, are typically detected in a subsequent image, except for the last spin. Since the pinning fidelity is mainly limited by spin depolarization, we neglect its contribution when estimating $\mathcal{O}_\text{tot}$, and instead focus on the hopping rate $\mathcal{H}$ and the vacuum losses during imaging. Similarly, the optical pumping fidelities $\mathcal{F}_\text{OP,1}$ and $\mathcal{F}_\text{OP,2}$ primarily reflect atoms that fail to reach the target spin state after optical pumping, rather than atoms leaving their lattice sites. Therefore, we neglect as well the optical pumping fidelities and estimate the total overlap as

\begin{align*}
    \mathcal{O}_\text{tot}=&f_{9/2}|{}_z\langle9/2,9/2 |\Psi(t)\rangle|^2 \\+& \sum_{m_F=-9/2}^{7/2} f'_{m_F}|{}_z\langle9/2,m_F |\Psi(t)\rangle|^2,
\end{align*}
with 
\begin{equation*}
    f'_{m_F}=(1-\mathcal{H}-\mathcal{L}_{\text{vac}})(1-\mathcal{L}_{\text{vac}})^{9/2+m_F}.
\end{equation*}
These estimates allow us to numerically compute the dynamics of the system, which we compare with our experimental results in Sec.~\ref{sec:sec5}.

\section{Detection-response matrix\label{app:confusion}}

Section~\ref{sec:sec6} of the main text summarizes the different sources of infidelity of our detection protocol and their associated error probabilities in the table of Fig.~\ref{fig:confusion}. It also introduces a detection-response matrix that gives the probability for a given input spin-occupation state to be detected as a different output state, including the empty state $\varnothing$. Double occupancies are not included, as we assume parity projection during the imaging process.

To gain further insight into the contributions to this matrix, we construct partial detection-response matrices that include only one of the processes listed in the table. Figure~\ref{fig:confusion_components}(a) shows the effect of vacuum losses, which becomes stronger towards higher $m_F$ due to the sequential nature of our imaging protocol. As discussed in Sec.~\ref{sec:sec5}, optical pumping imperfections primarily result in atoms failing to reach the imaging state rather than being lost from the trap. The atom is therefore likely to be recovered when imaging the next spin state, leading to the matrix shown in Fig.~\ref{fig:confusion_components}(b).

The effect of spin depolarization, measured from the data in Fig.~\ref{fig:SU(10)}, is shown in Fig.~\ref{fig:confusion_components}(c). The spin selectivity of the imaging affects the neighboring spin state $m_F=-7/2$, which can be detected as $m_F=-9/2$ with a finite probability, as shown in Fig.~\ref{fig:confusion_components}(d). Finally, reconstruction infidelity is shown in Fig.~\ref{fig:confusion_components}(e), including the effects of false positives (last column) and false negatives (last row). The combination of these processes yields the full detection-response matrix shown in Fig.~\ref{fig:confusion}.

Hopping during imaging involves two sites and therefore its effect depends strongly on the specific spin configuration and cannot be easily treated within this framework. Due to its low probability of $1.0(3)\%$ compared to other processes, we neglect its effect. Imaging losses $\mathcal{L}$ are also not included independently, as they are mostly accounted for through vacuum losses and spin depolarization. A small residual infidelity, on the order of $1\%$ and thus comparable in magnitude to the neglected hopping contribution, is neglected for the same reason.

Finally, we adapt the full detection-response matrix to the SU(3) case used in the numerical study of Sec.~\ref{sec:sec6}. Assuming the use of the spin states $m_F=-9/2$, $-7/2$, and $-5/2$, we obtain the reduced matrix shown in Fig.~\ref{fig:confusion_components}(f).

% \clearpage

%apsrev4-2.bst 2019-01-14 (MD) hand-edited version of apsrev4-1.bst
%Control: key (0)
%Control: author (8) initials jnrlst
%Control: editor formatted (1) identically to author
%Control: production of article title (0) allowed
%Control: page (0) single
%Control: year (1) truncated
%Control: production of eprint (0) enabled
%


\begin{thebibliography}{93}%
\makeatletter
\providecommand \@ifxundefined [1]{%
 \@ifx{#1\undefined}
}%
\providecommand \@ifnum [1]{%
 \ifnum #1\expandafter \@firstoftwo
 \else \expandafter \@secondoftwo
 \fi
}%
\providecommand \@ifx [1]{%
 \ifx #1\expandafter \@firstoftwo
 \else \expandafter \@secondoftwo
 \fi
}%
\providecommand \natexlab [1]{#1}%
\providecommand \enquote  [1]{``#1''}%
\providecommand \bibnamefont  [1]{#1}%
\providecommand \bibfnamefont [1]{#1}%
\providecommand \citenamefont [1]{#1}%
\providecommand \href@noop [0]{\@secondoftwo}%
\providecommand \href [0]{\begingroup \@sanitize@url \@href}%
\providecommand \@href[1]{\@@startlink{#1}\@@href}%
\providecommand \@@href[1]{\endgroup#1\@@endlink}%
\providecommand \@sanitize@url [0]{\catcode `\\12\catcode `\$12\catcode
  `\&12\catcode `\#12\catcode `\^12\catcode `\_12\catcode `\%12\relax}%
\providecommand \@@startlink[1]{}%
\providecommand \@@endlink[0]{}%
\providecommand \url  [0]{\begingroup\@sanitize@url \@url }%
\providecommand \@url [1]{\endgroup\@href {#1}{\urlprefix }}%
\providecommand \urlprefix  [0]{URL }%
\providecommand \Eprint [0]{\href }%
\providecommand \doibase [0]{https://doi.org/}%
\providecommand \selectlanguage [0]{\@gobble}%
\providecommand \bibinfo  [0]{\@secondoftwo}%
\providecommand \bibfield  [0]{\@secondoftwo}%
\providecommand \translation [1]{[#1]}%
\providecommand \BibitemOpen [0]{}%
\providecommand \bibitemStop [0]{}%
\providecommand \bibitemNoStop [0]{.\EOS\space}%
\providecommand \EOS [0]{\spacefactor3000\relax}%
\providecommand \BibitemShut  [1]{\csname bibitem#1\endcsname}%
\let\auto@bib@innerbib\@empty
%</preamble>
\bibitem [{\citenamefont {Browaeys}\ and\ \citenamefont
  {Lahaye}(2020)}]{Browaeys2020}%
  \BibitemOpen
  \bibfield  {author} {\bibinfo {author} {\bibfnamefont {A.}~\bibnamefont
  {Browaeys}}\ and\ \bibinfo {author} {\bibfnamefont {T.}~\bibnamefont
  {Lahaye}},\ }\bibfield  {title} {\bibinfo {title} {Many-body physics with
  individually controlled {R}ydberg atoms},\ }\href
  {https://doi.org/10.1038/s41567-019-0733-z} {\bibfield  {journal} {\bibinfo
  {journal} {Nat. Phys.}\ }\textbf {\bibinfo {volume} {16}},\ \bibinfo {pages}
  {132} (\bibinfo {year} {2020})}\BibitemShut {NoStop}%
\bibitem [{\citenamefont {Gross}\ and\ \citenamefont
  {Bakr}(2021)}]{gross_quantum_2021}%
  \BibitemOpen
  \bibfield  {author} {\bibinfo {author} {\bibfnamefont {C.}~\bibnamefont
  {Gross}}\ and\ \bibinfo {author} {\bibfnamefont {W.~S.}\ \bibnamefont
  {Bakr}},\ }\bibfield  {title} {\bibinfo {title} {Quantum gas microscopy for
  single atom and spin detection},\ }\href
  {https://doi.org/10.1038/s41567-021-01370-5} {\bibfield  {journal} {\bibinfo
  {journal} {Nat. Phys.}\ }\textbf {\bibinfo {volume} {17}},\ \bibinfo {pages}
  {1316} (\bibinfo {year} {2021})}\BibitemShut {NoStop}%
\bibitem [{\citenamefont {Parsons}\ \emph {et~al.}(2016)\citenamefont
  {Parsons}, \citenamefont {Mazurenko}, \citenamefont {Chiu}, \citenamefont
  {Ji}, \citenamefont {Greif},\ and\ \citenamefont
  {Greiner}}]{parsons_siteresolved_2016}%
  \BibitemOpen
  \bibfield  {author} {\bibinfo {author} {\bibfnamefont {M.~F.}\ \bibnamefont
  {Parsons}}, \bibinfo {author} {\bibfnamefont {A.}~\bibnamefont {Mazurenko}},
  \bibinfo {author} {\bibfnamefont {C.~S.}\ \bibnamefont {Chiu}}, \bibinfo
  {author} {\bibfnamefont {G.}~\bibnamefont {Ji}}, \bibinfo {author}
  {\bibfnamefont {D.}~\bibnamefont {Greif}},\ and\ \bibinfo {author}
  {\bibfnamefont {M.}~\bibnamefont {Greiner}},\ }\bibfield  {title} {\bibinfo
  {title} {Site-resolved measurement of the spin-correlation function in the
  {{Fermi-Hubbard}} model},\ }\href {https://doi.org/10.1126/science.aag1430}
  {\bibfield  {journal} {\bibinfo  {journal} {Science}\ }\textbf {\bibinfo
  {volume} {353}},\ \bibinfo {pages} {1253} (\bibinfo {year}
  {2016})}\BibitemShut {NoStop}%
\bibitem [{\citenamefont {Cheuk}\ \emph {et~al.}(2016)\citenamefont {Cheuk},
  \citenamefont {Nichols}, \citenamefont {Lawrence}, \citenamefont {Okan},
  \citenamefont {Zhang}, \citenamefont {Khatami}, \citenamefont {Trivedi},
  \citenamefont {Paiva}, \citenamefont {Rigol},\ and\ \citenamefont
  {Zwierlein}}]{cheuk_observation_2016a}%
  \BibitemOpen
  \bibfield  {author} {\bibinfo {author} {\bibfnamefont {L.~W.}\ \bibnamefont
  {Cheuk}}, \bibinfo {author} {\bibfnamefont {M.~A.}\ \bibnamefont {Nichols}},
  \bibinfo {author} {\bibfnamefont {K.~R.}\ \bibnamefont {Lawrence}}, \bibinfo
  {author} {\bibfnamefont {M.}~\bibnamefont {Okan}}, \bibinfo {author}
  {\bibfnamefont {H.}~\bibnamefont {Zhang}}, \bibinfo {author} {\bibfnamefont
  {E.}~\bibnamefont {Khatami}}, \bibinfo {author} {\bibfnamefont
  {N.}~\bibnamefont {Trivedi}}, \bibinfo {author} {\bibfnamefont
  {T.}~\bibnamefont {Paiva}}, \bibinfo {author} {\bibfnamefont
  {M.}~\bibnamefont {Rigol}},\ and\ \bibinfo {author} {\bibfnamefont {M.~W.}\
  \bibnamefont {Zwierlein}},\ }\bibfield  {title} {\bibinfo {title}
  {Observation of spatial charge and spin correlations in the {{2D
  Fermi-Hubbard}} model},\ }\href {https://doi.org/10.1126/science.aag3349}
  {\bibfield  {journal} {\bibinfo  {journal} {Science}\ }\textbf {\bibinfo
  {volume} {353}},\ \bibinfo {pages} {1260} (\bibinfo {year}
  {2016})}\BibitemShut {NoStop}%
\bibitem [{\citenamefont {Boll}\ \emph {et~al.}(2016)\citenamefont {Boll},
  \citenamefont {Hilker}, \citenamefont {Salomon}, \citenamefont {Omran},
  \citenamefont {Nespolo}, \citenamefont {Pollet}, \citenamefont {Bloch},\ and\
  \citenamefont {Gross}}]{boll_spin_2016}%
  \BibitemOpen
  \bibfield  {author} {\bibinfo {author} {\bibfnamefont {M.}~\bibnamefont
  {Boll}}, \bibinfo {author} {\bibfnamefont {T.~A.}\ \bibnamefont {Hilker}},
  \bibinfo {author} {\bibfnamefont {G.}~\bibnamefont {Salomon}}, \bibinfo
  {author} {\bibfnamefont {A.}~\bibnamefont {Omran}}, \bibinfo {author}
  {\bibfnamefont {J.}~\bibnamefont {Nespolo}}, \bibinfo {author} {\bibfnamefont
  {L.}~\bibnamefont {Pollet}}, \bibinfo {author} {\bibfnamefont
  {I.}~\bibnamefont {Bloch}},\ and\ \bibinfo {author} {\bibfnamefont
  {C.}~\bibnamefont {Gross}},\ }\bibfield  {title} {\bibinfo {title} {Spin- and
  density-resolved microscopy of antiferromagnetic correlations in
  {{Fermi-Hubbard}} chains},\ }\href {https://doi.org/10.1126/science.aag1635}
  {\bibfield  {journal} {\bibinfo  {journal} {Science}\ }\textbf {\bibinfo
  {volume} {353}},\ \bibinfo {pages} {1257} (\bibinfo {year}
  {2016})}\BibitemShut {NoStop}%
\bibitem [{\citenamefont {Fukuhara}\ \emph {et~al.}(2013)\citenamefont
  {Fukuhara}, \citenamefont {Kantian}, \citenamefont {Endres}, \citenamefont
  {Cheneau}, \citenamefont {Schau{\ss}}, \citenamefont {Hild}, \citenamefont
  {Bellem}, \citenamefont {Schollw{\"o}ck}, \citenamefont {Giamarchi},
  \citenamefont {Gross}, \citenamefont {Bloch},\ and\ \citenamefont
  {Kuhr}}]{fukuhara_quantum_2013}%
  \BibitemOpen
  \bibfield  {author} {\bibinfo {author} {\bibfnamefont {T.}~\bibnamefont
  {Fukuhara}}, \bibinfo {author} {\bibfnamefont {A.}~\bibnamefont {Kantian}},
  \bibinfo {author} {\bibfnamefont {M.}~\bibnamefont {Endres}}, \bibinfo
  {author} {\bibfnamefont {M.}~\bibnamefont {Cheneau}}, \bibinfo {author}
  {\bibfnamefont {P.}~\bibnamefont {Schau{\ss}}}, \bibinfo {author}
  {\bibfnamefont {S.}~\bibnamefont {Hild}}, \bibinfo {author} {\bibfnamefont
  {D.}~\bibnamefont {Bellem}}, \bibinfo {author} {\bibfnamefont
  {U.}~\bibnamefont {Schollw{\"o}ck}}, \bibinfo {author} {\bibfnamefont
  {T.}~\bibnamefont {Giamarchi}}, \bibinfo {author} {\bibfnamefont
  {C.}~\bibnamefont {Gross}}, \bibinfo {author} {\bibfnamefont
  {I.}~\bibnamefont {Bloch}},\ and\ \bibinfo {author} {\bibfnamefont
  {S.}~\bibnamefont {Kuhr}},\ }\bibfield  {title} {\bibinfo {title} {Quantum
  dynamics of a mobile spin impurity},\ }\href
  {https://doi.org/10.1038/nphys2561} {\bibfield  {journal} {\bibinfo
  {journal} {Nat. Phys.}\ }\textbf {\bibinfo {volume} {9}},\ \bibinfo {pages}
  {235} (\bibinfo {year} {2013})}\BibitemShut {NoStop}%
\bibitem [{\citenamefont {Wei}\ \emph {et~al.}(2022)\citenamefont {Wei},
  \citenamefont {{Rubio-Abadal}}, \citenamefont {Ye}, \citenamefont {Machado},
  \citenamefont {Kemp}, \citenamefont {Srakaew}, \citenamefont {Hollerith},
  \citenamefont {Rui}, \citenamefont {Gopalakrishnan}, \citenamefont {Yao},
  \citenamefont {Bloch},\ and\ \citenamefont {Zeiher}}]{wei_quantum_2022}%
  \BibitemOpen
  \bibfield  {author} {\bibinfo {author} {\bibfnamefont {D.}~\bibnamefont
  {Wei}}, \bibinfo {author} {\bibfnamefont {A.}~\bibnamefont {{Rubio-Abadal}}},
  \bibinfo {author} {\bibfnamefont {B.}~\bibnamefont {Ye}}, \bibinfo {author}
  {\bibfnamefont {F.}~\bibnamefont {Machado}}, \bibinfo {author} {\bibfnamefont
  {J.}~\bibnamefont {Kemp}}, \bibinfo {author} {\bibfnamefont {K.}~\bibnamefont
  {Srakaew}}, \bibinfo {author} {\bibfnamefont {S.}~\bibnamefont {Hollerith}},
  \bibinfo {author} {\bibfnamefont {J.}~\bibnamefont {Rui}}, \bibinfo {author}
  {\bibfnamefont {S.}~\bibnamefont {Gopalakrishnan}}, \bibinfo {author}
  {\bibfnamefont {N.~Y.}\ \bibnamefont {Yao}}, \bibinfo {author} {\bibfnamefont
  {I.}~\bibnamefont {Bloch}},\ and\ \bibinfo {author} {\bibfnamefont
  {J.}~\bibnamefont {Zeiher}},\ }\bibfield  {title} {\bibinfo {title} {Quantum
  gas microscopy of {{Kardar-Parisi-Zhang}} superdiffusion},\ }\href
  {https://doi.org/10.1126/science.abk2397} {\bibfield  {journal} {\bibinfo
  {journal} {Science}\ }\textbf {\bibinfo {volume} {376}},\ \bibinfo {pages}
  {716} (\bibinfo {year} {2022})}\BibitemShut {NoStop}%
\bibitem [{\citenamefont {Labuhn}\ \emph {et~al.}(2016)\citenamefont {Labuhn},
  \citenamefont {Barredo}, \citenamefont {Ravets}, \citenamefont
  {de~Léséleuc}, \citenamefont {Macrì}, \citenamefont {Lahaye},\ and\
  \citenamefont {Browaeys}}]{Labuhn2016Tunable}%
  \BibitemOpen
  \bibfield  {author} {\bibinfo {author} {\bibfnamefont {H.}~\bibnamefont
  {Labuhn}}, \bibinfo {author} {\bibfnamefont {D.}~\bibnamefont {Barredo}},
  \bibinfo {author} {\bibfnamefont {S.}~\bibnamefont {Ravets}}, \bibinfo
  {author} {\bibfnamefont {S.}~\bibnamefont {de~Léséleuc}}, \bibinfo {author}
  {\bibfnamefont {T.}~\bibnamefont {Macrì}}, \bibinfo {author} {\bibfnamefont
  {T.}~\bibnamefont {Lahaye}},\ and\ \bibinfo {author} {\bibfnamefont
  {A.}~\bibnamefont {Browaeys}},\ }\bibfield  {title} {\bibinfo {title}
  {Tunable two-dimensional arrays of single {Rydberg} atoms for realizing
  quantum {Ising} models},\ }\href {https://doi.org/10.1038/nature18274}
  {\bibfield  {journal} {\bibinfo  {journal} {Nature}\ }\textbf {\bibinfo
  {volume} {534}},\ \bibinfo {pages} {667–670} (\bibinfo {year}
  {2016})}\BibitemShut {NoStop}%
\bibitem [{\citenamefont {Bernien}\ \emph {et~al.}(2017)\citenamefont
  {Bernien}, \citenamefont {Schwartz}, \citenamefont {Keesling}, \citenamefont
  {Levine}, \citenamefont {Omran}, \citenamefont {Pichler}, \citenamefont
  {Choi}, \citenamefont {Zibrov}, \citenamefont {Endres}, \citenamefont
  {Greiner}, \citenamefont {Vuletić},\ and\ \citenamefont
  {Lukin}}]{Bernien2017Probing}%
  \BibitemOpen
  \bibfield  {author} {\bibinfo {author} {\bibfnamefont {H.}~\bibnamefont
  {Bernien}}, \bibinfo {author} {\bibfnamefont {S.}~\bibnamefont {Schwartz}},
  \bibinfo {author} {\bibfnamefont {A.}~\bibnamefont {Keesling}}, \bibinfo
  {author} {\bibfnamefont {H.}~\bibnamefont {Levine}}, \bibinfo {author}
  {\bibfnamefont {A.}~\bibnamefont {Omran}}, \bibinfo {author} {\bibfnamefont
  {H.}~\bibnamefont {Pichler}}, \bibinfo {author} {\bibfnamefont
  {S.}~\bibnamefont {Choi}}, \bibinfo {author} {\bibfnamefont {A.~S.}\
  \bibnamefont {Zibrov}}, \bibinfo {author} {\bibfnamefont {M.}~\bibnamefont
  {Endres}}, \bibinfo {author} {\bibfnamefont {M.}~\bibnamefont {Greiner}},
  \bibinfo {author} {\bibfnamefont {V.}~\bibnamefont {Vuletić}},\ and\
  \bibinfo {author} {\bibfnamefont {M.~D.}\ \bibnamefont {Lukin}},\ }\bibfield
  {title} {\bibinfo {title} {Probing many-body dynamics on a 51-atom quantum
  simulator},\ }\href {https://doi.org/10.1038/nature24622} {\bibfield
  {journal} {\bibinfo  {journal} {Nature}\ }\textbf {\bibinfo {volume} {551}},\
  \bibinfo {pages} {579–584} (\bibinfo {year} {2017})}\BibitemShut {NoStop}%
\bibitem [{\citenamefont {de~L\'es\'eleuc}\ \emph {et~al.}(2019)\citenamefont
  {de~L\'es\'eleuc}, \citenamefont {Lienhard}, \citenamefont {Scholl},
  \citenamefont {Barredo}, \citenamefont {Weber}, \citenamefont {Lang},
  \citenamefont {Büchler}, \citenamefont {Lahaye},\ and\ \citenamefont
  {Browaeys}}]{deLeseleuc2019Observation}%
  \BibitemOpen
  \bibfield  {author} {\bibinfo {author} {\bibfnamefont {S.}~\bibnamefont
  {de~L\'es\'eleuc}}, \bibinfo {author} {\bibfnamefont {V.}~\bibnamefont
  {Lienhard}}, \bibinfo {author} {\bibfnamefont {P.}~\bibnamefont {Scholl}},
  \bibinfo {author} {\bibfnamefont {D.}~\bibnamefont {Barredo}}, \bibinfo
  {author} {\bibfnamefont {S.}~\bibnamefont {Weber}}, \bibinfo {author}
  {\bibfnamefont {N.}~\bibnamefont {Lang}}, \bibinfo {author} {\bibfnamefont
  {H.~P.}\ \bibnamefont {Büchler}}, \bibinfo {author} {\bibfnamefont
  {T.}~\bibnamefont {Lahaye}},\ and\ \bibinfo {author} {\bibfnamefont
  {A.}~\bibnamefont {Browaeys}},\ }\bibfield  {title} {\bibinfo {title}
  {Observation of a symmetry-protected topological phase of interacting bosons
  with {Rydberg} atoms},\ }\href {https://doi.org/10.1126/science.aav9105}
  {\bibfield  {journal} {\bibinfo  {journal} {Science}\ }\textbf {\bibinfo
  {volume} {365}},\ \bibinfo {pages} {775} (\bibinfo {year}
  {2019})}\BibitemShut {NoStop}%
\bibitem [{\citenamefont {Koepsell}\ \emph {et~al.}(2020)\citenamefont
  {Koepsell}, \citenamefont {Hirthe}, \citenamefont {Bourgund}, \citenamefont
  {Sompet}, \citenamefont {Vijayan}, \citenamefont {Salomon}, \citenamefont
  {Gross},\ and\ \citenamefont {Bloch}}]{Koepsell2020}%
  \BibitemOpen
  \bibfield  {author} {\bibinfo {author} {\bibfnamefont {J.}~\bibnamefont
  {Koepsell}}, \bibinfo {author} {\bibfnamefont {S.}~\bibnamefont {Hirthe}},
  \bibinfo {author} {\bibfnamefont {D.}~\bibnamefont {Bourgund}}, \bibinfo
  {author} {\bibfnamefont {P.}~\bibnamefont {Sompet}}, \bibinfo {author}
  {\bibfnamefont {J.}~\bibnamefont {Vijayan}}, \bibinfo {author} {\bibfnamefont
  {G.}~\bibnamefont {Salomon}}, \bibinfo {author} {\bibfnamefont
  {C.}~\bibnamefont {Gross}},\ and\ \bibinfo {author} {\bibfnamefont
  {I.}~\bibnamefont {Bloch}},\ }\bibfield  {title} {\bibinfo {title} {Robust
  bilayer charge pumping for spin- and density-resolved quantum gas
  microscopy},\ }\href {https://doi.org/10.1103/PhysRevLett.125.010403}
  {\bibfield  {journal} {\bibinfo  {journal} {Phys. Rev. Lett.}\ }\textbf
  {\bibinfo {volume} {125}},\ \bibinfo {pages} {010403} (\bibinfo {year}
  {2020})}\BibitemShut {NoStop}%
\bibitem [{\citenamefont {Yan}\ \emph {et~al.}(2022)\citenamefont {Yan},
  \citenamefont {Spar}, \citenamefont {Prichard}, \citenamefont {Chi},
  \citenamefont {Wei}, \citenamefont {Ibarra-Garc\'{\i}a-Padilla},
  \citenamefont {Hazzard},\ and\ \citenamefont {Bakr}}]{Yan2022}%
  \BibitemOpen
  \bibfield  {author} {\bibinfo {author} {\bibfnamefont {Z.~Z.}\ \bibnamefont
  {Yan}}, \bibinfo {author} {\bibfnamefont {B.~M.}\ \bibnamefont {Spar}},
  \bibinfo {author} {\bibfnamefont {M.~L.}\ \bibnamefont {Prichard}}, \bibinfo
  {author} {\bibfnamefont {S.}~\bibnamefont {Chi}}, \bibinfo {author}
  {\bibfnamefont {H.-T.}\ \bibnamefont {Wei}}, \bibinfo {author} {\bibfnamefont
  {E.}~\bibnamefont {Ibarra-Garc\'{\i}a-Padilla}}, \bibinfo {author}
  {\bibfnamefont {K.~R.~A.}\ \bibnamefont {Hazzard}},\ and\ \bibinfo {author}
  {\bibfnamefont {W.~S.}\ \bibnamefont {Bakr}},\ }\bibfield  {title} {\bibinfo
  {title} {Two-dimensional programmable tweezer arrays of fermions},\ }\href
  {https://doi.org/10.1103/PhysRevLett.129.123201} {\bibfield  {journal}
  {\bibinfo  {journal} {Phys. Rev. Lett.}\ }\textbf {\bibinfo {volume} {129}},\
  \bibinfo {pages} {123201} (\bibinfo {year} {2022})}\BibitemShut {NoStop}%
\bibitem [{\citenamefont {Wu}\ \emph {et~al.}(2019)\citenamefont {Wu},
  \citenamefont {Kumar}, \citenamefont {Giraldo},\ and\ \citenamefont
  {Weiss}}]{Wu_SternGerlach_Qubit_2019}%
  \BibitemOpen
  \bibfield  {author} {\bibinfo {author} {\bibfnamefont {T.-Y.}\ \bibnamefont
  {Wu}}, \bibinfo {author} {\bibfnamefont {A.}~\bibnamefont {Kumar}}, \bibinfo
  {author} {\bibfnamefont {F.}~\bibnamefont {Giraldo}},\ and\ \bibinfo {author}
  {\bibfnamefont {D.~S.}\ \bibnamefont {Weiss}},\ }\bibfield  {title} {\bibinfo
  {title} {Stern–{G}erlach detection of neutral-atom qubits in a
  state-dependent optical lattice},\ }\href
  {https://doi.org/https://doi.org/10.1038/s41567-019-0478-8} {\bibfield
  {journal} {\bibinfo  {journal} {Nat. Phys.}\ }\textbf {\bibinfo {volume}
  {15}},\ \bibinfo {pages} {538} (\bibinfo {year} {2019})}\BibitemShut
  {NoStop}%
\bibitem [{\citenamefont {Bluvstein}\ \emph {et~al.}(2026)\citenamefont
  {Bluvstein}, \citenamefont {Geim}, \citenamefont {Li}, \citenamefont
  {Evered}, \citenamefont {Ataides}, \citenamefont {Baranes}, \citenamefont
  {Gu}, \citenamefont {Manovitz}, \citenamefont {Xu}, \citenamefont
  {Kalinowski}, \citenamefont {Majidy}, \citenamefont {Kokail}, \citenamefont
  {Maskara}, \citenamefont {Trapp}, \citenamefont {Stewart}, \citenamefont
  {Hollerith}, \citenamefont {Zhou}, \citenamefont {Gullans}, \citenamefont
  {Yelin}, \citenamefont {Greiner}, \citenamefont {Vuleti\'{c}}, \citenamefont
  {Cain},\ and\ \citenamefont {Lukin}}]{bluvstein2025architect}%
  \BibitemOpen
  \bibfield  {author} {\bibinfo {author} {\bibfnamefont {D.}~\bibnamefont
  {Bluvstein}}, \bibinfo {author} {\bibfnamefont {A.~A.}\ \bibnamefont {Geim}},
  \bibinfo {author} {\bibfnamefont {S.~H.}\ \bibnamefont {Li}}, \bibinfo
  {author} {\bibfnamefont {S.~J.}\ \bibnamefont {Evered}}, \bibinfo {author}
  {\bibfnamefont {J.~P.~B.}\ \bibnamefont {Ataides}}, \bibinfo {author}
  {\bibfnamefont {G.}~\bibnamefont {Baranes}}, \bibinfo {author} {\bibfnamefont
  {A.}~\bibnamefont {Gu}}, \bibinfo {author} {\bibfnamefont {T.}~\bibnamefont
  {Manovitz}}, \bibinfo {author} {\bibfnamefont {M.}~\bibnamefont {Xu}},
  \bibinfo {author} {\bibfnamefont {M.}~\bibnamefont {Kalinowski}}, \bibinfo
  {author} {\bibfnamefont {S.}~\bibnamefont {Majidy}}, \bibinfo {author}
  {\bibfnamefont {C.}~\bibnamefont {Kokail}}, \bibinfo {author} {\bibfnamefont
  {N.}~\bibnamefont {Maskara}}, \bibinfo {author} {\bibfnamefont {E.~C.}\
  \bibnamefont {Trapp}}, \bibinfo {author} {\bibfnamefont {L.~M.}\ \bibnamefont
  {Stewart}}, \bibinfo {author} {\bibfnamefont {S.}~\bibnamefont {Hollerith}},
  \bibinfo {author} {\bibfnamefont {H.}~\bibnamefont {Zhou}}, \bibinfo {author}
  {\bibfnamefont {M.~J.}\ \bibnamefont {Gullans}}, \bibinfo {author}
  {\bibfnamefont {S.~F.}\ \bibnamefont {Yelin}}, \bibinfo {author}
  {\bibfnamefont {M.}~\bibnamefont {Greiner}}, \bibinfo {author} {\bibfnamefont
  {V.}~\bibnamefont {Vuleti\'{c}}}, \bibinfo {author} {\bibfnamefont
  {M.}~\bibnamefont {Cain}},\ and\ \bibinfo {author} {\bibfnamefont {M.~D.}\
  \bibnamefont {Lukin}},\ }\bibfield  {title} {\bibinfo {title} {A
  fault-tolerant neutral-atom architecture for universal quantum computation},\
  }\href {https://doi.org/10.1038/s41586-025-09848-5} {\bibfield  {journal}
  {\bibinfo  {journal} {Nature}\ }\textbf {\bibinfo {volume} {649}},\ \bibinfo
  {pages} {39} (\bibinfo {year} {2026})}\BibitemShut {NoStop}%
\bibitem [{\citenamefont {Widera}\ \emph {et~al.}(2005)\citenamefont {Widera},
  \citenamefont {Gerbier}, \citenamefont {F\"olling}, \citenamefont {Gericke},
  \citenamefont {Mandel},\ and\ \citenamefont {Bloch}}]{Widera2005}%
  \BibitemOpen
  \bibfield  {author} {\bibinfo {author} {\bibfnamefont {A.}~\bibnamefont
  {Widera}}, \bibinfo {author} {\bibfnamefont {F.}~\bibnamefont {Gerbier}},
  \bibinfo {author} {\bibfnamefont {S.}~\bibnamefont {F\"olling}}, \bibinfo
  {author} {\bibfnamefont {T.}~\bibnamefont {Gericke}}, \bibinfo {author}
  {\bibfnamefont {O.}~\bibnamefont {Mandel}},\ and\ \bibinfo {author}
  {\bibfnamefont {I.}~\bibnamefont {Bloch}},\ }\bibfield  {title} {\bibinfo
  {title} {Coherent collisional spin dynamics in optical lattices},\ }\href
  {https://doi.org/10.1103/PhysRevLett.95.190405} {\bibfield  {journal}
  {\bibinfo  {journal} {Phys. Rev. Lett.}\ }\textbf {\bibinfo {volume} {95}},\
  \bibinfo {pages} {190405} (\bibinfo {year} {2005})}\BibitemShut {NoStop}%
\bibitem [{\citenamefont {Krauser}\ \emph {et~al.}(2012)\citenamefont
  {Krauser}, \citenamefont {Heinze}, \citenamefont {Fl{\"a}schner},
  \citenamefont {G{\"o}tze}, \citenamefont {Becker},\ and\ \citenamefont
  {Sengstock}}]{Krauser2012}%
  \BibitemOpen
  \bibfield  {author} {\bibinfo {author} {\bibfnamefont {J.~S.}\ \bibnamefont
  {Krauser}}, \bibinfo {author} {\bibfnamefont {J.}~\bibnamefont {Heinze}},
  \bibinfo {author} {\bibfnamefont {N.}~\bibnamefont {Fl{\"a}schner}}, \bibinfo
  {author} {\bibfnamefont {S.}~\bibnamefont {G{\"o}tze}}, \bibinfo {author}
  {\bibfnamefont {C.}~\bibnamefont {Becker}},\ and\ \bibinfo {author}
  {\bibfnamefont {K.}~\bibnamefont {Sengstock}},\ }\bibfield  {title} {\bibinfo
  {title} {Coherent multi-flavour spin dynamics in a fermionic quantum gas},\
  }\href {https://doi.org/10.1038/nphys2409} {\bibfield  {journal} {\bibinfo
  {journal} {Nat. Phys.}\ }\textbf {\bibinfo {volume} {8}},\ \bibinfo {pages}
  {813} (\bibinfo {year} {2012})}\BibitemShut {NoStop}%
\bibitem [{\citenamefont {de~Paz}\ \emph {et~al.}(2013)\citenamefont {de~Paz},
  \citenamefont {Chotia}, \citenamefont {Mar\'echal}, \citenamefont {Pedri},
  \citenamefont {Vernac}, \citenamefont {Gorceix},\ and\ \citenamefont
  {Laburthe-Tolra}}]{dePaz2013}%
  \BibitemOpen
  \bibfield  {author} {\bibinfo {author} {\bibfnamefont {A.}~\bibnamefont
  {de~Paz}}, \bibinfo {author} {\bibfnamefont {A.}~\bibnamefont {Chotia}},
  \bibinfo {author} {\bibfnamefont {E.}~\bibnamefont {Mar\'echal}}, \bibinfo
  {author} {\bibfnamefont {P.}~\bibnamefont {Pedri}}, \bibinfo {author}
  {\bibfnamefont {L.}~\bibnamefont {Vernac}}, \bibinfo {author} {\bibfnamefont
  {O.}~\bibnamefont {Gorceix}},\ and\ \bibinfo {author} {\bibfnamefont
  {B.}~\bibnamefont {Laburthe-Tolra}},\ }\bibfield  {title} {\bibinfo {title}
  {Resonant demagnetization of a dipolar {Bose-Einstein} condensate in a
  three-dimensional optical lattice},\ }\href
  {https://doi.org/10.1103/PhysRevA.87.051609} {\bibfield  {journal} {\bibinfo
  {journal} {Phys. Rev. A}\ }\textbf {\bibinfo {volume} {87}},\ \bibinfo
  {pages} {051609} (\bibinfo {year} {2013})}\BibitemShut {NoStop}%
\bibitem [{\citenamefont {Patscheider}\ \emph {et~al.}(2020)\citenamefont
  {Patscheider}, \citenamefont {Zhu}, \citenamefont {Chomaz}, \citenamefont
  {Petter}, \citenamefont {Baier}, \citenamefont {Rey}, \citenamefont
  {Ferlaino},\ and\ \citenamefont {Mark}}]{Patcheider2020}%
  \BibitemOpen
  \bibfield  {author} {\bibinfo {author} {\bibfnamefont {A.}~\bibnamefont
  {Patscheider}}, \bibinfo {author} {\bibfnamefont {B.}~\bibnamefont {Zhu}},
  \bibinfo {author} {\bibfnamefont {L.}~\bibnamefont {Chomaz}}, \bibinfo
  {author} {\bibfnamefont {D.}~\bibnamefont {Petter}}, \bibinfo {author}
  {\bibfnamefont {S.}~\bibnamefont {Baier}}, \bibinfo {author} {\bibfnamefont
  {A.-M.}\ \bibnamefont {Rey}}, \bibinfo {author} {\bibfnamefont
  {F.}~\bibnamefont {Ferlaino}},\ and\ \bibinfo {author} {\bibfnamefont
  {M.~J.}\ \bibnamefont {Mark}},\ }\bibfield  {title} {\bibinfo {title}
  {Controlling dipolar exchange interactions in a dense three-dimensional array
  of large-spin fermions},\ }\href
  {https://doi.org/10.1103/PhysRevResearch.2.023050} {\bibfield  {journal}
  {\bibinfo  {journal} {Phys. Rev. Res.}\ }\textbf {\bibinfo {volume} {2}},\
  \bibinfo {pages} {023050} (\bibinfo {year} {2020})}\BibitemShut {NoStop}%
\bibitem [{\citenamefont {Ottenstein}\ \emph {et~al.}(2008)\citenamefont
  {Ottenstein}, \citenamefont {Lompe}, \citenamefont {Kohnen}, \citenamefont
  {Wenz},\ and\ \citenamefont {Jochim}}]{Ottenstein2008}%
  \BibitemOpen
  \bibfield  {author} {\bibinfo {author} {\bibfnamefont {T.~B.}\ \bibnamefont
  {Ottenstein}}, \bibinfo {author} {\bibfnamefont {T.}~\bibnamefont {Lompe}},
  \bibinfo {author} {\bibfnamefont {M.}~\bibnamefont {Kohnen}}, \bibinfo
  {author} {\bibfnamefont {A.~N.}\ \bibnamefont {Wenz}},\ and\ \bibinfo
  {author} {\bibfnamefont {S.}~\bibnamefont {Jochim}},\ }\bibfield  {title}
  {\bibinfo {title} {Collisional stability of a three-component degenerate
  {Fermi} gas},\ }\href {https://doi.org/10.1103/PhysRevLett.101.203202}
  {\bibfield  {journal} {\bibinfo  {journal} {Phys. Rev. Lett.}\ }\textbf
  {\bibinfo {volume} {101}},\ \bibinfo {pages} {203202} (\bibinfo {year}
  {2008})}\BibitemShut {NoStop}%
\bibitem [{\citenamefont {Huckans}\ \emph {et~al.}(2009)\citenamefont
  {Huckans}, \citenamefont {Williams}, \citenamefont {Hazlett}, \citenamefont
  {Stites},\ and\ \citenamefont {O'Hara}}]{Huckans2009}%
  \BibitemOpen
  \bibfield  {author} {\bibinfo {author} {\bibfnamefont {J.~H.}\ \bibnamefont
  {Huckans}}, \bibinfo {author} {\bibfnamefont {J.~R.}\ \bibnamefont
  {Williams}}, \bibinfo {author} {\bibfnamefont {E.~L.}\ \bibnamefont
  {Hazlett}}, \bibinfo {author} {\bibfnamefont {R.~W.}\ \bibnamefont
  {Stites}},\ and\ \bibinfo {author} {\bibfnamefont {K.~M.}\ \bibnamefont
  {O'Hara}},\ }\bibfield  {title} {\bibinfo {title} {{Three-Body Recombination
  in a Three-State Fermi Gas with Widely Tunable Interactions}},\ }\href
  {https://doi.org/10.1103/PhysRevLett.102.165302} {\bibfield  {journal}
  {\bibinfo  {journal} {Phys. Rev. Lett.}\ }\textbf {\bibinfo {volume} {102}},\
  \bibinfo {pages} {165302} (\bibinfo {year} {2009})}\BibitemShut {NoStop}%
\bibitem [{\citenamefont {Mongkolkiattichai}\ \emph {et~al.}()\citenamefont
  {Mongkolkiattichai}, \citenamefont {Liu}, \citenamefont {Dasgupta},
  \citenamefont {Hazzard},\ and\ \citenamefont
  {Schauss}}]{mongkolkiattichai2025}%
  \BibitemOpen
  \bibfield  {author} {\bibinfo {author} {\bibfnamefont {J.}~\bibnamefont
  {Mongkolkiattichai}}, \bibinfo {author} {\bibfnamefont {L.}~\bibnamefont
  {Liu}}, \bibinfo {author} {\bibfnamefont {S.}~\bibnamefont {Dasgupta}},
  \bibinfo {author} {\bibfnamefont {K.~R.~A.}\ \bibnamefont {Hazzard}},\ and\
  \bibinfo {author} {\bibfnamefont {P.}~\bibnamefont {Schauss}},\ }\bibfield
  {title} {\bibinfo {title} {Quantum gas microscopy of three-flavor {Hubbard}
  systems},\ }\href {https://arxiv.org/abs/2503.05687} {\bibinfo  {journal}
  {arXiv:2503.05687}\ }\BibitemShut {NoStop}%
\bibitem [{\citenamefont {Bothwell}\ \emph {et~al.}(2022)\citenamefont
  {Bothwell}, \citenamefont {Kennedy}, \citenamefont {Aeppli}, \citenamefont
  {Kedar}, \citenamefont {Robinson}, \citenamefont {Oelker}, \citenamefont
  {Staron},\ and\ \citenamefont {Ye}}]{bothwell_resolving_2022}%
  \BibitemOpen
\bibfield  {journal} {  }\bibfield  {author} {\bibinfo {author} {\bibfnamefont
  {T.}~\bibnamefont {Bothwell}}, \bibinfo {author} {\bibfnamefont {C.~J.}\
  \bibnamefont {Kennedy}}, \bibinfo {author} {\bibfnamefont {A.}~\bibnamefont
  {Aeppli}}, \bibinfo {author} {\bibfnamefont {D.}~\bibnamefont {Kedar}},
  \bibinfo {author} {\bibfnamefont {J.~M.}\ \bibnamefont {Robinson}}, \bibinfo
  {author} {\bibfnamefont {E.}~\bibnamefont {Oelker}}, \bibinfo {author}
  {\bibfnamefont {A.}~\bibnamefont {Staron}},\ and\ \bibinfo {author}
  {\bibfnamefont {J.}~\bibnamefont {Ye}},\ }\bibfield  {title} {\bibinfo
  {title} {Resolving the gravitational redshift across a millimetre-scale
  atomic sample},\ }\href {https://doi.org/10.1038/s41586-021-04349-7}
  {\bibfield  {journal} {\bibinfo  {journal} {Nature}\ }\textbf {\bibinfo
  {volume} {602}},\ \bibinfo {pages} {420} (\bibinfo {year}
  {2022})}\BibitemShut {NoStop}%
\bibitem [{\citenamefont {Daley}\ \emph {et~al.}(2008)\citenamefont {Daley},
  \citenamefont {Boyd}, \citenamefont {Ye},\ and\ \citenamefont
  {Zoller}}]{daley_quantum_2008}%
  \BibitemOpen
  \bibfield  {author} {\bibinfo {author} {\bibfnamefont {A.~J.}\ \bibnamefont
  {Daley}}, \bibinfo {author} {\bibfnamefont {M.~M.}\ \bibnamefont {Boyd}},
  \bibinfo {author} {\bibfnamefont {J.}~\bibnamefont {Ye}},\ and\ \bibinfo
  {author} {\bibfnamefont {P.}~\bibnamefont {Zoller}},\ }\bibfield  {title}
  {\bibinfo {title} {Quantum {{Computing}} with {{Alkaline-Earth-Metal
  Atoms}}},\ }\href {https://doi.org/10.1103/PhysRevLett.101.170504} {\bibfield
   {journal} {\bibinfo  {journal} {Phys. Rev. Lett.}\ }\textbf {\bibinfo
  {volume} {101}},\ \bibinfo {pages} {170504} (\bibinfo {year}
  {2008})}\BibitemShut {NoStop}%
\bibitem [{\citenamefont {Barnes}\ \emph {et~al.}(2022)\citenamefont {Barnes},
  \citenamefont {Battaglino}, \citenamefont {Bloom}, \citenamefont {Cassella},
  \citenamefont {Coxe}, \citenamefont {Crisosto}, \citenamefont {King},
  \citenamefont {Kondov}, \citenamefont {Kotru}, \citenamefont {Larsen},
  \citenamefont {Lauigan}, \citenamefont {Lester}, \citenamefont {McDonald},
  \citenamefont {Megidish}, \citenamefont {Narayanaswami}, \citenamefont
  {Nishiguchi}, \citenamefont {Notermans}, \citenamefont {Peng}, \citenamefont
  {Ryou}, \citenamefont {Wu},\ and\ \citenamefont
  {Yarwood}}]{barnes_assembly_2022}%
  \BibitemOpen
  \bibfield  {author} {\bibinfo {author} {\bibfnamefont {K.}~\bibnamefont
  {Barnes}}, \bibinfo {author} {\bibfnamefont {P.}~\bibnamefont {Battaglino}},
  \bibinfo {author} {\bibfnamefont {B.~J.}\ \bibnamefont {Bloom}}, \bibinfo
  {author} {\bibfnamefont {K.}~\bibnamefont {Cassella}}, \bibinfo {author}
  {\bibfnamefont {R.}~\bibnamefont {Coxe}}, \bibinfo {author} {\bibfnamefont
  {N.}~\bibnamefont {Crisosto}}, \bibinfo {author} {\bibfnamefont {J.~P.}\
  \bibnamefont {King}}, \bibinfo {author} {\bibfnamefont {S.~S.}\ \bibnamefont
  {Kondov}}, \bibinfo {author} {\bibfnamefont {K.}~\bibnamefont {Kotru}},
  \bibinfo {author} {\bibfnamefont {S.~C.}\ \bibnamefont {Larsen}}, \bibinfo
  {author} {\bibfnamefont {J.}~\bibnamefont {Lauigan}}, \bibinfo {author}
  {\bibfnamefont {B.~J.}\ \bibnamefont {Lester}}, \bibinfo {author}
  {\bibfnamefont {M.}~\bibnamefont {McDonald}}, \bibinfo {author}
  {\bibfnamefont {E.}~\bibnamefont {Megidish}}, \bibinfo {author}
  {\bibfnamefont {S.}~\bibnamefont {Narayanaswami}}, \bibinfo {author}
  {\bibfnamefont {C.}~\bibnamefont {Nishiguchi}}, \bibinfo {author}
  {\bibfnamefont {R.}~\bibnamefont {Notermans}}, \bibinfo {author}
  {\bibfnamefont {L.~S.}\ \bibnamefont {Peng}}, \bibinfo {author}
  {\bibfnamefont {A.}~\bibnamefont {Ryou}}, \bibinfo {author} {\bibfnamefont
  {T.-Y.}\ \bibnamefont {Wu}},\ and\ \bibinfo {author} {\bibfnamefont
  {M.}~\bibnamefont {Yarwood}},\ }\bibfield  {title} {\bibinfo {title}
  {Assembly and coherent control of a register of nuclear spin qubits},\ }\href
  {https://doi.org/10.1038/s41467-022-29977-z} {\bibfield  {journal} {\bibinfo
  {journal} {Nat. Commun.}\ }\textbf {\bibinfo {volume} {13}},\ \bibinfo
  {pages} {2779} (\bibinfo {year} {2022})}\BibitemShut {NoStop}%
\bibitem [{\citenamefont {Huie}\ \emph {et~al.}(2023)\citenamefont {Huie},
  \citenamefont {Li}, \citenamefont {Chen}, \citenamefont {Hu}, \citenamefont
  {Jia}, \citenamefont {Sun},\ and\ \citenamefont {Covey}}]{Huie2023}%
  \BibitemOpen
  \bibfield  {author} {\bibinfo {author} {\bibfnamefont {W.}~\bibnamefont
  {Huie}}, \bibinfo {author} {\bibfnamefont {L.}~\bibnamefont {Li}}, \bibinfo
  {author} {\bibfnamefont {N.}~\bibnamefont {Chen}}, \bibinfo {author}
  {\bibfnamefont {X.}~\bibnamefont {Hu}}, \bibinfo {author} {\bibfnamefont
  {Z.}~\bibnamefont {Jia}}, \bibinfo {author} {\bibfnamefont {W.~K.~C.}\
  \bibnamefont {Sun}},\ and\ \bibinfo {author} {\bibfnamefont {J.~P.}\
  \bibnamefont {Covey}},\ }\bibfield  {title} {\bibinfo {title} {Repetitive
  readout and real-time control of nuclear spin qubits in ${}^{171}\mathrm{Yb}$
  atoms},\ }\href {https://doi.org/10.1103/PRXQuantum.4.030337} {\bibfield
  {journal} {\bibinfo  {journal} {PRX Quantum}\ }\textbf {\bibinfo {volume}
  {4}},\ \bibinfo {pages} {030337} (\bibinfo {year} {2023})}\BibitemShut
  {NoStop}%
\bibitem [{\citenamefont {Jenkins}\ \emph {et~al.}(2022)\citenamefont
  {Jenkins}, \citenamefont {Lis}, \citenamefont {Senoo}, \citenamefont
  {McGrew},\ and\ \citenamefont {Kaufman}}]{Jenkins2022}%
  \BibitemOpen
  \bibfield  {author} {\bibinfo {author} {\bibfnamefont {A.}~\bibnamefont
  {Jenkins}}, \bibinfo {author} {\bibfnamefont {J.~W.}\ \bibnamefont {Lis}},
  \bibinfo {author} {\bibfnamefont {A.}~\bibnamefont {Senoo}}, \bibinfo
  {author} {\bibfnamefont {W.~F.}\ \bibnamefont {McGrew}},\ and\ \bibinfo
  {author} {\bibfnamefont {A.~M.}\ \bibnamefont {Kaufman}},\ }\bibfield
  {title} {\bibinfo {title} {Ytterbium nuclear-spin qubits in an optical
  tweezer array},\ }\href {https://doi.org/10.1103/PhysRevX.12.021027}
  {\bibfield  {journal} {\bibinfo  {journal} {Phys. Rev. X}\ }\textbf {\bibinfo
  {volume} {12}},\ \bibinfo {pages} {021027} (\bibinfo {year}
  {2022})}\BibitemShut {NoStop}%
\bibitem [{\citenamefont {Ma}\ \emph {et~al.}(2022)\citenamefont {Ma},
  \citenamefont {Burgers}, \citenamefont {Liu}, \citenamefont {Wilson},
  \citenamefont {Zhang},\ and\ \citenamefont {Thompson}}]{Ma2022}%
  \BibitemOpen
  \bibfield  {author} {\bibinfo {author} {\bibfnamefont {S.}~\bibnamefont
  {Ma}}, \bibinfo {author} {\bibfnamefont {A.~P.}\ \bibnamefont {Burgers}},
  \bibinfo {author} {\bibfnamefont {G.}~\bibnamefont {Liu}}, \bibinfo {author}
  {\bibfnamefont {J.}~\bibnamefont {Wilson}}, \bibinfo {author} {\bibfnamefont
  {B.}~\bibnamefont {Zhang}},\ and\ \bibinfo {author} {\bibfnamefont {J.~D.}\
  \bibnamefont {Thompson}},\ }\bibfield  {title} {\bibinfo {title} {Universal
  gate operations on nuclear spin qubits in an optical tweezer array of
  $^{171}\mathrm{Yb}$ atoms},\ }\href
  {https://doi.org/10.1103/PhysRevX.12.021028} {\bibfield  {journal} {\bibinfo
  {journal} {Phys. Rev. X}\ }\textbf {\bibinfo {volume} {12}},\ \bibinfo
  {pages} {021028} (\bibinfo {year} {2022})}\BibitemShut {NoStop}%
\bibitem [{\citenamefont {Norcia}\ \emph {et~al.}(2023)\citenamefont {Norcia},
  \citenamefont {Cairncross}, \citenamefont {Barnes}, \citenamefont
  {Battaglino}, \citenamefont {Brown}, \citenamefont {Brown}, \citenamefont
  {Cassella}, \citenamefont {Chen}, \citenamefont {Coxe}, \citenamefont {Crow},
  \citenamefont {Epstein}, \citenamefont {Griger}, \citenamefont {Jones},
  \citenamefont {Kim}, \citenamefont {Kindem}, \citenamefont {King},
  \citenamefont {Kondov}, \citenamefont {Kotru}, \citenamefont {Lauigan},
  \citenamefont {Li}, \citenamefont {Lu}, \citenamefont {Megidish},
  \citenamefont {Marjanovic}, \citenamefont {McDonald}, \citenamefont
  {Mittiga}, \citenamefont {Muniz}, \citenamefont {Narayanaswami},
  \citenamefont {Nishiguchi}, \citenamefont {Notermans}, \citenamefont {Paule},
  \citenamefont {Pawlak}, \citenamefont {Peng}, \citenamefont {Ryou},
  \citenamefont {Smull}, \citenamefont {Stack}, \citenamefont {Stone},
  \citenamefont {Sucich}, \citenamefont {Urbanek}, \citenamefont {van~de
  Veerdonk}, \citenamefont {Vendeiro}, \citenamefont {Wilkason}, \citenamefont
  {Wu}, \citenamefont {Xie}, \citenamefont {Zhang},\ and\ \citenamefont
  {Bloom}}]{Norcia2023}%
  \BibitemOpen
  \bibfield  {author} {\bibinfo {author} {\bibfnamefont {M.~A.}\ \bibnamefont
  {Norcia}}, \bibinfo {author} {\bibfnamefont {W.~B.}\ \bibnamefont
  {Cairncross}}, \bibinfo {author} {\bibfnamefont {K.}~\bibnamefont {Barnes}},
  \bibinfo {author} {\bibfnamefont {P.}~\bibnamefont {Battaglino}}, \bibinfo
  {author} {\bibfnamefont {A.}~\bibnamefont {Brown}}, \bibinfo {author}
  {\bibfnamefont {M.~O.}\ \bibnamefont {Brown}}, \bibinfo {author}
  {\bibfnamefont {K.}~\bibnamefont {Cassella}}, \bibinfo {author}
  {\bibfnamefont {C.-A.}\ \bibnamefont {Chen}}, \bibinfo {author}
  {\bibfnamefont {R.}~\bibnamefont {Coxe}}, \bibinfo {author} {\bibfnamefont
  {D.}~\bibnamefont {Crow}}, \bibinfo {author} {\bibfnamefont {J.}~\bibnamefont
  {Epstein}}, \bibinfo {author} {\bibfnamefont {C.}~\bibnamefont {Griger}},
  \bibinfo {author} {\bibfnamefont {A.~M.~W.}\ \bibnamefont {Jones}}, \bibinfo
  {author} {\bibfnamefont {H.}~\bibnamefont {Kim}}, \bibinfo {author}
  {\bibfnamefont {J.~M.}\ \bibnamefont {Kindem}}, \bibinfo {author}
  {\bibfnamefont {J.}~\bibnamefont {King}}, \bibinfo {author} {\bibfnamefont
  {S.~S.}\ \bibnamefont {Kondov}}, \bibinfo {author} {\bibfnamefont
  {K.}~\bibnamefont {Kotru}}, \bibinfo {author} {\bibfnamefont
  {J.}~\bibnamefont {Lauigan}}, \bibinfo {author} {\bibfnamefont
  {M.}~\bibnamefont {Li}}, \bibinfo {author} {\bibfnamefont {M.}~\bibnamefont
  {Lu}}, \bibinfo {author} {\bibfnamefont {E.}~\bibnamefont {Megidish}},
  \bibinfo {author} {\bibfnamefont {J.}~\bibnamefont {Marjanovic}}, \bibinfo
  {author} {\bibfnamefont {M.}~\bibnamefont {McDonald}}, \bibinfo {author}
  {\bibfnamefont {T.}~\bibnamefont {Mittiga}}, \bibinfo {author} {\bibfnamefont
  {J.~A.}\ \bibnamefont {Muniz}}, \bibinfo {author} {\bibfnamefont
  {S.}~\bibnamefont {Narayanaswami}}, \bibinfo {author} {\bibfnamefont
  {C.}~\bibnamefont {Nishiguchi}}, \bibinfo {author} {\bibfnamefont
  {R.}~\bibnamefont {Notermans}}, \bibinfo {author} {\bibfnamefont
  {T.}~\bibnamefont {Paule}}, \bibinfo {author} {\bibfnamefont {K.~A.}\
  \bibnamefont {Pawlak}}, \bibinfo {author} {\bibfnamefont {L.~S.}\
  \bibnamefont {Peng}}, \bibinfo {author} {\bibfnamefont {A.}~\bibnamefont
  {Ryou}}, \bibinfo {author} {\bibfnamefont {A.}~\bibnamefont {Smull}},
  \bibinfo {author} {\bibfnamefont {D.}~\bibnamefont {Stack}}, \bibinfo
  {author} {\bibfnamefont {M.}~\bibnamefont {Stone}}, \bibinfo {author}
  {\bibfnamefont {A.}~\bibnamefont {Sucich}}, \bibinfo {author} {\bibfnamefont
  {M.}~\bibnamefont {Urbanek}}, \bibinfo {author} {\bibfnamefont {R.~J.~M.}\
  \bibnamefont {van~de Veerdonk}}, \bibinfo {author} {\bibfnamefont
  {Z.}~\bibnamefont {Vendeiro}}, \bibinfo {author} {\bibfnamefont
  {T.}~\bibnamefont {Wilkason}}, \bibinfo {author} {\bibfnamefont {T.-Y.}\
  \bibnamefont {Wu}}, \bibinfo {author} {\bibfnamefont {X.}~\bibnamefont
  {Xie}}, \bibinfo {author} {\bibfnamefont {X.}~\bibnamefont {Zhang}},\ and\
  \bibinfo {author} {\bibfnamefont {B.~J.}\ \bibnamefont {Bloom}},\ }\bibfield
  {title} {\bibinfo {title} {Midcircuit qubit measurement and rearrangement in
  a $^{171}\mathrm{Yb}$ atomic array},\ }\href
  {https://doi.org/10.1103/PhysRevX.13.041034} {\bibfield  {journal} {\bibinfo
  {journal} {Phys. Rev. X}\ }\textbf {\bibinfo {volume} {13}},\ \bibinfo
  {pages} {041034} (\bibinfo {year} {2023})}\BibitemShut {NoStop}%
\bibitem [{\citenamefont {González-Cuadra}\ \emph {et~al.}(2023)\citenamefont
  {González-Cuadra}, \citenamefont {Bluvstein}, \citenamefont {Kalinowski},
  \citenamefont {Kaubruegger}, \citenamefont {Maskara}, \citenamefont
  {Naldesi}, \citenamefont {Zache}, \citenamefont {Kaufman}, \citenamefont
  {Lukin}, \citenamefont {Pichler}, \citenamefont {Vermersch}, \citenamefont
  {Ye},\ and\ \citenamefont {Zoller}}]{GonzalezCuadra2023}%
  \BibitemOpen
  \bibfield  {author} {\bibinfo {author} {\bibfnamefont {D.}~\bibnamefont
  {González-Cuadra}}, \bibinfo {author} {\bibfnamefont {D.}~\bibnamefont
  {Bluvstein}}, \bibinfo {author} {\bibfnamefont {M.}~\bibnamefont
  {Kalinowski}}, \bibinfo {author} {\bibfnamefont {R.}~\bibnamefont
  {Kaubruegger}}, \bibinfo {author} {\bibfnamefont {N.}~\bibnamefont
  {Maskara}}, \bibinfo {author} {\bibfnamefont {P.}~\bibnamefont {Naldesi}},
  \bibinfo {author} {\bibfnamefont {T.~V.}\ \bibnamefont {Zache}}, \bibinfo
  {author} {\bibfnamefont {A.~M.}\ \bibnamefont {Kaufman}}, \bibinfo {author}
  {\bibfnamefont {M.~D.}\ \bibnamefont {Lukin}}, \bibinfo {author}
  {\bibfnamefont {H.}~\bibnamefont {Pichler}}, \bibinfo {author} {\bibfnamefont
  {B.}~\bibnamefont {Vermersch}}, \bibinfo {author} {\bibfnamefont
  {J.}~\bibnamefont {Ye}},\ and\ \bibinfo {author} {\bibfnamefont
  {P.}~\bibnamefont {Zoller}},\ }\bibfield  {title} {\bibinfo {title}
  {Fermionic quantum processing with programmable neutral atom arrays},\ }\href
  {https://doi.org/10.1073/pnas.2304294120} {\bibfield  {journal} {\bibinfo
  {journal} {Proc. Natl. Acad. Sci. U.S.A.}\ }\textbf {\bibinfo {volume}
  {120}},\ \bibinfo {pages} {e2304294120} (\bibinfo {year} {2023})}\BibitemShut
  {NoStop}%
\bibitem [{\citenamefont {Zache}\ \emph {et~al.}(2023)\citenamefont {Zache},
  \citenamefont {Gonz{\'a}lez-Cuadra},\ and\ \citenamefont
  {Zoller}}]{Zache2023}%
  \BibitemOpen
  \bibfield  {author} {\bibinfo {author} {\bibfnamefont {T.~V.}\ \bibnamefont
  {Zache}}, \bibinfo {author} {\bibfnamefont {D.}~\bibnamefont
  {Gonz{\'a}lez-Cuadra}},\ and\ \bibinfo {author} {\bibfnamefont
  {P.}~\bibnamefont {Zoller}},\ }\bibfield  {title} {\bibinfo {title}
  {Fermion-qudit quantum processors for simulating lattice gauge theories with
  matter},\ }\href {https://doi.org/10.22331/q-2023-10-16-1140} {\bibfield
  {journal} {\bibinfo  {journal} {Quantum}\ }\textbf {\bibinfo {volume} {7}},\
  \bibinfo {pages} {1140} (\bibinfo {year} {2023})}\BibitemShut {NoStop}%
\bibitem [{\citenamefont {Omanakuttan}\ \emph {et~al.}(2021)\citenamefont
  {Omanakuttan}, \citenamefont {Mitra}, \citenamefont {Martin},\ and\
  \citenamefont {Deutsch}}]{Omanakuttan2021}%
  \BibitemOpen
  \bibfield  {author} {\bibinfo {author} {\bibfnamefont {S.}~\bibnamefont
  {Omanakuttan}}, \bibinfo {author} {\bibfnamefont {A.}~\bibnamefont {Mitra}},
  \bibinfo {author} {\bibfnamefont {M.~J.}\ \bibnamefont {Martin}},\ and\
  \bibinfo {author} {\bibfnamefont {I.~H.}\ \bibnamefont {Deutsch}},\
  }\bibfield  {title} {\bibinfo {title} {Quantum optimal control of ten-level
  nuclear spin qudits in $^{87}\mathrm{Sr}$},\ }\href
  {https://doi.org/10.1103/PhysRevA.104.L060401} {\bibfield  {journal}
  {\bibinfo  {journal} {Phys. Rev. A}\ }\textbf {\bibinfo {volume} {104}},\
  \bibinfo {pages} {L060401} (\bibinfo {year} {2021})}\BibitemShut {NoStop}%
\bibitem [{\citenamefont {Ahmed}\ \emph {et~al.}(2025)\citenamefont {Ahmed},
  \citenamefont {Litvinov}, \citenamefont {Guesdon}, \citenamefont
  {Mar\'echal}, \citenamefont {Huckans}, \citenamefont {Pasquiou},
  \citenamefont {Laburthe-Tolra},\ and\ \citenamefont {Robert-de
  Saint-Vincent}}]{Ahmed2025}%
  \BibitemOpen
  \bibfield  {author} {\bibinfo {author} {\bibfnamefont {H.}~\bibnamefont
  {Ahmed}}, \bibinfo {author} {\bibfnamefont {A.}~\bibnamefont {Litvinov}},
  \bibinfo {author} {\bibfnamefont {P.}~\bibnamefont {Guesdon}}, \bibinfo
  {author} {\bibfnamefont {E.}~\bibnamefont {Mar\'echal}}, \bibinfo {author}
  {\bibfnamefont {J.}~\bibnamefont {Huckans}}, \bibinfo {author} {\bibfnamefont
  {B.}~\bibnamefont {Pasquiou}}, \bibinfo {author} {\bibfnamefont
  {B.}~\bibnamefont {Laburthe-Tolra}},\ and\ \bibinfo {author} {\bibfnamefont
  {M.}~\bibnamefont {Robert-de Saint-Vincent}},\ }\bibfield  {title} {\bibinfo
  {title} {Coherent control over the high-dimensional space of the nuclear spin
  of alkaline-earth atoms},\ }\href
  {https://doi.org/10.1103/PRXQuantum.6.020352} {\bibfield  {journal} {\bibinfo
   {journal} {PRX Quantum}\ }\textbf {\bibinfo {volume} {6}},\ \bibinfo {pages}
  {020352} (\bibinfo {year} {2025})}\BibitemShut {NoStop}%
\bibitem [{\citenamefont {Honerkamp}\ and\ \citenamefont
  {Hofstetter}(2004)}]{Hofstetter2004}%
  \BibitemOpen
  \bibfield  {author} {\bibinfo {author} {\bibfnamefont {C.}~\bibnamefont
  {Honerkamp}}\ and\ \bibinfo {author} {\bibfnamefont {W.}~\bibnamefont
  {Hofstetter}},\ }\bibfield  {title} {\bibinfo {title} {Ultracold fermions and
  the $\mathrm{SU}({N})$ {H}ubbard model},\ }\href
  {https://doi.org/10.1103/PhysRevLett.92.170403} {\bibfield  {journal}
  {\bibinfo  {journal} {Phys. Rev. Lett.}\ }\textbf {\bibinfo {volume} {92}},\
  \bibinfo {pages} {170403} (\bibinfo {year} {2004})}\BibitemShut {NoStop}%
\bibitem [{\citenamefont {Cazalilla}\ and\ \citenamefont
  {Rey}(2014)}]{cazalilla_ultracold_2014}%
  \BibitemOpen
  \bibfield  {author} {\bibinfo {author} {\bibfnamefont {M.~A.}\ \bibnamefont
  {Cazalilla}}\ and\ \bibinfo {author} {\bibfnamefont {A.~M.}\ \bibnamefont
  {Rey}},\ }\bibfield  {title} {\bibinfo {title} {Ultracold {Fermi} gases with
  emergent {SU($N$)} symmetry},\ }\href
  {https://doi.org/10.1088/0034-4885/77/12/124401} {\bibfield  {journal}
  {\bibinfo  {journal} {Rep. Prog. Phys.}\ }\textbf {\bibinfo {volume} {77}},\
  \bibinfo {pages} {124401} (\bibinfo {year} {2014})}\BibitemShut {NoStop}%
\bibitem [{\citenamefont {Ibarra-García-Padilla}\ and\ \citenamefont
  {Choudhury}(2024)}]{Ibarra-Garcia-Padilla_2025}%
  \BibitemOpen
  \bibfield  {author} {\bibinfo {author} {\bibfnamefont {E.}~\bibnamefont
  {Ibarra-García-Padilla}}\ and\ \bibinfo {author} {\bibfnamefont
  {S.}~\bibnamefont {Choudhury}},\ }\bibfield  {title} {\bibinfo {title}
  {Many-body physics of ultracold alkaline-earth atoms with
  {SU}(${N}$)-symmetric interactions},\ }\href
  {https://doi.org/10.1088/1361-648X/ad9658} {\bibfield  {journal} {\bibinfo
  {journal} {J. Phys.: Condens. Matter}\ }\textbf {\bibinfo {volume} {37}},\
  \bibinfo {pages} {083003} (\bibinfo {year} {2024})}\BibitemShut {NoStop}%
\bibitem [{\citenamefont {Stellmer}\ \emph {et~al.}(2011)\citenamefont
  {Stellmer}, \citenamefont {Grimm},\ and\ \citenamefont
  {Schreck}}]{StellmerDetection2011}%
  \BibitemOpen
  \bibfield  {author} {\bibinfo {author} {\bibfnamefont {S.}~\bibnamefont
  {Stellmer}}, \bibinfo {author} {\bibfnamefont {R.}~\bibnamefont {Grimm}},\
  and\ \bibinfo {author} {\bibfnamefont {F.}~\bibnamefont {Schreck}},\
  }\bibfield  {title} {\bibinfo {title} {Detection and manipulation of nuclear
  spin states in fermionic strontium},\ }\href
  {https://doi.org/10.1103/PhysRevA.84.043611} {\bibfield  {journal} {\bibinfo
  {journal} {Phys. Rev. A}\ }\textbf {\bibinfo {volume} {84}},\ \bibinfo
  {pages} {043611} (\bibinfo {year} {2011})}\BibitemShut {NoStop}%
\bibitem [{\citenamefont {Hermele}\ \emph {et~al.}(2009)\citenamefont
  {Hermele}, \citenamefont {Gurarie},\ and\ \citenamefont
  {Rey}}]{hermele_mott_2009}%
  \BibitemOpen
  \bibfield  {author} {\bibinfo {author} {\bibfnamefont {M.}~\bibnamefont
  {Hermele}}, \bibinfo {author} {\bibfnamefont {V.}~\bibnamefont {Gurarie}},\
  and\ \bibinfo {author} {\bibfnamefont {A.~M.}\ \bibnamefont {Rey}},\
  }\bibfield  {title} {\bibinfo {title} {Mott {{Insulators}} of {{Ultracold
  Fermionic Alkaline Earth Atoms}}: {{Underconstrained Magnetism}} and {{Chiral
  Spin Liquid}}},\ }\href {https://doi.org/10.1103/PhysRevLett.103.135301}
  {\bibfield  {journal} {\bibinfo  {journal} {Phys. Rev. Lett.}\ }\textbf
  {\bibinfo {volume} {103}},\ \bibinfo {pages} {135301} (\bibinfo {year}
  {2009})}\BibitemShut {NoStop}%
\bibitem [{\citenamefont {T\'oth}\ \emph {et~al.}(2010)\citenamefont {T\'oth},
  \citenamefont {L\"auchli}, \citenamefont {Mila},\ and\ \citenamefont
  {Penc}}]{Toth2010}%
  \BibitemOpen
  \bibfield  {author} {\bibinfo {author} {\bibfnamefont {T.~A.}\ \bibnamefont
  {T\'oth}}, \bibinfo {author} {\bibfnamefont {A.~M.}\ \bibnamefont
  {L\"auchli}}, \bibinfo {author} {\bibfnamefont {F.}~\bibnamefont {Mila}},\
  and\ \bibinfo {author} {\bibfnamefont {K.}~\bibnamefont {Penc}},\ }\bibfield
  {title} {\bibinfo {title} {Three-sublattice ordering of the {SU}(3)
  {H}eisenberg model of three-flavor fermions on the square and cubic
  lattices},\ }\href {https://doi.org/10.1103/PhysRevLett.105.265301}
  {\bibfield  {journal} {\bibinfo  {journal} {Phys. Rev. Lett.}\ }\textbf
  {\bibinfo {volume} {105}},\ \bibinfo {pages} {265301} (\bibinfo {year}
  {2010})}\BibitemShut {NoStop}%
\bibitem [{\citenamefont {Corboz}\ \emph {et~al.}(2011)\citenamefont {Corboz},
  \citenamefont {L{\"a}uchli}, \citenamefont {Penc}, \citenamefont {Troyer},\
  and\ \citenamefont {Mila}}]{corboz_simultaneous_2011}%
  \BibitemOpen
  \bibfield  {author} {\bibinfo {author} {\bibfnamefont {P.}~\bibnamefont
  {Corboz}}, \bibinfo {author} {\bibfnamefont {A.~M.}\ \bibnamefont
  {L{\"a}uchli}}, \bibinfo {author} {\bibfnamefont {K.}~\bibnamefont {Penc}},
  \bibinfo {author} {\bibfnamefont {M.}~\bibnamefont {Troyer}},\ and\ \bibinfo
  {author} {\bibfnamefont {F.}~\bibnamefont {Mila}},\ }\bibfield  {title}
  {\bibinfo {title} {Simultaneous {{Dimerization}} and {{SU}}(4) {{Symmetry
  Breaking}} of 4-{{Color Fermions}} on the {{Square Lattice}}},\ }\href
  {https://doi.org/10.1103/PhysRevLett.107.215301} {\bibfield  {journal}
  {\bibinfo  {journal} {Phys. Rev. Lett.}\ }\textbf {\bibinfo {volume} {107}},\
  \bibinfo {pages} {215301} (\bibinfo {year} {2011})}\BibitemShut {NoStop}%
\bibitem [{\citenamefont {Nataf}\ and\ \citenamefont {Mila}(2014)}]{Nataf2014}%
  \BibitemOpen
  \bibfield  {author} {\bibinfo {author} {\bibfnamefont {P.}~\bibnamefont
  {Nataf}}\ and\ \bibinfo {author} {\bibfnamefont {F.}~\bibnamefont {Mila}},\
  }\bibfield  {title} {\bibinfo {title} {Exact diagonalization of {H}eisenberg
  $\mathrm{SU}({N})$ models},\ }\href
  {https://doi.org/10.1103/PhysRevLett.113.127204} {\bibfield  {journal}
  {\bibinfo  {journal} {Phys. Rev. Lett.}\ }\textbf {\bibinfo {volume} {113}},\
  \bibinfo {pages} {127204} (\bibinfo {year} {2014})}\BibitemShut {NoStop}%
\bibitem [{\citenamefont {Nataf}\ \emph {et~al.}(2016)\citenamefont {Nataf},
  \citenamefont {Lajk\'o}, \citenamefont {Corboz}, \citenamefont {L\"auchli},
  \citenamefont {Penc},\ and\ \citenamefont {Mila}}]{Nataf2016}%
  \BibitemOpen
  \bibfield  {author} {\bibinfo {author} {\bibfnamefont {P.}~\bibnamefont
  {Nataf}}, \bibinfo {author} {\bibfnamefont {M.}~\bibnamefont {Lajk\'o}},
  \bibinfo {author} {\bibfnamefont {P.}~\bibnamefont {Corboz}}, \bibinfo
  {author} {\bibfnamefont {A.~M.}\ \bibnamefont {L\"auchli}}, \bibinfo {author}
  {\bibfnamefont {K.}~\bibnamefont {Penc}},\ and\ \bibinfo {author}
  {\bibfnamefont {F.}~\bibnamefont {Mila}},\ }\bibfield  {title} {\bibinfo
  {title} {Plaquette order in the {SU}(6) {H}eisenberg model on the honeycomb
  lattice},\ }\href {https://doi.org/10.1103/PhysRevB.93.201113} {\bibfield
  {journal} {\bibinfo  {journal} {Phys. Rev. B}\ }\textbf {\bibinfo {volume}
  {93}},\ \bibinfo {pages} {201113} (\bibinfo {year} {2016})}\BibitemShut
  {NoStop}%
\bibitem [{\citenamefont {Romen}\ and\ \citenamefont
  {L\"auchli}(2020)}]{Romen2020}%
  \BibitemOpen
  \bibfield  {author} {\bibinfo {author} {\bibfnamefont {C.}~\bibnamefont
  {Romen}}\ and\ \bibinfo {author} {\bibfnamefont {A.~M.}\ \bibnamefont
  {L\"auchli}},\ }\bibfield  {title} {\bibinfo {title} {Structure of spin
  correlations in high-temperature $\mathrm{SU}({N})$ quantum magnets},\ }\href
  {https://doi.org/10.1103/PhysRevResearch.2.043009} {\bibfield  {journal}
  {\bibinfo  {journal} {Phys. Rev. Res.}\ }\textbf {\bibinfo {volume} {2}},\
  \bibinfo {pages} {043009} (\bibinfo {year} {2020})}\BibitemShut {NoStop}%
\bibitem [{\citenamefont {Feng}\ \emph {et~al.}(2023)\citenamefont {Feng},
  \citenamefont {Ibarra-Garc\'{\i}a-Padilla}, \citenamefont {Hazzard},
  \citenamefont {Scalettar}, \citenamefont {Zhang},\ and\ \citenamefont
  {Vitali}}]{Feng2023}%
  \BibitemOpen
  \bibfield  {author} {\bibinfo {author} {\bibfnamefont {C.}~\bibnamefont
  {Feng}}, \bibinfo {author} {\bibfnamefont {E.}~\bibnamefont
  {Ibarra-Garc\'{\i}a-Padilla}}, \bibinfo {author} {\bibfnamefont {K.~R.~A.}\
  \bibnamefont {Hazzard}}, \bibinfo {author} {\bibfnamefont {R.}~\bibnamefont
  {Scalettar}}, \bibinfo {author} {\bibfnamefont {S.}~\bibnamefont {Zhang}},\
  and\ \bibinfo {author} {\bibfnamefont {E.}~\bibnamefont {Vitali}},\
  }\bibfield  {title} {\bibinfo {title} {Metal-insulator transition and quantum
  magnetism in the {SU(3)} {F}ermi-{H}ubbard model},\ }\href
  {https://doi.org/10.1103/PhysRevResearch.5.043267} {\bibfield  {journal}
  {\bibinfo  {journal} {Phys. Rev. Res.}\ }\textbf {\bibinfo {volume} {5}},\
  \bibinfo {pages} {043267} (\bibinfo {year} {2023})}\BibitemShut {NoStop}%
\bibitem [{\citenamefont {Bird}\ \emph {et~al.}(2025)\citenamefont {Bird},
  \citenamefont {Huber},\ and\ \citenamefont {Nys}}]{Bird2025}%
  \BibitemOpen
  \bibfield  {author} {\bibinfo {author} {\bibfnamefont {S.}~\bibnamefont
  {Bird}}, \bibinfo {author} {\bibfnamefont {S.}~\bibnamefont {Huber}},\ and\
  \bibinfo {author} {\bibfnamefont {J.}~\bibnamefont {Nys}},\ }\bibfield
  {title} {\bibinfo {title} {Partial suppression of magnetism in the square
  lattice {SU}(3) {H}ubbard model},\ }\href {https://doi.org/10.1103/jfxt-9r1c}
  {\bibfield  {journal} {\bibinfo  {journal} {Phys. Rev. B}\ }\textbf {\bibinfo
  {volume} {112}},\ \bibinfo {pages} {L161115} (\bibinfo {year}
  {2025})}\BibitemShut {NoStop}%
\bibitem [{\citenamefont {Kleijweg}\ and\ \citenamefont
  {Corboz}()}]{kleijweg_corboz_2025}%
  \BibitemOpen
  \bibfield  {author} {\bibinfo {author} {\bibfnamefont {S.~V.}\ \bibnamefont
  {Kleijweg}}\ and\ \bibinfo {author} {\bibfnamefont {P.}~\bibnamefont
  {Corboz}},\ }\bibfield  {title} {\bibinfo {title} {Zigzag antiferromagnets in
  the {SU}(3) {Hubbard} model on the square lattice},\ }\href
  {https://arxiv.org/abs/2506.14703} {\bibinfo  {journal} {arXiv:2506.14703}\
  }\BibitemShut {NoStop}%
\bibitem [{\citenamefont {Schl\"omer}\ \emph {et~al.}(2024)\citenamefont
  {Schl\"omer}, \citenamefont {Grusdt}, \citenamefont {Schollw\"ock},
  \citenamefont {Hazzard},\ and\ \citenamefont {Bohrdt}}]{schlomer_tJ_SU(3)}%
  \BibitemOpen
\bibfield  {journal} {  }\bibfield  {author} {\bibinfo {author} {\bibfnamefont
  {H.}~\bibnamefont {Schl\"omer}}, \bibinfo {author} {\bibfnamefont
  {F.}~\bibnamefont {Grusdt}}, \bibinfo {author} {\bibfnamefont
  {U.}~\bibnamefont {Schollw\"ock}}, \bibinfo {author} {\bibfnamefont
  {K.~R.~A.}\ \bibnamefont {Hazzard}},\ and\ \bibinfo {author} {\bibfnamefont
  {A.}~\bibnamefont {Bohrdt}},\ }\bibfield  {title} {\bibinfo {title}
  {Subdimensional magnetic polarons in the one-hole doped {SU}(3)
  $t\text{\ensuremath{-}}{J}$ model},\ }\href
  {https://doi.org/10.1103/PhysRevB.110.125134} {\bibfield  {journal} {\bibinfo
   {journal} {Phys. Rev. B}\ }\textbf {\bibinfo {volume} {110}},\ \bibinfo
  {pages} {125134} (\bibinfo {year} {2024})}\BibitemShut {NoStop}%
\bibitem [{\citenamefont {Fukuhara}\ \emph {et~al.}(2007)\citenamefont
  {Fukuhara}, \citenamefont {Takasu}, \citenamefont {Kumakura},\ and\
  \citenamefont {Takahashi}}]{FukuharaYb}%
  \BibitemOpen
  \bibfield  {author} {\bibinfo {author} {\bibfnamefont {T.}~\bibnamefont
  {Fukuhara}}, \bibinfo {author} {\bibfnamefont {Y.}~\bibnamefont {Takasu}},
  \bibinfo {author} {\bibfnamefont {M.}~\bibnamefont {Kumakura}},\ and\
  \bibinfo {author} {\bibfnamefont {Y.}~\bibnamefont {Takahashi}},\ }\bibfield
  {title} {\bibinfo {title} {Degenerate {F}ermi {G}ases of {Y}tterbium},\ }\href
  {https://doi.org/10.1103/PhysRevLett.98.030401} {\bibfield  {journal}
  {\bibinfo  {journal} {Phys. Rev. Lett.}\ }\textbf {\bibinfo {volume} {98}},\
  \bibinfo {pages} {030401} (\bibinfo {year} {2007})}\BibitemShut {NoStop}%
\bibitem [{\citenamefont {DeSalvo}\ \emph {et~al.}(2010)\citenamefont
  {DeSalvo}, \citenamefont {Yan}, \citenamefont {Mickelson}, \citenamefont
  {Martinez~de Escobar},\ and\ \citenamefont
  {Killian}}]{desalvo_degenerate_2010}%
  \BibitemOpen
  \bibfield  {author} {\bibinfo {author} {\bibfnamefont {B.~J.}\ \bibnamefont
  {DeSalvo}}, \bibinfo {author} {\bibfnamefont {M.}~\bibnamefont {Yan}},
  \bibinfo {author} {\bibfnamefont {P.~G.}\ \bibnamefont {Mickelson}}, \bibinfo
  {author} {\bibfnamefont {Y.~N.}\ \bibnamefont {Martinez~de Escobar}},\ and\
  \bibinfo {author} {\bibfnamefont {T.~C.}\ \bibnamefont {Killian}},\
  }\bibfield  {title} {\bibinfo {title} {Degenerate {Fermi} gas of
  $^{87}\mathrm{Sr}$},\ }\href {https://doi.org/10.1103/PhysRevLett.105.030402}
  {\bibfield  {journal} {\bibinfo  {journal} {Phys. Rev. Lett.}\ }\textbf
  {\bibinfo {volume} {105}},\ \bibinfo {pages} {030402} (\bibinfo {year}
  {2010})}\BibitemShut {NoStop}%
\bibitem [{\citenamefont {Zhang}\ \emph {et~al.}(2014)\citenamefont {Zhang},
  \citenamefont {Bishof}, \citenamefont {Bromley}, \citenamefont {Kraus},
  \citenamefont {Safronova}, \citenamefont {Zoller}, \citenamefont {Rey},\ and\
  \citenamefont {Ye}}]{Zhang2014SrSUN}%
  \BibitemOpen
  \bibfield  {author} {\bibinfo {author} {\bibfnamefont {X.}~\bibnamefont
  {Zhang}}, \bibinfo {author} {\bibfnamefont {M.}~\bibnamefont {Bishof}},
  \bibinfo {author} {\bibfnamefont {S.~L.}\ \bibnamefont {Bromley}}, \bibinfo
  {author} {\bibfnamefont {C.~V.}\ \bibnamefont {Kraus}}, \bibinfo {author}
  {\bibfnamefont {M.~S.}\ \bibnamefont {Safronova}}, \bibinfo {author}
  {\bibfnamefont {P.}~\bibnamefont {Zoller}}, \bibinfo {author} {\bibfnamefont
  {A.~M.}\ \bibnamefont {Rey}},\ and\ \bibinfo {author} {\bibfnamefont
  {J.}~\bibnamefont {Ye}},\ }\bibfield  {title} {\bibinfo {title}
  {Spectroscopic observation of {SU}(${N}$)-symmetric interactions in {Sr}
  orbital magnetism},\ }\href {https://doi.org/10.1126/science.1254978}
  {\bibfield  {journal} {\bibinfo  {journal} {Science}\ }\textbf {\bibinfo
  {volume} {345}},\ \bibinfo {pages} {1467} (\bibinfo {year}
  {2014})}\BibitemShut {NoStop}%
\bibitem [{\citenamefont {Pagano}\ \emph {et~al.}(2014)\citenamefont {Pagano},
  \citenamefont {Mancini}, \citenamefont {Cappellini}, \citenamefont
  {Lombardi}, \citenamefont {Schäfer}, \citenamefont {Hu}, \citenamefont
  {Liu}, \citenamefont {Catani}, \citenamefont {Sias}, \citenamefont
  {Inguscio},\ and\ \citenamefont {Fallani}}]{Pagano2014OneDimensional}%
  \BibitemOpen
  \bibfield  {author} {\bibinfo {author} {\bibfnamefont {G.}~\bibnamefont
  {Pagano}}, \bibinfo {author} {\bibfnamefont {M.}~\bibnamefont {Mancini}},
  \bibinfo {author} {\bibfnamefont {G.}~\bibnamefont {Cappellini}}, \bibinfo
  {author} {\bibfnamefont {P.}~\bibnamefont {Lombardi}}, \bibinfo {author}
  {\bibfnamefont {F.}~\bibnamefont {Schäfer}}, \bibinfo {author}
  {\bibfnamefont {H.}~\bibnamefont {Hu}}, \bibinfo {author} {\bibfnamefont
  {X.-J.}\ \bibnamefont {Liu}}, \bibinfo {author} {\bibfnamefont
  {J.}~\bibnamefont {Catani}}, \bibinfo {author} {\bibfnamefont
  {C.}~\bibnamefont {Sias}}, \bibinfo {author} {\bibfnamefont {M.}~\bibnamefont
  {Inguscio}},\ and\ \bibinfo {author} {\bibfnamefont {L.}~\bibnamefont
  {Fallani}},\ }\bibfield  {title} {\bibinfo {title} {A one-dimensional liquid
  of fermions with tunable spin},\ }\href {https://doi.org/10.1038/nphys2878}
  {\bibfield  {journal} {\bibinfo  {journal} {Nat. Phys.}\ }\textbf {\bibinfo
  {volume} {10}},\ \bibinfo {pages} {198} (\bibinfo {year} {2014})}\BibitemShut
  {NoStop}%
\bibitem [{\citenamefont {Song}\ \emph {et~al.}(2020)\citenamefont {Song},
  \citenamefont {Yan}, \citenamefont {He}, \citenamefont {Ren}, \citenamefont
  {Zhou},\ and\ \citenamefont {Jo}}]{Song2020}%
  \BibitemOpen
  \bibfield  {author} {\bibinfo {author} {\bibfnamefont {B.}~\bibnamefont
  {Song}}, \bibinfo {author} {\bibfnamefont {Y.}~\bibnamefont {Yan}}, \bibinfo
  {author} {\bibfnamefont {C.}~\bibnamefont {He}}, \bibinfo {author}
  {\bibfnamefont {Z.}~\bibnamefont {Ren}}, \bibinfo {author} {\bibfnamefont
  {Q.}~\bibnamefont {Zhou}},\ and\ \bibinfo {author} {\bibfnamefont {G.-B.}\
  \bibnamefont {Jo}},\ }\bibfield  {title} {\bibinfo {title} {Evidence for
  bosonization in a three-dimensional gas of $\mathrm{SU}({N})$ fermions},\
  }\href {https://doi.org/10.1103/PhysRevX.10.041053} {\bibfield  {journal}
  {\bibinfo  {journal} {Phys. Rev. X}\ }\textbf {\bibinfo {volume} {10}},\
  \bibinfo {pages} {041053} (\bibinfo {year} {2020})}\BibitemShut {NoStop}%
\bibitem [{\citenamefont {Sonderhouse}\ \emph {et~al.}(2020)\citenamefont
  {Sonderhouse}, \citenamefont {Sanner}, \citenamefont {Hutson}, \citenamefont
  {Goban}, \citenamefont {Bilitewski}, \citenamefont {Yan}, \citenamefont
  {Milner}, \citenamefont {Rey},\ and\ \citenamefont
  {Ye}}]{sonderhouse_thermodynamics_2020}%
  \BibitemOpen
  \bibfield  {author} {\bibinfo {author} {\bibfnamefont {L.}~\bibnamefont
  {Sonderhouse}}, \bibinfo {author} {\bibfnamefont {C.}~\bibnamefont {Sanner}},
  \bibinfo {author} {\bibfnamefont {R.~B.}\ \bibnamefont {Hutson}}, \bibinfo
  {author} {\bibfnamefont {A.}~\bibnamefont {Goban}}, \bibinfo {author}
  {\bibfnamefont {T.}~\bibnamefont {Bilitewski}}, \bibinfo {author}
  {\bibfnamefont {L.}~\bibnamefont {Yan}}, \bibinfo {author} {\bibfnamefont
  {W.~R.}\ \bibnamefont {Milner}}, \bibinfo {author} {\bibfnamefont {A.~M.}\
  \bibnamefont {Rey}},\ and\ \bibinfo {author} {\bibfnamefont {J.}~\bibnamefont
  {Ye}},\ }\bibfield  {title} {\bibinfo {title} {Thermodynamics of a deeply
  degenerate {{SU}}({{$N$}})-symmetric {{Fermi}} gas},\ }\href
  {https://doi.org/10.1038/s41567-020-0986-6} {\bibfield  {journal} {\bibinfo
  {journal} {Nat. Phys.}\ }\textbf {\bibinfo {volume} {16}},\ \bibinfo {pages}
  {1216} (\bibinfo {year} {2020})}\BibitemShut {NoStop}%
\bibitem [{\citenamefont {Taie}\ \emph {et~al.}(2012)\citenamefont {Taie},
  \citenamefont {Yamazaki}, \citenamefont {Sugawa},\ and\ \citenamefont
  {Takahashi}}]{taie_su_2012}%
  \BibitemOpen
  \bibfield  {author} {\bibinfo {author} {\bibfnamefont {S.}~\bibnamefont
  {Taie}}, \bibinfo {author} {\bibfnamefont {R.}~\bibnamefont {Yamazaki}},
  \bibinfo {author} {\bibfnamefont {S.}~\bibnamefont {Sugawa}},\ and\ \bibinfo
  {author} {\bibfnamefont {Y.}~\bibnamefont {Takahashi}},\ }\bibfield  {title}
  {\bibinfo {title} {An {{SU}}(6) {{Mott}} insulator of an atomic {{Fermi}} gas
  realized by large-spin {{Pomeranchuk}} cooling},\ }\href
  {https://doi.org/10.1038/nphys2430} {\bibfield  {journal} {\bibinfo
  {journal} {Nat. Phys.}\ }\textbf {\bibinfo {volume} {8}},\ \bibinfo {pages}
  {825} (\bibinfo {year} {2012})}\BibitemShut {NoStop}%
\bibitem [{\citenamefont {Mancini}\ \emph {et~al.}(2015)\citenamefont
  {Mancini}, \citenamefont {Pagano}, \citenamefont {Cappellini}, \citenamefont
  {Livi}, \citenamefont {Rider}, \citenamefont {Catani}, \citenamefont {Sias},
  \citenamefont {Zoller}, \citenamefont {Inguscio}, \citenamefont {Dalmonte},\
  and\ \citenamefont {Fallani}}]{mancini_observation_2015}%
  \BibitemOpen
  \bibfield  {author} {\bibinfo {author} {\bibfnamefont {M.}~\bibnamefont
  {Mancini}}, \bibinfo {author} {\bibfnamefont {G.}~\bibnamefont {Pagano}},
  \bibinfo {author} {\bibfnamefont {G.}~\bibnamefont {Cappellini}}, \bibinfo
  {author} {\bibfnamefont {L.}~\bibnamefont {Livi}}, \bibinfo {author}
  {\bibfnamefont {M.}~\bibnamefont {Rider}}, \bibinfo {author} {\bibfnamefont
  {J.}~\bibnamefont {Catani}}, \bibinfo {author} {\bibfnamefont
  {C.}~\bibnamefont {Sias}}, \bibinfo {author} {\bibfnamefont {P.}~\bibnamefont
  {Zoller}}, \bibinfo {author} {\bibfnamefont {M.}~\bibnamefont {Inguscio}},
  \bibinfo {author} {\bibfnamefont {M.}~\bibnamefont {Dalmonte}},\ and\
  \bibinfo {author} {\bibfnamefont {L.}~\bibnamefont {Fallani}},\ }\bibfield
  {title} {\bibinfo {title} {Observation of chiral edge states with neutral
  fermions in synthetic {{Hall}} ribbons},\ }\href
  {https://doi.org/10.1126/science.aaa8736} {\bibfield  {journal} {\bibinfo
  {journal} {Science}\ }\textbf {\bibinfo {volume} {349}},\ \bibinfo {pages}
  {1510} (\bibinfo {year} {2015})}\BibitemShut {NoStop}%
\bibitem [{\citenamefont {Han}\ \emph {et~al.}(2019)\citenamefont {Han},
  \citenamefont {Kang},\ and\ \citenamefont {Shin}}]{han_band_2019}%
  \BibitemOpen
  \bibfield  {author} {\bibinfo {author} {\bibfnamefont {J.~H.}\ \bibnamefont
  {Han}}, \bibinfo {author} {\bibfnamefont {J.~H.}\ \bibnamefont {Kang}},\ and\
  \bibinfo {author} {\bibfnamefont {Y.}~\bibnamefont {Shin}},\ }\bibfield
  {title} {\bibinfo {title} {Band {G}ap {C}losing in a {S}ynthetic {H}all
  {T}ube of {N}eutral {F}ermions},\ }\href
  {https://doi.org/10.1103/PhysRevLett.122.065303} {\bibfield  {journal}
  {\bibinfo  {journal} {Phys. Rev. Lett.}\ }\textbf {\bibinfo {volume} {122}},\
  \bibinfo {pages} {065303} (\bibinfo {year} {2019})}\BibitemShut {NoStop}%
\bibitem [{\citenamefont {Hofrichter}\ \emph {et~al.}(2016)\citenamefont
  {Hofrichter}, \citenamefont {Riegger}, \citenamefont {Scazza}, \citenamefont
  {Höfer}, \citenamefont {Fernandes}, \citenamefont {Bloch},\ and\
  \citenamefont {Fölling}}]{hofrichter_direct_2016}%
  \BibitemOpen
  \bibfield  {author} {\bibinfo {author} {\bibfnamefont {C.}~\bibnamefont
  {Hofrichter}}, \bibinfo {author} {\bibfnamefont {L.}~\bibnamefont {Riegger}},
  \bibinfo {author} {\bibfnamefont {F.}~\bibnamefont {Scazza}}, \bibinfo
  {author} {\bibfnamefont {M.}~\bibnamefont {Höfer}}, \bibinfo {author}
  {\bibfnamefont {D.~R.}\ \bibnamefont {Fernandes}}, \bibinfo {author}
  {\bibfnamefont {I.}~\bibnamefont {Bloch}},\ and\ \bibinfo {author}
  {\bibfnamefont {S.}~\bibnamefont {Fölling}},\ }\bibfield  {title} {\bibinfo
  {title} {Direct probing of the {Mott} crossover in the {$\mathrm{SU}(N)$}
  {Fermi-Hubbard} model},\ }\href {https://doi.org/10.1103/PhysRevX.6.021030}
  {\bibfield  {journal} {\bibinfo  {journal} {Phys. Rev. X}\ }\textbf {\bibinfo
  {volume} {6}},\ \bibinfo {pages} {021030} (\bibinfo {year}
  {2016})}\BibitemShut {NoStop}%
\bibitem [{\citenamefont {Pasqualetti}\ \emph {et~al.}(2024)\citenamefont
  {Pasqualetti}, \citenamefont {Bettermann}, \citenamefont {Darkwah~Oppong},
  \citenamefont {Ibarra-Garc\'{\i}a-Padilla}, \citenamefont {Dasgupta},
  \citenamefont {Scalettar}, \citenamefont {Hazzard}, \citenamefont {Bloch},\
  and\ \citenamefont {F\"olling}}]{pasqualetti_equation_2023}%
  \BibitemOpen
  \bibfield  {author} {\bibinfo {author} {\bibfnamefont {G.}~\bibnamefont
  {Pasqualetti}}, \bibinfo {author} {\bibfnamefont {O.}~\bibnamefont
  {Bettermann}}, \bibinfo {author} {\bibfnamefont {N.}~\bibnamefont
  {Darkwah~Oppong}}, \bibinfo {author} {\bibfnamefont {E.}~\bibnamefont
  {Ibarra-Garc\'{\i}a-Padilla}}, \bibinfo {author} {\bibfnamefont
  {S.}~\bibnamefont {Dasgupta}}, \bibinfo {author} {\bibfnamefont {R.~T.}\
  \bibnamefont {Scalettar}}, \bibinfo {author} {\bibfnamefont {K.~R.~A.}\
  \bibnamefont {Hazzard}}, \bibinfo {author} {\bibfnamefont {I.}~\bibnamefont
  {Bloch}},\ and\ \bibinfo {author} {\bibfnamefont {S.}~\bibnamefont
  {F\"olling}},\ }\bibfield  {title} {\bibinfo {title} {Equation of {S}tate and
  {T}hermometry of the {2D} $\mathrm{SU}({N})$ {F}ermi-{H}ubbard {M}odel},\
  }\href {https://doi.org/10.1103/PhysRevLett.132.083401} {\bibfield  {journal}
  {\bibinfo  {journal} {Phys. Rev. Lett.}\ }\textbf {\bibinfo {volume} {132}},\
  \bibinfo {pages} {083401} (\bibinfo {year} {2024})}\BibitemShut {NoStop}%
\bibitem [{\citenamefont {Tusi}\ \emph {et~al.}(2022)\citenamefont {Tusi},
  \citenamefont {Franchi}, \citenamefont {Livi}, \citenamefont {Baumann},
  \citenamefont {Orenes}, \citenamefont {Re}, \citenamefont {Barfknecht},
  \citenamefont {Zhou}, \citenamefont {Inguscio}, \citenamefont {Cappellini},
  \citenamefont {Capone}, \citenamefont {Catani},\ and\ \citenamefont
  {Fallani}}]{Tusi2022}%
  \BibitemOpen
  \bibfield  {author} {\bibinfo {author} {\bibfnamefont {D.}~\bibnamefont
  {Tusi}}, \bibinfo {author} {\bibfnamefont {L.}~\bibnamefont {Franchi}},
  \bibinfo {author} {\bibfnamefont {L.~F.}\ \bibnamefont {Livi}}, \bibinfo
  {author} {\bibfnamefont {K.}~\bibnamefont {Baumann}}, \bibinfo {author}
  {\bibfnamefont {D.~B.}\ \bibnamefont {Orenes}}, \bibinfo {author}
  {\bibfnamefont {L.~D.}\ \bibnamefont {Re}}, \bibinfo {author} {\bibfnamefont
  {R.~E.}\ \bibnamefont {Barfknecht}}, \bibinfo {author} {\bibfnamefont
  {T.-W.}\ \bibnamefont {Zhou}}, \bibinfo {author} {\bibfnamefont
  {M.}~\bibnamefont {Inguscio}}, \bibinfo {author} {\bibfnamefont
  {G.}~\bibnamefont {Cappellini}}, \bibinfo {author} {\bibfnamefont
  {M.}~\bibnamefont {Capone}}, \bibinfo {author} {\bibfnamefont
  {J.}~\bibnamefont {Catani}},\ and\ \bibinfo {author} {\bibfnamefont
  {L.}~\bibnamefont {Fallani}},\ }\bibfield  {title} {\bibinfo {title}
  {Flavour-selective localization in interacting lattice fermions},\ }\href
  {https://doi.org/10.1038/s41567-022-01726-5} {\bibfield  {journal} {\bibinfo
  {journal} {Nat. Phys.}\ }\textbf {\bibinfo {volume} {18}},\ \bibinfo {pages}
  {1201} (\bibinfo {year} {2022})}\BibitemShut {NoStop}%
\bibitem [{\citenamefont {Taie}\ \emph {et~al.}(2022)\citenamefont {Taie},
  \citenamefont {Ibarra-García-Padilla}, \citenamefont {Nishizawa},
  \citenamefont {Takasu}, \citenamefont {Kuno}, \citenamefont {Wei},
  \citenamefont {Scalettar}, \citenamefont {Hazzard},\ and\ \citenamefont
  {Takahashi}}]{taie_observation_2022}%
  \BibitemOpen
  \bibfield  {author} {\bibinfo {author} {\bibfnamefont {S.}~\bibnamefont
  {Taie}}, \bibinfo {author} {\bibfnamefont {E.}~\bibnamefont
  {Ibarra-García-Padilla}}, \bibinfo {author} {\bibfnamefont {N.}~\bibnamefont
  {Nishizawa}}, \bibinfo {author} {\bibfnamefont {Y.}~\bibnamefont {Takasu}},
  \bibinfo {author} {\bibfnamefont {Y.}~\bibnamefont {Kuno}}, \bibinfo {author}
  {\bibfnamefont {H.-T.}\ \bibnamefont {Wei}}, \bibinfo {author} {\bibfnamefont
  {R.~T.}\ \bibnamefont {Scalettar}}, \bibinfo {author} {\bibfnamefont
  {K.~R.~A.}\ \bibnamefont {Hazzard}},\ and\ \bibinfo {author} {\bibfnamefont
  {Y.}~\bibnamefont {Takahashi}},\ }\bibfield  {title} {\bibinfo {title}
  {Observation of antiferromagnetic correlations in an ultracold {SU($N$)}
  {Hubbard} model},\ }\href {https://doi.org/10.1038/s41567-022-01725-6}
  {\bibfield  {journal} {\bibinfo  {journal} {Nat. Phys.}\ }\textbf {\bibinfo
  {volume} {18}},\ \bibinfo {pages} {1356} (\bibinfo {year}
  {2022})}\BibitemShut {NoStop}%
\bibitem [{\citenamefont {Miranda}\ \emph {et~al.}(2015)\citenamefont
  {Miranda}, \citenamefont {Inoue}, \citenamefont {Okuyama}, \citenamefont
  {Nakamoto},\ and\ \citenamefont {Kozuma}}]{miranda_siteresolved_2015}%
  \BibitemOpen
  \bibfield  {author} {\bibinfo {author} {\bibfnamefont {M.}~\bibnamefont
  {Miranda}}, \bibinfo {author} {\bibfnamefont {R.}~\bibnamefont {Inoue}},
  \bibinfo {author} {\bibfnamefont {Y.}~\bibnamefont {Okuyama}}, \bibinfo
  {author} {\bibfnamefont {A.}~\bibnamefont {Nakamoto}},\ and\ \bibinfo
  {author} {\bibfnamefont {M.}~\bibnamefont {Kozuma}},\ }\bibfield  {title}
  {\bibinfo {title} {Site-resolved imaging of ytterbium atoms in a
  two-dimensional optical lattice},\ }\href
  {https://doi.org/10.1103/PhysRevA.91.063414} {\bibfield  {journal} {\bibinfo
  {journal} {Phys. Rev. A}\ }\textbf {\bibinfo {volume} {91}},\ \bibinfo
  {pages} {063414} (\bibinfo {year} {2015})}\BibitemShut {NoStop}%
\bibitem [{\citenamefont {Yamamoto}\ \emph {et~al.}(2016)\citenamefont
  {Yamamoto}, \citenamefont {Kobayashi}, \citenamefont {Kuno}, \citenamefont
  {Kato},\ and\ \citenamefont {Takahashi}}]{yamamoto_ytterbium_2016}%
  \BibitemOpen
  \bibfield  {author} {\bibinfo {author} {\bibfnamefont {R.}~\bibnamefont
  {Yamamoto}}, \bibinfo {author} {\bibfnamefont {J.}~\bibnamefont {Kobayashi}},
  \bibinfo {author} {\bibfnamefont {T.}~\bibnamefont {Kuno}}, \bibinfo {author}
  {\bibfnamefont {K.}~\bibnamefont {Kato}},\ and\ \bibinfo {author}
  {\bibfnamefont {Y.}~\bibnamefont {Takahashi}},\ }\bibfield  {title} {\bibinfo
  {title} {An ytterbium quantum gas microscope with narrow-line laser
  cooling},\ }\href {https://doi.org/10.1088/1367-2630/18/2/023016} {\bibfield
  {journal} {\bibinfo  {journal} {New J. Phys.}\ }\textbf {\bibinfo {volume}
  {18}},\ \bibinfo {pages} {023016} (\bibinfo {year} {2016})}\BibitemShut
  {NoStop}%
\bibitem [{\citenamefont {Buob}\ \emph {et~al.}(2024)\citenamefont {Buob},
  \citenamefont {H\"oschele}, \citenamefont {Makhalov}, \citenamefont
  {Rubio-Abadal},\ and\ \citenamefont {Tarruell}}]{buob_strontium_2024}%
  \BibitemOpen
  \bibfield  {author} {\bibinfo {author} {\bibfnamefont {S.}~\bibnamefont
  {Buob}}, \bibinfo {author} {\bibfnamefont {J.}~\bibnamefont {H\"oschele}},
  \bibinfo {author} {\bibfnamefont {V.}~\bibnamefont {Makhalov}}, \bibinfo
  {author} {\bibfnamefont {A.}~\bibnamefont {Rubio-Abadal}},\ and\ \bibinfo
  {author} {\bibfnamefont {L.}~\bibnamefont {Tarruell}},\ }\bibfield  {title}
  {\bibinfo {title} {A {S}trontium {Q}uantum-{G}as {M}icroscope},\ }\href
  {https://doi.org/10.1103/PRXQuantum.5.020316} {\bibfield  {journal} {\bibinfo
   {journal} {PRX Quantum}\ }\textbf {\bibinfo {volume} {5}},\ \bibinfo {pages}
  {020316} (\bibinfo {year} {2024})}\BibitemShut {NoStop}%
\bibitem [{\citenamefont {Chiu}\ \emph {et~al.}(2018)\citenamefont {Chiu},
  \citenamefont {Ji}, \citenamefont {Mazurenko}, \citenamefont {Greif},\ and\
  \citenamefont {Greiner}}]{ChiuEngineering2018}%
  \BibitemOpen
  \bibfield  {author} {\bibinfo {author} {\bibfnamefont {C.~S.}\ \bibnamefont
  {Chiu}}, \bibinfo {author} {\bibfnamefont {G.}~\bibnamefont {Ji}}, \bibinfo
  {author} {\bibfnamefont {A.}~\bibnamefont {Mazurenko}}, \bibinfo {author}
  {\bibfnamefont {D.}~\bibnamefont {Greif}},\ and\ \bibinfo {author}
  {\bibfnamefont {M.}~\bibnamefont {Greiner}},\ }\bibfield  {title} {\bibinfo
  {title} {Quantum state engineering of a {H}ubbard system with ultracold
  fermions},\ }\href {https://doi.org/10.1103/PhysRevLett.120.243201}
  {\bibfield  {journal} {\bibinfo  {journal} {Phys. Rev. Lett.}\ }\textbf
  {\bibinfo {volume} {120}},\ \bibinfo {pages} {243201} (\bibinfo {year}
  {2018})}\BibitemShut {NoStop}%
\bibitem [{\citenamefont {Mazurenko}\ \emph {et~al.}(2017)\citenamefont
  {Mazurenko}, \citenamefont {Chiu}, \citenamefont {Ji}, \citenamefont
  {Parsons}, \citenamefont {{Kan{\'a}sz-Nagy}}, \citenamefont {Schmidt},
  \citenamefont {Grusdt}, \citenamefont {Demler}, \citenamefont {Greif},\ and\
  \citenamefont {Greiner}}]{mazurenko_coldatom_2017}%
  \BibitemOpen
  \bibfield  {author} {\bibinfo {author} {\bibfnamefont {A.}~\bibnamefont
  {Mazurenko}}, \bibinfo {author} {\bibfnamefont {C.~S.}\ \bibnamefont {Chiu}},
  \bibinfo {author} {\bibfnamefont {G.}~\bibnamefont {Ji}}, \bibinfo {author}
  {\bibfnamefont {M.~F.}\ \bibnamefont {Parsons}}, \bibinfo {author}
  {\bibfnamefont {M.}~\bibnamefont {{Kan{\'a}sz-Nagy}}}, \bibinfo {author}
  {\bibfnamefont {R.}~\bibnamefont {Schmidt}}, \bibinfo {author} {\bibfnamefont
  {F.}~\bibnamefont {Grusdt}}, \bibinfo {author} {\bibfnamefont
  {E.}~\bibnamefont {Demler}}, \bibinfo {author} {\bibfnamefont
  {D.}~\bibnamefont {Greif}},\ and\ \bibinfo {author} {\bibfnamefont
  {M.}~\bibnamefont {Greiner}},\ }\bibfield  {title} {\bibinfo {title} {A
  cold-atom {{Fermi}}{\textendash}{{Hubbard}} antiferromagnet},\ }\href
  {https://doi.org/10.1038/nature22362} {\bibfield  {journal} {\bibinfo
  {journal} {Nature}\ }\textbf {\bibinfo {volume} {545}},\ \bibinfo {pages}
  {462} (\bibinfo {year} {2017})}\BibitemShut {NoStop}%
\bibitem [{\citenamefont {Urech}\ \emph {et~al.}(2022)\citenamefont {Urech},
  \citenamefont {Knottnerus}, \citenamefont {Spreeuw},\ and\ \citenamefont
  {Schreck}}]{urech_narrowline_2022}%
  \BibitemOpen
  \bibfield  {author} {\bibinfo {author} {\bibfnamefont {A.}~\bibnamefont
  {Urech}}, \bibinfo {author} {\bibfnamefont {I.~H.~A.}\ \bibnamefont
  {Knottnerus}}, \bibinfo {author} {\bibfnamefont {R.~J.~C.}\ \bibnamefont
  {Spreeuw}},\ and\ \bibinfo {author} {\bibfnamefont {F.}~\bibnamefont
  {Schreck}},\ }\bibfield  {title} {\bibinfo {title} {Narrow-line imaging of
  single strontium atoms in shallow optical tweezers},\ }\href
  {https://doi.org/10.1103/PhysRevResearch.4.023245} {\bibfield  {journal}
  {\bibinfo  {journal} {Phys. Rev. Res.}\ }\textbf {\bibinfo {volume} {4}},\
  \bibinfo {pages} {023245} (\bibinfo {year} {2022})}\BibitemShut {NoStop}%
\bibitem [{\citenamefont {Mukaiyama}\ \emph {et~al.}(2003)\citenamefont
  {Mukaiyama}, \citenamefont {Katori}, \citenamefont {Ido}, \citenamefont
  {Li},\ and\ \citenamefont {Kuwata-Gonokami}}]{Mukaiyama_redMOT_2003}%
  \BibitemOpen
  \bibfield  {author} {\bibinfo {author} {\bibfnamefont {T.}~\bibnamefont
  {Mukaiyama}}, \bibinfo {author} {\bibfnamefont {H.}~\bibnamefont {Katori}},
  \bibinfo {author} {\bibfnamefont {T.}~\bibnamefont {Ido}}, \bibinfo {author}
  {\bibfnamefont {Y.}~\bibnamefont {Li}},\ and\ \bibinfo {author}
  {\bibfnamefont {M.}~\bibnamefont {Kuwata-Gonokami}},\ }\bibfield  {title}
  {\bibinfo {title} {Recoil-limited laser cooling of
  $^{87}\mathrm{S}\mathrm{r}$ atoms near the {F}ermi temperature},\ }\href
  {https://doi.org/10.1103/PhysRevLett.90.113002} {\bibfield  {journal}
  {\bibinfo  {journal} {Phys. Rev. Lett.}\ }\textbf {\bibinfo {volume} {90}},\
  \bibinfo {pages} {113002} (\bibinfo {year} {2003})}\BibitemShut {NoStop}%
\bibitem [{\citenamefont {Cooper}\ \emph {et~al.}(2018)\citenamefont {Cooper},
  \citenamefont {Covey}, \citenamefont {Madjarov}, \citenamefont {Porsev},
  \citenamefont {Safronova},\ and\ \citenamefont
  {Endres}}]{cooper_alkalineearth_2018}%
  \BibitemOpen
  \bibfield  {author} {\bibinfo {author} {\bibfnamefont {A.}~\bibnamefont
  {Cooper}}, \bibinfo {author} {\bibfnamefont {J.~P.}\ \bibnamefont {Covey}},
  \bibinfo {author} {\bibfnamefont {I.~S.}\ \bibnamefont {Madjarov}}, \bibinfo
  {author} {\bibfnamefont {S.~G.}\ \bibnamefont {Porsev}}, \bibinfo {author}
  {\bibfnamefont {M.~S.}\ \bibnamefont {Safronova}},\ and\ \bibinfo {author}
  {\bibfnamefont {M.}~\bibnamefont {Endres}},\ }\bibfield  {title} {\bibinfo
  {title} {Alkaline-{{Earth Atoms}} in {{Optical Tweezers}}},\ }\href
  {https://doi.org/10.1103/PhysRevX.8.041055} {\bibfield  {journal} {\bibinfo
  {journal} {Phys. Rev. X}\ }\textbf {\bibinfo {volume} {8}},\ \bibinfo {pages}
  {041055} (\bibinfo {year} {2018})}\BibitemShut {NoStop}%
\bibitem [{\citenamefont {Norcia}\ \emph {et~al.}(2018)\citenamefont {Norcia},
  \citenamefont {Young},\ and\ \citenamefont
  {Kaufman}}]{norcia_microscopic_2018}%
  \BibitemOpen
  \bibfield  {author} {\bibinfo {author} {\bibfnamefont {M.~A.}\ \bibnamefont
  {Norcia}}, \bibinfo {author} {\bibfnamefont {A.~W.}\ \bibnamefont {Young}},\
  and\ \bibinfo {author} {\bibfnamefont {A.~M.}\ \bibnamefont {Kaufman}},\
  }\bibfield  {title} {\bibinfo {title} {Microscopic {{Control}} and
  {{Detection}} of {{Ultracold Strontium}} in {{Optical-Tweezer Arrays}}},\
  }\href {https://doi.org/10.1103/PhysRevX.8.041054} {\bibfield  {journal}
  {\bibinfo  {journal} {Phys. Rev. X}\ }\textbf {\bibinfo {volume} {8}},\
  \bibinfo {pages} {041054} (\bibinfo {year} {2018})}\BibitemShut {NoStop}%
\bibitem [{\citenamefont {Tao}\ \emph {et~al.}(2024)\citenamefont {Tao},
  \citenamefont {Ammenwerth}, \citenamefont {Gyger}, \citenamefont {Bloch},\
  and\ \citenamefont {Zeiher}}]{tao_highfidelity_2024}%
  \BibitemOpen
  \bibfield  {author} {\bibinfo {author} {\bibfnamefont {R.}~\bibnamefont
  {Tao}}, \bibinfo {author} {\bibfnamefont {M.}~\bibnamefont {Ammenwerth}},
  \bibinfo {author} {\bibfnamefont {F.}~\bibnamefont {Gyger}}, \bibinfo
  {author} {\bibfnamefont {I.}~\bibnamefont {Bloch}},\ and\ \bibinfo {author}
  {\bibfnamefont {J.}~\bibnamefont {Zeiher}},\ }\bibfield  {title} {\bibinfo
  {title} {High-fidelity detection of large-scale atom arrays in an optical
  lattice},\ }\href {https://doi.org/10.1103/PhysRevLett.133.013401} {\bibfield
   {journal} {\bibinfo  {journal} {Phys. Rev. Lett.}\ }\textbf {\bibinfo
  {volume} {133}},\ \bibinfo {pages} {013401} (\bibinfo {year}
  {2024})}\BibitemShut {NoStop}%
\bibitem [{\citenamefont {Tao}\ \emph {et~al.}(2026)\citenamefont {Tao},
  \citenamefont {Lib}, \citenamefont {Gyger}, \citenamefont {Timme},
  \citenamefont {Ammenwerth}, \citenamefont {Bloch},\ and\ \citenamefont
  {Zeiher}}]{tao_universalgates_2025}%
  \BibitemOpen
  \bibfield  {author} {\bibinfo {author} {\bibfnamefont {R.}~\bibnamefont
  {Tao}}, \bibinfo {author} {\bibfnamefont {O.}~\bibnamefont {Lib}}, \bibinfo
  {author} {\bibfnamefont {F.}~\bibnamefont {Gyger}}, \bibinfo {author}
  {\bibfnamefont {H.}~\bibnamefont {Timme}}, \bibinfo {author} {\bibfnamefont
  {M.}~\bibnamefont {Ammenwerth}}, \bibinfo {author} {\bibfnamefont
  {I.}~\bibnamefont {Bloch}},\ and\ \bibinfo {author} {\bibfnamefont
  {J.}~\bibnamefont {Zeiher}},\ }\bibfield  {title} {\bibinfo {title}
  {Universal {G}lobal {G}ates for a {F}ine-{S}tructure {Q}ubit in
  {S}trontium-88},\ }\href {https://doi.org/10.1103/gj3b-5ngl} {\bibfield
  {journal} {\bibinfo  {journal} {Phys. Rev. Lett.}\ }\textbf {\bibinfo
  {volume} {136}},\ \bibinfo {pages} {153602} (\bibinfo {year}
  {2026})}\BibitemShut {NoStop}%
\bibitem [{\citenamefont {Covey}\ \emph {et~al.}(2019)\citenamefont {Covey},
  \citenamefont {Madjarov}, \citenamefont {Cooper},\ and\ \citenamefont
  {Endres}}]{covey_2000times_2019}%
  \BibitemOpen
  \bibfield  {author} {\bibinfo {author} {\bibfnamefont {J.~P.}\ \bibnamefont
  {Covey}}, \bibinfo {author} {\bibfnamefont {I.~S.}\ \bibnamefont {Madjarov}},
  \bibinfo {author} {\bibfnamefont {A.}~\bibnamefont {Cooper}},\ and\ \bibinfo
  {author} {\bibfnamefont {M.}~\bibnamefont {Endres}},\ }\bibfield  {title}
  {\bibinfo {title} {2000-{{Times Repeated Imaging}} of {{Strontium Atoms}} in
  {{Clock-Magic Tweezer Arrays}}},\ }\href
  {https://doi.org/10.1103/PhysRevLett.122.173201} {\bibfield  {journal}
  {\bibinfo  {journal} {Phys. Rev. Lett.}\ }\textbf {\bibinfo {volume} {122}},\
  \bibinfo {pages} {173201} (\bibinfo {year} {2019})}\BibitemShut {NoStop}%
\bibitem [{\citenamefont {Chalopin}\ \emph
  {et~al.}(2026{\natexlab{a}})\citenamefont {Chalopin}, \citenamefont
  {Ferrier-Barbut}, \citenamefont {Lahaye}, \citenamefont {Browaeys},\ and\
  \citenamefont {Cl\'ement}}]{ChalopinCRAS2026}%
  \BibitemOpen
  \bibfield  {author} {\bibinfo {author} {\bibfnamefont {T.}~\bibnamefont
  {Chalopin}}, \bibinfo {author} {\bibfnamefont {I.}~\bibnamefont
  {Ferrier-Barbut}}, \bibinfo {author} {\bibfnamefont {T.}~\bibnamefont
  {Lahaye}}, \bibinfo {author} {\bibfnamefont {A.}~\bibnamefont {Browaeys}},\
  and\ \bibinfo {author} {\bibfnamefont {D.}~\bibnamefont {Cl\'ement}},\
  }\bibfield  {title} {\bibinfo {title} {Connected correlations in cold atom
  experiments},\ }\href {https://doi.org/10.5802/crphys.274} {\bibfield
  {journal} {\bibinfo  {journal} {C. R. Phys.}\ }\textbf {\bibinfo {volume}
  {27}},\ \bibinfo {pages} {65} (\bibinfo {year}
  {2026}{\natexlab{a}})}\BibitemShut {NoStop}%
\bibitem [{\citenamefont {Chalopin}\ \emph
  {et~al.}(2026{\natexlab{b}})\citenamefont {Chalopin}, \citenamefont
  {Bojović}, \citenamefont {Wang}, \citenamefont {Franz}, \citenamefont
  {Sinha}, \citenamefont {Wang}, \citenamefont {Bourgund}, \citenamefont
  {Obermeyer}, \citenamefont {Grusdt}, \citenamefont {Bohrdt}, \citenamefont
  {Pollet}, \citenamefont {Wietek}, \citenamefont {Georges}, \citenamefont
  {Hilker},\ and\ \citenamefont {Bloch}}]{Chalopin2024}%
  \BibitemOpen
  \bibfield  {author} {\bibinfo {author} {\bibfnamefont {T.}~\bibnamefont
  {Chalopin}}, \bibinfo {author} {\bibfnamefont {P.}~\bibnamefont {Bojović}},
  \bibinfo {author} {\bibfnamefont {S.}~\bibnamefont {Wang}}, \bibinfo {author}
  {\bibfnamefont {T.}~\bibnamefont {Franz}}, \bibinfo {author} {\bibfnamefont
  {A.}~\bibnamefont {Sinha}}, \bibinfo {author} {\bibfnamefont
  {Z.}~\bibnamefont {Wang}}, \bibinfo {author} {\bibfnamefont {D.}~\bibnamefont
  {Bourgund}}, \bibinfo {author} {\bibfnamefont {J.}~\bibnamefont {Obermeyer}},
  \bibinfo {author} {\bibfnamefont {F.}~\bibnamefont {Grusdt}}, \bibinfo
  {author} {\bibfnamefont {A.}~\bibnamefont {Bohrdt}}, \bibinfo {author}
  {\bibfnamefont {L.}~\bibnamefont {Pollet}}, \bibinfo {author} {\bibfnamefont
  {A.}~\bibnamefont {Wietek}}, \bibinfo {author} {\bibfnamefont
  {A.}~\bibnamefont {Georges}}, \bibinfo {author} {\bibfnamefont
  {T.}~\bibnamefont {Hilker}},\ and\ \bibinfo {author} {\bibfnamefont
  {I.}~\bibnamefont {Bloch}},\ }\bibfield  {title} {\bibinfo {title}
  {Observation of emergent scaling of spin–charge correlations at the onset
  of the pseudogap},\ }\href {https://doi.org/10.1073/pnas.2525539123}
  {\bibfield  {journal} {\bibinfo  {journal} {Proc. Natl. Acad. Sci. U.S.A.}\
  }\textbf {\bibinfo {volume} {123}},\ \bibinfo {pages} {e2525539123} (\bibinfo
  {year} {2026}{\natexlab{b}})}\BibitemShut {NoStop}%
\bibitem [{\citenamefont {Hartke}\ \emph {et~al.}()\citenamefont {Hartke},
  \citenamefont {Oreg}, \citenamefont {Feng}, \citenamefont {Turnbaugh},
  \citenamefont {Hertkorn}, \citenamefont {He}, \citenamefont {Jia},
  \citenamefont {Khatami}, \citenamefont {Zhang},\ and\ \citenamefont
  {Zwierlein}}]{Hartke2025}%
  \BibitemOpen
  \bibfield  {author} {\bibinfo {author} {\bibfnamefont {T.}~\bibnamefont
  {Hartke}}, \bibinfo {author} {\bibfnamefont {B.}~\bibnamefont {Oreg}},
  \bibinfo {author} {\bibfnamefont {C.}~\bibnamefont {Feng}}, \bibinfo {author}
  {\bibfnamefont {C.}~\bibnamefont {Turnbaugh}}, \bibinfo {author}
  {\bibfnamefont {J.}~\bibnamefont {Hertkorn}}, \bibinfo {author}
  {\bibfnamefont {Y.-Y.}\ \bibnamefont {He}}, \bibinfo {author} {\bibfnamefont
  {N.}~\bibnamefont {Jia}}, \bibinfo {author} {\bibfnamefont {E.}~\bibnamefont
  {Khatami}}, \bibinfo {author} {\bibfnamefont {S.}~\bibnamefont {Zhang}},\
  and\ \bibinfo {author} {\bibfnamefont {M.}~\bibnamefont {Zwierlein}},\
  }\bibfield  {title} {\bibinfo {title} {Competition of fermion pairing,
  magnetism, and charge order in the spin-doped attractive {Hubbard} gas},\
  }\href {https://doi.org/10.48550/arXiv.2511.10605} {\bibinfo  {journal}
  {arXiv:2511.10605}\ }\BibitemShut {NoStop}%
\bibitem [{\citenamefont {Jain}\ \emph {et~al.}(2026)\citenamefont {Jain},
  \citenamefont {Zhang}, \citenamefont {Culemann},\ and\ \citenamefont
  {Preiss}}]{jain2025}%
  \BibitemOpen
\bibfield  {journal} {  }\bibfield  {author} {\bibinfo {author} {\bibfnamefont
  {N.}~\bibnamefont {Jain}}, \bibinfo {author} {\bibfnamefont {J.}~\bibnamefont
  {Zhang}}, \bibinfo {author} {\bibfnamefont {M.}~\bibnamefont {Culemann}},\
  and\ \bibinfo {author} {\bibfnamefont {P.~M.}\ \bibnamefont {Preiss}},\
  }\bibfield  {title} {\bibinfo {title} {Programmable {A}ssembly of {G}round
  {S}tate {F}ermionic {T}weezer {A}rrays},\ }\href
  {https://doi.org/10.1103/fsmh-dz71} {\bibfield  {journal} {\bibinfo
  {journal} {Phys. Rev. Lett.}\ }\textbf {\bibinfo {volume} {136}},\ \bibinfo
  {pages} {223402} (\bibinfo {year} {2026})}\BibitemShut {NoStop}%
\bibitem [{\citenamefont {Hammel}\ \emph {et~al.}()\citenamefont {Hammel},
  \citenamefont {Kaiser}, \citenamefont {Dux}, \citenamefont {Weidemüller},\
  and\ \citenamefont {Jochim}}]{hammel2025}%
  \BibitemOpen
  \bibfield  {author} {\bibinfo {author} {\bibfnamefont {T.}~\bibnamefont
  {Hammel}}, \bibinfo {author} {\bibfnamefont {M.}~\bibnamefont {Kaiser}},
  \bibinfo {author} {\bibfnamefont {D.}~\bibnamefont {Dux}}, \bibinfo {author}
  {\bibfnamefont {M.}~\bibnamefont {Weidemüller}},\ and\ \bibinfo {author}
  {\bibfnamefont {S.}~\bibnamefont {Jochim}},\ }\bibfield  {title} {\bibinfo
  {title} {Atom and spin resolved imaging in a single shot},\ }\href
  {https://arxiv.org/abs/2512.09865} {\bibinfo  {journal} {arXiv:2512.09865}\
  }\BibitemShut {NoStop}%
\bibitem [{\citenamefont {Manmana}\ \emph {et~al.}(2011)\citenamefont
  {Manmana}, \citenamefont {Hazzard}, \citenamefont {Chen}, \citenamefont
  {Feiguin},\ and\ \citenamefont {Rey}}]{ManmanaPRA2011}%
  \BibitemOpen
\bibfield  {journal} {  }\bibfield  {author} {\bibinfo {author} {\bibfnamefont
  {S.~R.}\ \bibnamefont {Manmana}}, \bibinfo {author} {\bibfnamefont
  {K.~R.~A.}\ \bibnamefont {Hazzard}}, \bibinfo {author} {\bibfnamefont
  {G.}~\bibnamefont {Chen}}, \bibinfo {author} {\bibfnamefont {A.~E.}\
  \bibnamefont {Feiguin}},\ and\ \bibinfo {author} {\bibfnamefont {A.~M.}\
  \bibnamefont {Rey}},\ }\bibfield  {title} {\bibinfo {title}
  {$\mathrm{SU}({N})$ magnetism in chains of ultracold alkaline-earth-metal
  atoms: Mott transitions and quantum correlations},\ }\href
  {https://doi.org/10.1103/PhysRevA.84.043601} {\bibfield  {journal} {\bibinfo
  {journal} {Phys. Rev. A}\ }\textbf {\bibinfo {volume} {84}},\ \bibinfo
  {pages} {043601} (\bibinfo {year} {2011})}\BibitemShut {NoStop}%
\bibitem [{\citenamefont {Abdel~Karim}\ \emph {et~al.}(2025)\citenamefont
  {Abdel~Karim}, \citenamefont {Muzi~Falconi}, \citenamefont {Panza},
  \citenamefont {Liu},\ and\ \citenamefont {Scazza}}]{Abdelkarim2025}%
  \BibitemOpen
  \bibfield  {author} {\bibinfo {author} {\bibfnamefont {O.}~\bibnamefont
  {Abdel~Karim}}, \bibinfo {author} {\bibfnamefont {A.}~\bibnamefont
  {Muzi~Falconi}}, \bibinfo {author} {\bibfnamefont {R.}~\bibnamefont {Panza}},
  \bibinfo {author} {\bibfnamefont {W.}~\bibnamefont {Liu}},\ and\ \bibinfo
  {author} {\bibfnamefont {F.}~\bibnamefont {Scazza}},\ }\bibfield  {title}
  {\bibinfo {title} {Single-atom imaging of $^{173}${Yb} in optical tweezers
  loaded by a five-beam magneto-optical trap},\ }\href
  {https://doi.org/10.1088/2058-9565/adf7cf} {\bibfield  {journal} {\bibinfo
  {journal} {Quantum Sci. Technol.}\ }\textbf {\bibinfo {volume} {10}},\
  \bibinfo {pages} {045019} (\bibinfo {year} {2025})}\BibitemShut {NoStop}%
\bibitem [{\citenamefont {Ibarra-Garc\'{\i}a-Padilla}\ \emph
  {et~al.}(2023)\citenamefont {Ibarra-Garc\'{\i}a-Padilla}, \citenamefont
  {Feng}, \citenamefont {Pasqualetti}, \citenamefont {F\"olling}, \citenamefont
  {Scalettar}, \citenamefont {Khatami},\ and\ \citenamefont
  {Hazzard}}]{Ibarra_metalinsulator_2023}%
  \BibitemOpen
  \bibfield  {author} {\bibinfo {author} {\bibfnamefont {E.}~\bibnamefont
  {Ibarra-Garc\'{\i}a-Padilla}}, \bibinfo {author} {\bibfnamefont
  {C.}~\bibnamefont {Feng}}, \bibinfo {author} {\bibfnamefont {G.}~\bibnamefont
  {Pasqualetti}}, \bibinfo {author} {\bibfnamefont {S.}~\bibnamefont
  {F\"olling}}, \bibinfo {author} {\bibfnamefont {R.~T.}\ \bibnamefont
  {Scalettar}}, \bibinfo {author} {\bibfnamefont {E.}~\bibnamefont {Khatami}},\
  and\ \bibinfo {author} {\bibfnamefont {K.~R.~A.}\ \bibnamefont {Hazzard}},\
  }\bibfield  {title} {\bibinfo {title} {Metal-insulator transition and
  magnetism of {SU}(3) fermions in the square lattice},\ }\href
  {https://doi.org/10.1103/PhysRevA.108.053312} {\bibfield  {journal} {\bibinfo
   {journal} {Phys. Rev. A}\ }\textbf {\bibinfo {volume} {108}},\ \bibinfo
  {pages} {053312} (\bibinfo {year} {2023})}\BibitemShut {NoStop}%
\bibitem [{\citenamefont {D\"orscher}\ \emph {et~al.}(2018)\citenamefont
  {D\"orscher}, \citenamefont {Schwarz}, \citenamefont {Al-Masoudi},
  \citenamefont {Falke}, \citenamefont {Sterr},\ and\ \citenamefont
  {Lisdat}}]{Dorscher_2018}%
  \BibitemOpen
  \bibfield  {author} {\bibinfo {author} {\bibfnamefont {S.}~\bibnamefont
  {D\"orscher}}, \bibinfo {author} {\bibfnamefont {R.}~\bibnamefont {Schwarz}},
  \bibinfo {author} {\bibfnamefont {A.}~\bibnamefont {Al-Masoudi}}, \bibinfo
  {author} {\bibfnamefont {S.}~\bibnamefont {Falke}}, \bibinfo {author}
  {\bibfnamefont {U.}~\bibnamefont {Sterr}},\ and\ \bibinfo {author}
  {\bibfnamefont {C.}~\bibnamefont {Lisdat}},\ }\bibfield  {title} {\bibinfo
  {title} {Lattice-induced photon scattering in an optical lattice clock},\
  }\href {https://doi.org/10.1103/PhysRevA.97.063419} {\bibfield  {journal}
  {\bibinfo  {journal} {Phys. Rev. A}\ }\textbf {\bibinfo {volume} {97}},\
  \bibinfo {pages} {063419} (\bibinfo {year} {2018})}\BibitemShut {NoStop}%
\bibitem [{\citenamefont {Olschewski}(1972)}]{olschewski_messung_1972}%
  \BibitemOpen
  \bibfield  {author} {\bibinfo {author} {\bibfnamefont {L.}~\bibnamefont
  {Olschewski}},\ }\bibfield  {title} {\bibinfo {title} {Messung der
  magnetischen {{Kerndipolmomente}} an freien
  $^{43}${Ca}-,$^{87}${Sr}-,$^{135}${Ba}-,$^{137}${Ba}-,$^{171}${Yb}- und
  $^{173}${Yb}- {{Atomen}} mit optischem {{Pumpen}}},\ }\href
  {https://doi.org/10.1007/BF01400226} {\bibfield  {journal} {\bibinfo
  {journal} {Z. Phys.}\ }\textbf {\bibinfo {volume} {249}},\ \bibinfo {pages}
  {205} (\bibinfo {year} {1972})}\BibitemShut {NoStop}%
\bibitem [{\citenamefont {Thekkeppatt}\ \emph {et~al.}(2025)\citenamefont
  {Thekkeppatt}, \citenamefont {Digvijay}, \citenamefont {Urech}, \citenamefont
  {Schreck},\ and\ \citenamefont {van Druten}}]{thekkeppatt2025}%
  \BibitemOpen
  \bibfield  {author} {\bibinfo {author} {\bibfnamefont {P.}~\bibnamefont
  {Thekkeppatt}}, \bibinfo {author} {\bibnamefont {Digvijay}}, \bibinfo
  {author} {\bibfnamefont {A.}~\bibnamefont {Urech}}, \bibinfo {author}
  {\bibfnamefont {F.}~\bibnamefont {Schreck}},\ and\ \bibinfo {author}
  {\bibfnamefont {K.}~\bibnamefont {van Druten}},\ }\bibfield  {title}
  {\bibinfo {title} {Measurement of the $g$ factor of ground-state
  $^{87}\mathrm{Sr}$ at the parts-per-million level using co-trapped ultracold
  atoms},\ }\href {https://doi.org/10.1103/cjks-9hlp} {\bibfield  {journal}
  {\bibinfo  {journal} {Phys. Rev. Lett.}\ }\textbf {\bibinfo {volume} {135}},\
  \bibinfo {pages} {193001} (\bibinfo {year} {2025})}\BibitemShut {NoStop}%
\bibitem [{\citenamefont {Bonnes}\ \emph {et~al.}(2012)\citenamefont {Bonnes},
  \citenamefont {Hazzard}, \citenamefont {Manmana}, \citenamefont {Rey},\ and\
  \citenamefont {Wessel}}]{BonnesPRL2012}%
  \BibitemOpen
  \bibfield  {author} {\bibinfo {author} {\bibfnamefont {L.}~\bibnamefont
  {Bonnes}}, \bibinfo {author} {\bibfnamefont {K.~R.~A.}\ \bibnamefont
  {Hazzard}}, \bibinfo {author} {\bibfnamefont {S.~R.}\ \bibnamefont
  {Manmana}}, \bibinfo {author} {\bibfnamefont {A.~M.}\ \bibnamefont {Rey}},\
  and\ \bibinfo {author} {\bibfnamefont {S.}~\bibnamefont {Wessel}},\
  }\bibfield  {title} {\bibinfo {title} {Adiabatic loading of one-dimensional
  $\mathrm{SU}({N})$ alkaline-earth-atom fermions in optical lattices},\ }\href
  {https://doi.org/10.1103/PhysRevLett.109.205305} {\bibfield  {journal}
  {\bibinfo  {journal} {Phys. Rev. Lett.}\ }\textbf {\bibinfo {volume} {109}},\
  \bibinfo {pages} {205305} (\bibinfo {year} {2012})}\BibitemShut {NoStop}%
\bibitem [{\citenamefont {Celi}\ \emph {et~al.}(2014)\citenamefont {Celi},
  \citenamefont {Massignan}, \citenamefont {Ruseckas}, \citenamefont {Goldman},
  \citenamefont {Spielman}, \citenamefont {Juzeli\ifmmode~\bar{u}\else
  \={u}\fi{}nas},\ and\ \citenamefont {Lewenstein}}]{Celi2014}%
  \BibitemOpen
  \bibfield  {author} {\bibinfo {author} {\bibfnamefont {A.}~\bibnamefont
  {Celi}}, \bibinfo {author} {\bibfnamefont {P.}~\bibnamefont {Massignan}},
  \bibinfo {author} {\bibfnamefont {J.}~\bibnamefont {Ruseckas}}, \bibinfo
  {author} {\bibfnamefont {N.}~\bibnamefont {Goldman}}, \bibinfo {author}
  {\bibfnamefont {I.~B.}\ \bibnamefont {Spielman}}, \bibinfo {author}
  {\bibfnamefont {G.}~\bibnamefont {Juzeli\ifmmode~\bar{u}\else
  \={u}\fi{}nas}},\ and\ \bibinfo {author} {\bibfnamefont {M.}~\bibnamefont
  {Lewenstein}},\ }\bibfield  {title} {\bibinfo {title} {Synthetic gauge fields
  in synthetic dimensions},\ }\href
  {https://doi.org/10.1103/PhysRevLett.112.043001} {\bibfield  {journal}
  {\bibinfo  {journal} {Phys. Rev. Lett.}\ }\textbf {\bibinfo {volume} {112}},\
  \bibinfo {pages} {043001} (\bibinfo {year} {2014})}\BibitemShut {NoStop}%
\bibitem [{\citenamefont {Zhou}\ \emph {et~al.}(2023)\citenamefont {Zhou},
  \citenamefont {Cappellini}, \citenamefont {Tusi}, \citenamefont {Franchi},
  \citenamefont {Parravicini}, \citenamefont {Repellin}, \citenamefont
  {Greschner}, \citenamefont {Inguscio}, \citenamefont {Giamarchi},
  \citenamefont {Filippone}, \citenamefont {Catani},\ and\ \citenamefont
  {Fallani}}]{zhou_observation_2023}%
  \BibitemOpen
  \bibfield  {author} {\bibinfo {author} {\bibfnamefont {T.-W.}\ \bibnamefont
  {Zhou}}, \bibinfo {author} {\bibfnamefont {G.}~\bibnamefont {Cappellini}},
  \bibinfo {author} {\bibfnamefont {D.}~\bibnamefont {Tusi}}, \bibinfo {author}
  {\bibfnamefont {L.}~\bibnamefont {Franchi}}, \bibinfo {author} {\bibfnamefont
  {J.}~\bibnamefont {Parravicini}}, \bibinfo {author} {\bibfnamefont
  {C.}~\bibnamefont {Repellin}}, \bibinfo {author} {\bibfnamefont
  {S.}~\bibnamefont {Greschner}}, \bibinfo {author} {\bibfnamefont
  {M.}~\bibnamefont {Inguscio}}, \bibinfo {author} {\bibfnamefont
  {T.}~\bibnamefont {Giamarchi}}, \bibinfo {author} {\bibfnamefont
  {M.}~\bibnamefont {Filippone}}, \bibinfo {author} {\bibfnamefont
  {J.}~\bibnamefont {Catani}},\ and\ \bibinfo {author} {\bibfnamefont
  {L.}~\bibnamefont {Fallani}},\ }\bibfield  {title} {\bibinfo {title}
  {Observation of universal {{Hall}} response in strongly interacting
  {{Fermions}}},\ }\href {https://doi.org/10.1126/science.add1969} {\bibfield
  {journal} {\bibinfo  {journal} {Science}\ }\textbf {\bibinfo {volume}
  {381}},\ \bibinfo {pages} {427} (\bibinfo {year} {2023})}\BibitemShut
  {NoStop}%
\bibitem [{\citenamefont {Zhou}\ \emph {et~al.}(2025)\citenamefont {Zhou},
  \citenamefont {Beller}, \citenamefont {Masini}, \citenamefont {Parravicini},
  \citenamefont {Cappellini}, \citenamefont {Repellin}, \citenamefont
  {Giamarchi}, \citenamefont {Catani}, \citenamefont {Filippone},\ and\
  \citenamefont {Fallani}}]{Zhou2025_measuring_hall_quantum_simulator}%
  \BibitemOpen
  \bibfield  {author} {\bibinfo {author} {\bibfnamefont {T.}~\bibnamefont
  {Zhou}}, \bibinfo {author} {\bibfnamefont {T.}~\bibnamefont {Beller}},
  \bibinfo {author} {\bibfnamefont {G.}~\bibnamefont {Masini}}, \bibinfo
  {author} {\bibfnamefont {J.}~\bibnamefont {Parravicini}}, \bibinfo {author}
  {\bibfnamefont {G.}~\bibnamefont {Cappellini}}, \bibinfo {author}
  {\bibfnamefont {C.}~\bibnamefont {Repellin}}, \bibinfo {author}
  {\bibfnamefont {T.}~\bibnamefont {Giamarchi}}, \bibinfo {author}
  {\bibfnamefont {J.}~\bibnamefont {Catani}}, \bibinfo {author} {\bibfnamefont
  {M.}~\bibnamefont {Filippone}},\ and\ \bibinfo {author} {\bibfnamefont
  {L.}~\bibnamefont {Fallani}},\ }\bibfield  {title} {\bibinfo {title}
  {Measuring {H}all voltage and {H}all resistance in an atom-based quantum
  simulator},\ }\href {https://doi.org/10.1038/s41467-025-65083-6} {\bibfield
  {journal} {\bibinfo  {journal} {Nat. Commun.}\ }\textbf {\bibinfo {volume}
  {16}},\ \bibinfo {pages} {10247} (\bibinfo {year} {2025})}\BibitemShut
  {NoStop}%
\bibitem [{\citenamefont {Tarruell}\ \emph {et~al.}(2012)\citenamefont
  {Tarruell}, \citenamefont {Greif}, \citenamefont {Uehlinger}, \citenamefont
  {Jotzu},\ and\ \citenamefont {Esslinger}}]{tarruell_creating_2012}%
  \BibitemOpen
  \bibfield  {author} {\bibinfo {author} {\bibfnamefont {L.}~\bibnamefont
  {Tarruell}}, \bibinfo {author} {\bibfnamefont {D.}~\bibnamefont {Greif}},
  \bibinfo {author} {\bibfnamefont {T.}~\bibnamefont {Uehlinger}}, \bibinfo
  {author} {\bibfnamefont {G.}~\bibnamefont {Jotzu}},\ and\ \bibinfo {author}
  {\bibfnamefont {T.}~\bibnamefont {Esslinger}},\ }\bibfield  {title} {\bibinfo
  {title} {Creating, moving and merging {{Dirac}} points with a {{Fermi}} gas
  in a tunable honeycomb lattice},\ }\href
  {https://doi.org/10.1038/nature10871} {\bibfield  {journal} {\bibinfo
  {journal} {Nature}\ }\textbf {\bibinfo {volume} {483}},\ \bibinfo {pages}
  {302} (\bibinfo {year} {2012})}\BibitemShut {NoStop}%
\bibitem [{\citenamefont {Wei}\ \emph {et~al.}(2023)\citenamefont {Wei},
  \citenamefont {Adler}, \citenamefont {Srakaew}, \citenamefont {Agrawal},
  \citenamefont {Weckesser}, \citenamefont {Bloch},\ and\ \citenamefont
  {Zeiher}}]{wei_observation_2023}%
  \BibitemOpen
  \bibfield  {author} {\bibinfo {author} {\bibfnamefont {D.}~\bibnamefont
  {Wei}}, \bibinfo {author} {\bibfnamefont {D.}~\bibnamefont {Adler}}, \bibinfo
  {author} {\bibfnamefont {K.}~\bibnamefont {Srakaew}}, \bibinfo {author}
  {\bibfnamefont {S.}~\bibnamefont {Agrawal}}, \bibinfo {author} {\bibfnamefont
  {P.}~\bibnamefont {Weckesser}}, \bibinfo {author} {\bibfnamefont
  {I.}~\bibnamefont {Bloch}},\ and\ \bibinfo {author} {\bibfnamefont
  {J.}~\bibnamefont {Zeiher}},\ }\bibfield  {title} {\bibinfo {title}
  {Observation of {{Brane Parity Order}} in {{Programmable Optical
  Lattices}}},\ }\href {https://doi.org/10.1103/PhysRevX.13.021042} {\bibfield
  {journal} {\bibinfo  {journal} {Phys. Rev. X}\ }\textbf {\bibinfo {volume}
  {13}},\ \bibinfo {pages} {021042} (\bibinfo {year} {2023})}\BibitemShut
  {NoStop}%
\bibitem [{\citenamefont {Ma}\ \emph {et~al.}(2025)\citenamefont {Ma},
  \citenamefont {Dolde}, \citenamefont {Zheng}, \citenamefont {Ganapathy},
  \citenamefont {Shtov}, \citenamefont {Chen}, \citenamefont {St\"oltzel},
  \citenamefont {Christensen},\ and\ \citenamefont {Kolkowitz}}]{ma2025}%
  \BibitemOpen
  \bibfield  {author} {\bibinfo {author} {\bibfnamefont {S.}~\bibnamefont
  {Ma}}, \bibinfo {author} {\bibfnamefont {J.}~\bibnamefont {Dolde}}, \bibinfo
  {author} {\bibfnamefont {X.}~\bibnamefont {Zheng}}, \bibinfo {author}
  {\bibfnamefont {D.}~\bibnamefont {Ganapathy}}, \bibinfo {author}
  {\bibfnamefont {A.}~\bibnamefont {Shtov}}, \bibinfo {author} {\bibfnamefont
  {J.}~\bibnamefont {Chen}}, \bibinfo {author} {\bibfnamefont {A.}~\bibnamefont
  {St\"oltzel}}, \bibinfo {author} {\bibfnamefont {B.~J.}\ \bibnamefont
  {Christensen}},\ and\ \bibinfo {author} {\bibfnamefont {S.}~\bibnamefont
  {Kolkowitz}},\ }\bibfield  {title} {\bibinfo {title} {Enhancing optical
  lattice clock coherence times with erasure conversion},\ }\href
  {https://doi.org/10.1103/2rqf-r9gw} {\bibfield  {journal} {\bibinfo
  {journal} {PRX Quantum}\ }\textbf {\bibinfo {volume} {6}},\ \bibinfo {pages}
  {040340} (\bibinfo {year} {2025})}\BibitemShut {NoStop}%
\bibitem [{\citenamefont {Plassmann}\ \emph {et~al.}()\citenamefont
  {Plassmann}, \citenamefont {Schaefer}, \citenamefont {Menashes},\ and\
  \citenamefont {Salomon}}]{plassmann_2026}%
  \BibitemOpen
  \bibfield  {author} {\bibinfo {author} {\bibfnamefont {T.}~\bibnamefont
  {Plassmann}}, \bibinfo {author} {\bibfnamefont {L.}~\bibnamefont {Schaefer}},
  \bibinfo {author} {\bibfnamefont {M.}~\bibnamefont {Menashes}},\ and\
  \bibinfo {author} {\bibfnamefont {G.}~\bibnamefont {Salomon}},\ }\bibfield
  {title} {\bibinfo {title} {Rapid state-resolved single-atom imaging of
  alkaline-earth fermions},\ }\href {https://arxiv.org/abs/2602.19876}
  {\bibinfo  {journal} {arXiv:2602.19876}\ }\BibitemShut {NoStop}%
\bibitem [{\citenamefont {Gas-Ferrer}\ \emph {et~al.}()\citenamefont
  {Gas-Ferrer}, \citenamefont {Rubio-Abadal}, \citenamefont {Buob}, \citenamefont {Bezzo}, \citenamefont {H{\"o}schele}, \ and\
  \citenamefont {Tarruell}}]{OnlineData}%
  \BibitemOpen
  \bibfield  {author} {\bibinfo {author} {\bibfnamefont {C.}~\bibnamefont
  {Gas-Ferrer}}, \bibinfo {author} {\bibfnamefont {A.}~\bibnamefont {Rubio-Abadal}},
  \bibinfo {author} {\bibfnamefont {S.}~\bibnamefont {Buob}}, \bibinfo {author} {\bibfnamefont {L.}~\bibnamefont {Bezzo}}, \bibinfo {author} {\bibfnamefont {J.}~\bibnamefont {H{\"o}schele}},\ and\
  \bibinfo {author} {\bibfnamefont {L.}~\bibnamefont {Tarruell}},\ }\bibfield
  {title} {\bibinfo {title} {Replication data for Spin-resolved microscopy of \textsuperscript{87}Sr SU(\textit{N}) Fermi-Hubbard systems},\ }\href {https://doi.org/10.34810/DATA3667}
  {\bibinfo  {journal} {https://doi.org/10.34810/DATA3667}\ }\BibitemShut {NoStop}%
\bibitem [{\citenamefont {H{\"o}schele}\ \emph {et~al.}(2023)\citenamefont
  {H{\"o}schele}, \citenamefont {Buob}, \citenamefont {{Rubio-Abadal}},
  \citenamefont {Makhalov},\ and\ \citenamefont
  {Tarruell}}]{hoschele_atomnumber_2023}%
  \BibitemOpen
\bibfield  {journal} {  }\bibfield  {author} {\bibinfo {author} {\bibfnamefont
  {J.}~\bibnamefont {H{\"o}schele}}, \bibinfo {author} {\bibfnamefont
  {S.}~\bibnamefont {Buob}}, \bibinfo {author} {\bibfnamefont {A.}~\bibnamefont
  {{Rubio-Abadal}}}, \bibinfo {author} {\bibfnamefont {V.}~\bibnamefont
  {Makhalov}},\ and\ \bibinfo {author} {\bibfnamefont {L.}~\bibnamefont
  {Tarruell}},\ }\bibfield  {title} {\bibinfo {title} {Atom-{{Number
  Enhancement}} by {{Shielding Atoms From Losses}} in {{Strontium
  Magneto-Optical Traps}}},\ }\href
  {https://doi.org/10.1103/PhysRevApplied.19.064011} {\bibfield  {journal}
  {\bibinfo  {journal} {Phys. Rev. Appl.}\ }\textbf {\bibinfo {volume} {19}},\
  \bibinfo {pages} {064011} (\bibinfo {year} {2023})}\BibitemShut {NoStop}%
\bibitem [{\citenamefont {Richardson}(1972)}]{richardson_bayesianbased_1972}%
  \BibitemOpen
  \bibfield  {author} {\bibinfo {author} {\bibfnamefont {W.~H.}\ \bibnamefont
  {Richardson}},\ }\bibfield  {title} {\bibinfo {title} {Bayesian-{{Based
  Iterative Method}} of {{Image Restoration}}},\ }\href
  {https://doi.org/10.1364/JOSA.62.000055} {\bibfield  {journal} {\bibinfo
  {journal} {J. Opt. Soc. Am.}\ }\textbf {\bibinfo {volume} {62}},\ \bibinfo
  {pages} {55} (\bibinfo {year} {1972})}\BibitemShut {NoStop}%
\bibitem [{\citenamefont {Cheneau}\ \emph {et~al.}(2025)\citenamefont
  {Cheneau}, \citenamefont {Journet}, \citenamefont {Boffety}, \citenamefont
  {Goudail}, \citenamefont {Kulcs\'ar},\ and\ \citenamefont
  {Trouv\'e-Peloux}}]{cheneau_reconstruction_2025}%
  \BibitemOpen
  \bibfield  {author} {\bibinfo {author} {\bibfnamefont {M.}~\bibnamefont
  {Cheneau}}, \bibinfo {author} {\bibfnamefont {R.}~\bibnamefont {Journet}},
  \bibinfo {author} {\bibfnamefont {M.}~\bibnamefont {Boffety}}, \bibinfo
  {author} {\bibfnamefont {F.}~\bibnamefont {Goudail}}, \bibinfo {author}
  {\bibfnamefont {C.}~\bibnamefont {Kulcs\'ar}},\ and\ \bibinfo {author}
  {\bibfnamefont {P.}~\bibnamefont {Trouv\'e-Peloux}},\ }\bibfield  {title}
  {\bibinfo {title} {Fast, accurate, and predictive method for atom detection
  in site-resolved images of microtrap arrays},\ }\href
  {https://doi.org/10.1103/ymkm-frhw} {\bibfield  {journal} {\bibinfo
  {journal} {Phys. Rev. Appl.}\ }\textbf {\bibinfo {volume} {24}},\ \bibinfo
  {pages} {064039} (\bibinfo {year} {2025})}\BibitemShut {NoStop}%
\end{thebibliography}
\end{document}